\documentclass{cernrep}

\usepackage{graphicx}
\def\beq{\begin{equation}}
\def\eeq{\end{equation}}
\def\bea{\begin{eqnarray}}
\def\beaa{\begin{eqnarray*}}
\def\eea{\end{eqnarray}}
\def\eeaa{\end{eqnarray*}}
\def\bq{\begin{quote}}
\def\eq{\end{quote}}
\def\gappeq{\mathrel{\rlap {\raise.5ex\hbox{$>$}}
{\lower.5ex\hbox{$\sim$}}}}
\def\lappeq{\mathrel{\rlap{\raise.5ex\hbox{$<$}}
{\lower.5ex\hbox{$\sim$}}}}

\def\sm{Standard Model}

\def\MPL{{\it Mod.Phys.Lett.} }

\def\NP{{\it Nucl.Phys.} }
\def\PL{{\it Phys.Lett.} }
\def\PR{{\it Phys.Rev.} }
\def\PRL{{\it Phys.Rev.Lett.} }

\begin{document}
\pagestyle{plain}

\title{BEYOND THE STANDARD MODEL FOR HILLWALKERS}

\author{John ELLIS}

\institute{Theoretical Physics Division, CERN\\
CH - 1211 Geneva 23}

\maketitle

\begin{flushright}
CERN-TH/98-329 \\
hep-ph/9812235 \\
\end{flushright}

\begin{abstract}
In the first lecture, the Standard Model is reviewed, with the aim of
seeing how its
successes constrain possible extensions, the significance of the
apparently low Higgs mass indicated by precision electroweak experiments
is discussed, and {\it defects of the Standard Model} are examined.
The second lecture includes a general discussion of the electroweak
vacuum and an {\it introduction to supersymmetry}, motivated by the gauge
hierarchy problem. In the third lecture, the {\it phenomenology of
supersymmetric models} is discussed in more detail, with emphasis on
the information provided by LEP data. The fourth lecture introduces
{\it Grand Unified Theories}, with emphases on general principles and on
neutrino masses and mixing. Finally, the last lecture contains short
discussions of some {\it further topics}, including supersymmetry
breaking,
gauge-mediated messenger models, supergravity, strings and $M$
phenomenology.
\end{abstract}

\section{GETTING MOTIVATED}
\label{sec1:GG}

There have been many reviews of different subjects in particle
physics `for pedestrians'. At this school many of us had the fun
experience of walking in the Scottish hills, which is more strenuous
than a stroll across the Old Course at St Andrews, though less
dangerous than mountain climbing in the Alps. The spirit of these
lectures is similar: an invigorating introduction to modern
phenomenological trends, but not too close to the theoretical
precipices.

\subsection{Recap of the Standard Model}
\label{subsec:ROTSM}

The \sm~ continues to survive all experimental tests at accelerators.
However, despite its tremendous successes, no-one finds the \sm~\cite{GSW}
satisfactory, and many present and future experiments are being aimed at
some of the Big Issues raised by the \sm ~: is there a Higgs boson? is
there supersymmetry? why are there only six quarks and six leptons? what
is the origin of flavour mixing and CP violation? can the different
interactions be unified? does the proton decay? are there neutrino
masses? For the first time, clear evidence for new physics beyond the \sm~
may be emerging from non-accelerator neutrino physics~\cite{SK}.
Nevertheless the
\sm ~remains the rock on which our quest for new physics must be built, so
let us start by reviewing its basic features and examine whether its
successes offer any hint of the direction in which to search for new
physics.

We first review the electroweak gauge bosons and the Higgs mechanism of
spontaneous symmetry breaking by which we believe they acquire
masses~\cite{SSB}.
The gauge bosons are described by the action
\beq
{\cal L} = -\frac{1}{4}~G^i_{\mu\nu} G^{i\mu\nu} - {1\over 4}
F_{\mu\nu}F^{\mu\nu}
\label{oneone}
\eeq
where $G^i_{\mu\nu} \equiv \partial_\mu W^i_\nu - \partial_\nu W^i_\mu +
ig \epsilon_{ijk}W^j_\mu W^k_\nu$ is the field strength for the $SU(2)$
gauge boson $W^i_\mu$, and $F_{\mu\nu} \equiv \partial_\mu W^i_\nu - 
\partial_\nu W^i_\mu$ is the field strength for the $U(1)$ gauge boson
$B_\mu$. The action (\ref{oneone}) contains bilinear terms that yield the
boson propagators, and also trilinear and quartic gauge-boson interactions. The
gauge bosons couple to quarks and leptons via
\beq
{\cal L}_F = -\sum_f i~~\left[ \bar f_L \gamma^\mu D_\mu f_L +
\bar f_R \gamma^\mu D_\mu f_R \right]
\label{onetwo}
\eeq
where the $D_\mu$ are covariant derivatives:
\beq
D_\mu \equiv \partial_\mu - i~ g ~\sigma_i~ W^i_\mu - i~g^\prime~ Y~ B_\mu
\label{onethree}
\eeq
The $SU(2)$ piece appears only for the left-handed fermions $f_L$, which are isospin
doublets, while the right-handed fermions $f_R$ are isospin singlets, and hence couple only
to the
$U(1)$ gauge boson $B_\mu$, via hypercharges $Y$.

The origin of all the masses in the \sm~ is an isodoublet scalar Higgs
field, whose kinetic term in the action is
\beq
{\cal L}_\phi = -\vert D_\mu \phi\vert^2
\label{onefour}
\eeq
and which has the magic potential:
\beq
{\cal L}_V = -V(\phi ) : V(\phi ) = -\mu^2\phi^{\dagger}\phi +
{\lambda \over 2} (\phi^{\dagger}\phi)^2
\label{onefive}
\eeq
Because of the negative sign for the quadratic term in (\ref{onefive}),
the symmetric solution $<0\vert\phi\vert 0> = 0$ is unstable, and if
$\lambda > 0$ the favoured solution has a non-zero vacuum expectation
value which we may write in the form:
\beq
<0\vert\phi\vert 0> = <0\vert\phi^{\dagger}\vert 0> = v\left(\matrix{0\cr
{1\over
\sqrt{2}}}\right) : v^2 = {\mu^2\over 2\lambda}
\label{onesix}
\eeq
corresponding to spontaneous breakdown of the electroweak gauge symmetry.

Expanding around the vacuum: $\phi = <0\vert\phi\vert 0> + \,\hat\phi$,
the
kinetic term (\ref{onefour}) for the Higgs field yields mass terms for
the gauge bosons:
\beq
{\cal L}_\phi \ni -{g^2v^2\over 2}~~W^+_\mu ~W^{\mu -} - g^{\prime 2}
~{v^2\over 2}~B_\mu~B^\mu + g~g^\prime  v^2~B_\mu~W^{\mu 3} -
g^2~{v^2 \over 2}~W^3_\mu~W^{\mu 3}
\label{oneseven}
\eeq
There are also bilinear derivative couplings of the gauge bosons to the
massless Goldstone bosons $\eta$, e.g., in the charged-boson sector we
have
\beq
- \partial_\mu ~\eta^+~\partial_\mu~\eta^- + ~\left({ig v\over 2} 
\partial_\mu~\eta^+~W^{\mu -} + h.c.\right)
\label{oneseven1}
\eeq
Combining these with the first term in (\ref{oneseven}), we see a
quadratic mass term for the combination
\beq
W^+_\mu - 2 i~~{\partial_\mu ~\eta^+\over gv}
\label{oneseven2}
\eeq
of charged bosons. This clearly gives a mass to the $W^\pm$ bosons:
\beq
m_{W^\pm} = {gv\over 2}
\label{oneeight}
\eeq
whilst the neutral gauge bosons $(W^3_\mu , B_\mu)$ have a 2$\times$2
mass-squared matrix:
\beq
\left(\matrix{
{g^2\over 2} & {-gg^\prime\over 2} \cr \cr
{-gg^\prime\over 2} & {g^{\prime 2}\over 2}}\right) v^2
\label{onenine}
\eeq
This is easily diagonalized to yield the mass eigenstates:
\beq
Z_\mu = {gW^3_\mu - g^\prime B_\mu\over \sqrt{g^2+g^{\prime 2}}}~ :~~ m_Z
= {1\over 2} \sqrt{g^2+g^{\prime 2}} v~;~~
A_\mu = {g^\prime W^3_\mu + g B_\mu\over \sqrt{g^2+g^{\prime 2}}} ~:~~
m_A = 0
\label{oneten}
\eeq
that we identify with the $Z$ and $\gamma$, respectively. It is useful to
introduce the electroweak mixing angle $\theta_W$ defined by
\beq
\sin\theta_W = {g^\prime\over \sqrt{g^2+g^{\prime 2}}} 
\label{oneeleven}
\eeq
in terms of the $SU(2)$ gauge coupling $g$ and $g^\prime$. Many other
quantities can be expressed in terms of $\sin\theta_W$ (\ref{oneeleven}):
for example, $m^2_W/ m^2_Z = \cos^2\theta_W$. The charged-current
interactions are of the current-current form:
\beq
{1 \over 4}{\cal L}_{cc} = {G_F\over
\sqrt{2}}~~J^+_\mu~J^{-\mu}~~:~~{G_F\over\sqrt{2}}
\equiv {g^2\over 8 m^2_W}
\label{onetwelve}
\eeq
as are the neutral-current interactions:
\beq
{1 \over 4} {\cal L}_{NC} = {G^{NC}_F\over \sqrt{2}}~~J^0_\mu~J^{\mu
0}~~:~~
J^0_\mu \equiv J^3_\mu - \sin^2\theta_W~J^{em}_\mu~,~~G^{NC}_F \equiv
{g^2+g^{\prime 2}\over 8 m^2_Z}
\label{onethirteen}
\eeq
The ratio of neutral- and charged-current interaction strengths is often
expressed as
\beq
\rho = {G^{NC}_F\over G_F} = {m^2_W\over m^2_Z \cos^2\theta_W}
\label{oneforteen}
\eeq
which takes the value unity in the \sm ~ with only Higgs
doublets~\cite{RV}, as assumed
here. However, this and the other tree-level relations given above are
modified by quantum corrections (loop effects), as we discuss later.

Figures 1 and 2 compile the most important precision
electroweak measurements~\cite{LEPEWWG}. It is striking that $m_Z$ (Fig.
1) is now known
more accurately than the muon decay constant. Precision measurements of
$Z$ decays also restrict possible extensions of the \sm. For example, the
number of effective equivalent light-neutrino species is measured very
accurately:
\beq
N_\nu = 2.994 \pm 0.011
\label{oneforteenone}
\eeq
I had always hoped that $N_\nu$ might turn out to be non-integer: $N_\nu
= \pi$ would have been good, and $N_\nu = e$ would have been even better, but this was not
to be! The constraint (\ref{oneforteenone}) is also important for possible physics
beyond the \sm, such as supersymmetry as we discuss later. The
measurement  (\ref{oneforteenone}) implies, by extension, that there can
only be three charged leptons and hence, in order to keep triangle
anomalies cancelled, no more quarks~\cite{BIM}. Hence a fourth
conventional matter
generation is not a possible extension of the \sm.

\begin{figure}
\centerline{\includegraphics[height=3in]{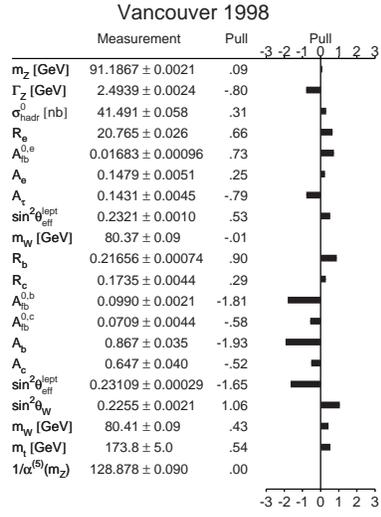}}
\caption[]{Precision electroweak measurements and the pulls they exert in a
global fit~\cite{LEPEWWG}.}
\end{figure}

\begin{figure}
\centerline{\includegraphics[height=3in]{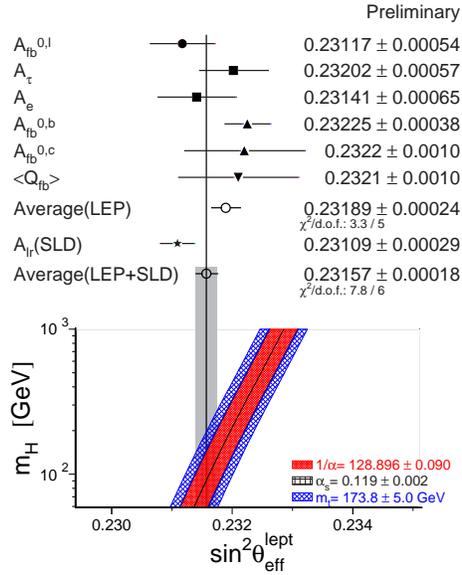}}
\caption[]{Precision determinations of $\sin^2\theta_W$~\cite{LEPEWWG}.}
\end{figure}

There are by now many precision meaaurements of $\sin^2\theta_W$ (Fig.
2): this is a free parameter in the \sm, whose value~\cite{GQW} is a
suggestive
hint for grand unification~\cite{GG} and supersymmetry~\cite{DRW}, as we
discuss later.
Notice also in Fig. 2 that consistency of the data seems to prefer a
relatively low value for the Higgs mass, which is another possible
suggestion of supersymmetry, as we also discuss later.

The previous field-theoretical discussion of the Higgs mechanism can be
rephrased in more physical language. It is well known that a massless
vector boson such as the photon $\gamma$ or gluon $g$ has just two
polarization states: $\lambda = \pm 1$. However, a massive vector boson
such as the $\rho$ has three polarization states: $\lambda = 0, \pm 1$.
This third polarization state is provided by a spin-0 field as seen in 
 (\ref{oneseven2}). In order to make $m_{W^\pm,Z^0} \not= 0$, this
should have non-zero electroweak isospin $I \not= 0$,  and the simplest
possibility is a complex  isodoublet $(\phi^+,\phi^0)$, as assumed above.
This has four degrees of freedom, three of which are eaten by the $W^\pm$
amd $Z^0$ as their third polarization states, leaving us with one
physical Higgs boson $H$. Once the vacuum expectation value $\vert <
0\vert\phi\vert 0 > \vert = v/ \sqrt{2}~:~~v = \mu /
\sqrt{2\lambda}$ is fixed, the mass of the remaining physical Higgs boson is given
by
\beq
m^2_H = 2\mu^2 = 4 \lambda v^2
\label{onefifteen}
\eeq
which is a free parameter in the \sm.

The necessity for such a physical
Higgs boson may be further demonstrated by considering the scattering amplitude for
$\bar ff \rightarrow W^+W^-$~\cite{unitarity}. By unitarity this
contributes to elastic
$\bar ff \rightarrow\bar ff$ scttering at the one-loop level. This
contribution would be divergent and unrenormalizable in the absence of a
direct-channel spin-0 contribution to cancel mass-dependent contributions
from the established $\nu, \gamma$ and $Z^0$ exchanges. If these spin-0
contributions to $\bar ff\rightarrow W^+W^-$  and analogously
$W^+W^-\rightarrow W^+W^-$ are due to a single Higgs boson, as  in the
\sm, its couplings to fermions and gauge bosons are completely determined:
\beq
g_{H\bar ff} = {g \over 2} {m_f \over m_W}~,~~ g_{HW^+W^-} =
g m_W~,~~g_{HZ^0Z^0} = g m_Z
\label{onesixteen}
\eeq
Thus the Higgs production and decay rates are completely fixed as
functions of the unknown mass $M_H$ (\ref{onefifteen})~\cite{EGN}. This
unitarity
argument actually requires that $m_H \leq$ 1 TeV in order to accomplish
its one-loop cancellation mission~\cite{V,LQT,DL}.

The search for the Higgs boson is one of the main objectives of the LEP~2
experimental programme. The dominant production mechanism is
$e^+e^-\rightarrow Z^0+H$~\cite{EGN,IK}, which has the tree-level cross
section~\cite{IK,LQT}
\beq
\sigma_{ZM} = {G^2_F m^4_Z\over
96\pi s}~~(1+(1-4\sin^2\theta_W)^2)~~\lambda^{1/2}~~
{\lambda + 12 m^2_Z/s\over (1-m^2_Z/s)^2}
\label{oneseventeen}
\eeq
the prefactor comes from the known $HZ^0Z^0$ vertex (\ref{onesixteen}),
and the phase-space factor
\beq
\lambda\equiv \left( 1-({m^2_H\over s}) - ({m^2_Z\over s})\right)^2 -
{4m^2_H m^2_Z\over s^2}
\label{oneeighteen}
\eeq
which gives us sensitivity to $m_H \lappeq E_{cm} - M_Z - $.
With the current LEP~2 running at 189 GeV, each individual LEP experiment
has established a lower limit $m_H \gappeq$ 96 GeV, and the four
experiments could probably be combined to yield 
$m_H \gappeq$ 98 GeV~\cite{LEPC}. The next two years of LEP~2 running at
energies $\sim 200$ GeV should enable the Higgs to be discovered if 
$m_H \lappeq$ 110 GeV, if as much luminosity is accumulated as in 1998. As
we see shortly, this covers the range of $m_H$ where the precision
electroweak data~\cite{LEPEWWG} indicate the highest probability density.
Hence, the
integrated probability that LEP~2 discovers the Higgs boson is not
negligible, though we must brace ourselves for the likelihood that it is
too heavy to be discovered at LEP.

\subsection{Interpretation of the Precision Electroweak Data}

The precision of the electroweak data shown in Figure 1 is so high -- of
order 0.1 \% in some cases -- that quantum corrections are crucial for
their interpretation~\cite{loops}. At the one-loop level, these include
vacuum-polarization, vector and box diagrams. The dominant contributions
from two- and higher-loop diagrams must also be taken into account. These
loop diagrams must be renormalized, and this is achieved by fixing three
quantities at  their physical values:
$m_Z = 91.1867 \pm 0.0021$ GeV, $\alpha^{-1}_{em} = 137.035 999
59(38)13)$,
$G_\mu = 1.166389(22)\times 10^{-5}$ GeV$^{-2}$. In the case of experiments
at the $Z^0$ peak, one needs to calculate the renormalization of
$\alpha_{em}$ over scales $m_e \lappeq Q \lappeq m_Z$ due to vacuum
polarization diagrams. The principal uncertainty in this renormalization
is due to hadronic diagrams in the range of $Q$ where perturbative QCD
calculations are not directly applicable. The renormalized value used by
the LEP electroweak working group is
\beq
\alpha^{-1}_{em} (m_Z) = 128.878 \pm 0.090
\label{onenineteen}
\eeq
However, this may be refined to $\alpha^{-1}_{em} (m_Z) = 128.933 \pm
0.021$ by more complete use of constraints from perturbative QCD and
data on $\tau$ decays~\cite{DH}. Beyond the tree level, the parameter
$\sin^2\theta_W$ may be defined in several different ways. One option is
the ``on-shell" definition  $\sin^2\theta_W \equiv 1 = m^2_W/
m^2_Z$~\cite{onshell}. The LEP experiments often use another physical
definition more
closely related to their experimental observables, as in Fig. 2, but
theorists often favour the $\overline{\rm MS}$ definition~\cite{onshell},
which is more
convenient for comparison with QCD and  GUT calculations.

Consistency between the different measurements shown in Fig. 1 -- e.g.,
the $\sin^2\theta_W$ measurements shown displayed in Fig. 2 -- imposes
constraints on the masses of heavy virtual particles that appear in loop
diagrams, such as the top quark and the Higgs boson~\cite{Vt,VH}. As
examples of this,
consider their contributions to $m_W$ and $m_Z$ in the ``on-shell"
renormalization scheme:
\beq
m^2_W \sin^2\theta_W = m^2_Z \cos^2\theta_W\sin^2\theta_W =
{\pi\alpha\over \sqrt{2} G_\mu} ~~(1+\Delta r)
\label{onetwenty}
\eeq
In the absence of the top quark, the gauge symmetry of the \sm ~ would be
lost, since the $b$ quark would occupy an incomplete  doublet of weak
isospin, destroying the renormalizability of the theory at the
one-loop level. This is reflected in the contributions of the one-loop
vacuum-polarization diagrams~\cite{Vt}:
\beq
\Delta r \ni {3G_\mu\over 8\pi^2 \sqrt{2}} ~~m^2_t + \ldots
\label{onetwentyone}
\eeq
in the limit $m_t \gg m_b$. Likewise, the \sm ~ would be
non-renormalizable in the absence of a physical Higgs boson, so $\Delta
r$ must also blow up as $m_H\rightarrow\infty$. As pointed out by
Veltman~\cite{VH}, a screening theorem restricts this to a logarithmic
dependence
at the one-loop level
\beq
\Delta r \ni {\sqrt{2}G_\mu\over 16\pi^2}~~m^2_W~~\left\{{4\over 3} \ln
{m^2_H\over m^2_W} + \ldots\right\}
\label{onetwentytwo}
\eeq
for $m_H \gg m_W$, though there is a quadratic dependence at the two-loop
level.

Comparing (\ref{onetwentyone}) and (\ref{onetwentytwo}), we see that the
dependence on $m_t$ is much greater than that on $m_H$. A measurement of
$\Delta r$ gives in principle an estimate of $m_t$, though with some
uncertainty if $m_H$ is allowed to vary between 10 GeV and 1 TeV. Before
the start-up of LEP, we gave the upper bound $m_t \lappeq$ 170
GeV~\cite{Amaldi,Costa,EFL}. By
combining several different types of precision electroweak measurement, it
is in principle possible to estimate independently both $m_t$ and $m_H$.
The present world data set implies~\cite{LEPEWWG}
\beq
m_t = 161^{+9}_{-8}~{\rm GeV}
\label{onetwentythree}
\eeq
which is compatible with both the pre-LEP estimate and the direct
measurements by CDF and $D\phi$~\cite{CDFD0}:
\beq
m_t = 173.8 \pm 5.0~{\rm GeV}
\label{onetwentyfour}
\eeq
Combining this with the precision electroweak data enables a more precise
estimate of $m_H$ to be made.

A key r\^ole in this estimate is being played by direct measurements of
$m_W$. Until now, the most precise of these has been that from $\bar pp$
colliders, dominated by the Fermilab Tevatron collider~\cite{CDFD0}:
\beq
m_W = 80.41 \pm 0.09~{\rm GeV}
\label{onetwentyfive}
\eeq
with an honourable mention for the indirect determinations from
deep-inelastic $\nu$ scattering:
\beq
m_W = 80.25 \pm 0.11~{\rm GeV}
\label{onetwentyfivehalf}
\eeq
with some slight dependence on $m_t$ and $m_H$. These values
can be compared with the indirect prediction based on other
precision electroweak data~\cite{LEPEWWG}, within the framework of the \sm
:
\beq
m_W = 80.329 \pm 0.029 ~{\rm GeV}~.
\label{onetwentysix}
\eeq
Reducing the error in the direct measurement (\ref{onetwentyfive}) would
constrain further the estimate of $m_H$ within the \sm , and could
constrain significantly its  possible extensions, such as
supersymmetry,
with the error in (\ref{onetwentysix}) providing a relevant target for
the experimental precision. This is also demonstrated by the
implications for the error in the estimate of $m_H$ corresponding to a
given error in $m_W$:
\beq
\matrix{
\Delta m_W = \quad\quad& ~25 & ~50 & {\rm MeV} \cr\cr
m_H~ =~ 100 & ^{+~86} _{-~54} & ^{+ 140} _{-~72}&{\rm GeV} \cr\cr
m_H~ =~ 300 & ^{+196} _{-126} & ^{+323} _{-168} &{\rm GeV} }
\label{onetwentyseven}
\eeq
It is a major goal of the LEP 2 experimental programme to achieve such
precision~\cite{LEP2YB}.

The measured cross-section for $e^+e^-\rightarrow W^+W^-$ is shown in
Fig. 3~\cite{LEPEWWG}. We see that $\nu$ exchange alone does not fit the
data: one also
needs to include both the $\gamma W^+W^-$ and $Z^0W^+W^-$ vertices
present in the \sm \footnote{It is surely too soon to cry ``new physics"
on the basis of the cross-section measurement at $E_{cm}$ = 189 GeV,
particularly since the more recent data shown at the LEPC~\cite{LEPC}
indicate a lesser discrepancy!}.
The first LEP 2 measurement of $m_W$ was obtained by measuring the cross
section at $E_{cm} = 161$ GeV,  close to the threshold, but this has now
been surpassed in accuracy by the direct reconstruction of $W^\pm$ decays
at higher $E_{cm}$. The current LEP 2 average is~\cite{LEPEWWG}
\beq
m_W = 80.37 \pm 0.90~{\rm GeV}
\label{onetwentyeight}
\eeq
which now matches the $\bar pp$ measurement error (\ref{onetwentyfive}).

\begin{figure}
\centerline{\includegraphics[height=3in]{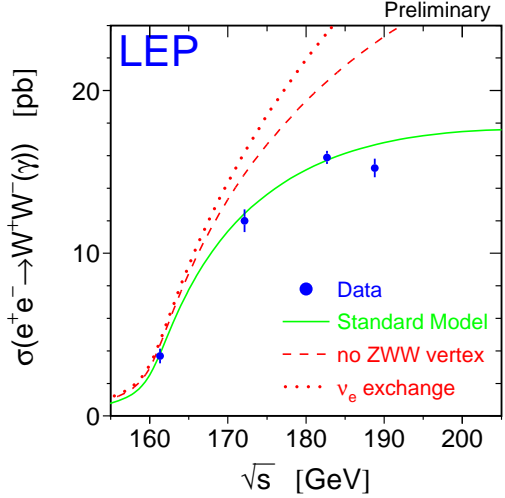}}
\caption[]{Measurements of $\sigma(e^+e^-\rightarrow
W^+W^-$)~\cite{LEPEWWG}.}
\end{figure}

We see in Fig. 4 that the $m_W$ measurements favour qualitatively $m_H <$
300 GeV, though not at a high level of significance~\cite{LEPEWWG}.
Stronger evidence
for a light Higgs boson~\cite{EFL} is provided by the lower energy LEP 1,
SLD
and $\nu$N data, as also seen in Fig. 4. Combining all the precision
electroweak data, one finds
\beq
m_H = 76^{+85}_{-47} \pm 10~{\rm GeV}
\label{onetwentynine}
\eeq
as seen in Fig. 5, corresponding to $m_H < 260$~GeV at the $95\%$
confidence level, if one
uses a conservative error in $\alpha_{em}(m_Z)$
and makes due allowance for unknown higher-loop uncertainties in the
analysis~\cite{LEPEWWG}.

\begin{figure}
\centerline{\includegraphics[height=3in]{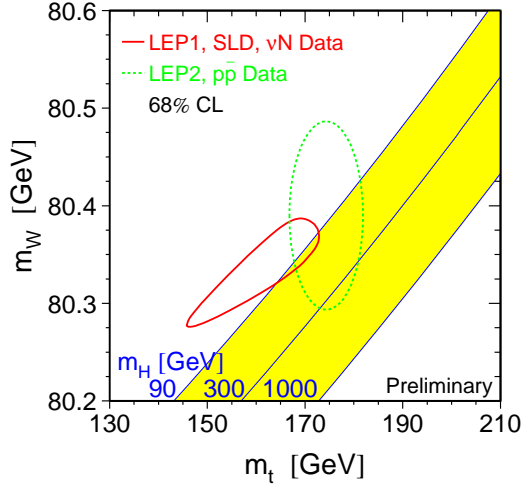}}
\caption[]{Contours of $m_W$ and $m_t$ from direct measurements (dotted
ellipse) compared with predictions based on the precision electroweak data
analyzed within the \sm ~(solid curve)~\cite{LEPEWWG}. Note the
sensitivity to $m_H$.}
\end{figure}

\begin{figure}
\centerline{\includegraphics[height=3in]{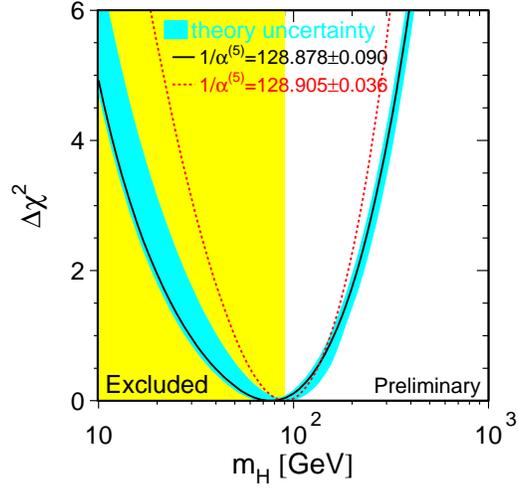}}
\caption[]{The $\chi^2$ function for a global fit to the precision electroweak
data prefers $m_H \sim$ 100 GeV~\cite{LEPEWWG}.}
\end{figure}

The range (\ref{onetwentynine}) may be compared with upper and lower
bounds derived within the \sm . The tree-level unitarity limit $m_H
\lappeq$ 1 TeV~\cite{V,LQT,DL} may be strengthened by including loop
effects via
renormalization-group calculations~\cite{loopMH}. We see in Fig. 6 the
upper bound on
$m_H$ that is obtained by requiring the \sm~ couplings to remain finite at
all energies up to some cutoff $\Lambda : m_H \lappeq$ 200 GeV if
$\Lambda \simeq m_P$ and $m_H \lappeq$ 700 GeV if $\Lambda \simeq m_H$,
corresponding to upper limits from lattice calculations~\cite{latticeMH}.
Also shown in
Fig. 6 are lower limits on $m_H$ obtained by requiring that the effective
Higgs potential remain positive for $\vert\phi\vert \lappeq \Lambda : m_H
\gappeq$ 140 GeV if $\Lambda \simeq m_P$~\cite{loopMH}.

\begin{figure}
\centerline{\includegraphics[height=3in]{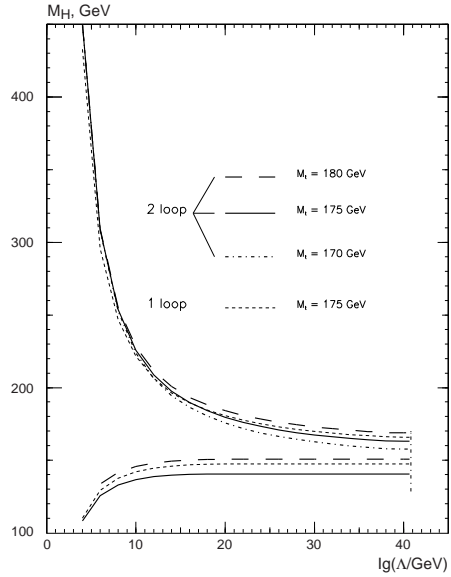}}
\caption[]{The range of $m_H$ compatible with the \sm~ remaining valid up to a
high scale $\Lambda$~\cite{loopMH}.}
\end{figure}

It is depressing to note that the range (\ref{onetwentynine}) of $m_H$
estimated on the basis of the precision electroweak data is compatible
with the \sm ~remaining valid all the way up to the Planck scale: $\Lambda
\simeq m_P$ \footnote{Nevertheless, the range (\ref{onetwentynine}) is
even more compatible with supersymmetry, which is one possible example
physics of new physics at $\Lambda \lappeq$ 1 TeV.}. Moreover, the  range 
(\ref{onetwentynine}) also imposes strong constraints on possible
extensions of the \sm . For example, Fig. 7 shows that it effectively
excludes a fourth generation~\cite{loopMH}. Note that this
renormalization-group
argument is independent of the neutrino-counting argument
(\ref{oneforteenone}) given earlier. In particular, this argument still
holds if $m_{\nu_4} > m_{Z/2}$: in fact, it even becomes slightly
stronger!

\begin{figure}
\centerline{\includegraphics[height=3in]{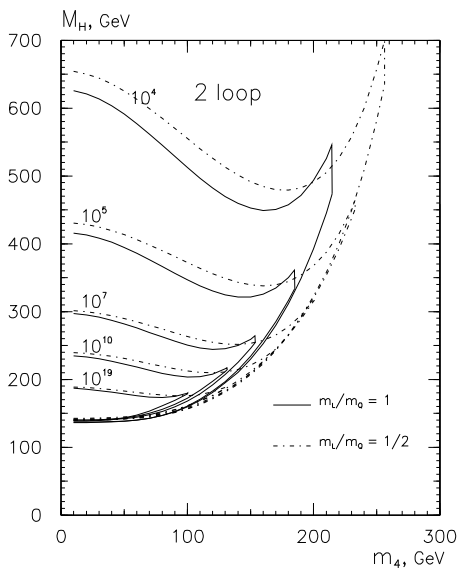}}
\caption[]{The regions of fourth-generation mass $m_4$ and Higgs mass $m_H$
compatible with the \sm~remaining valid up to a high scale
$\Lambda$~\cite{loopMH}.}
\end{figure}

\subsection{Defects of the \sm}

It has been said repeatedly that there is no confirmed experimental
evidence from accelerators against the \sm , and several possible
extensions have been ruled out. Nevertheless, no thinking physicist
could imagine that the \sm ~is the end of physics. Even if one accepts the
rather bizarre set of group representations and hypercharges that it
requires, the \sm~ contains at least 19 parameters: 3 gauge couplings
$g_{1,2,3}$ and 1 CP-violating non-perturbative vacuum angle $\theta_3$, 6
quark and 3 charged-lepton masses with 3 charged weak mixing angles and 1
CP-violating phase $\delta$, and 2 parameters: $(\mu ,\lambda )$ or
$(m_H,m_W)$ to characterize the Higgs sector. Moreover, many more
parameters are required if one wishes to accommodate non-accelerator
observations. For example, neutrino masses and mixing introduce at least
7 parameters: 3 masses, 3 mixing angles and 1 CP-violating phase,
cosmological inflation introduces at least 1 new mass scale of order
10$^{16}$ GeV, the cosmological baryon asymmetry is not explicable within
the \sm , so one or more additional parameters are needed, and the
cosmological constant may be non-zero. The ultimate ``Theory of
Everything" should explain all these as well as the parameters of the \sm.

It is convenient to organize the questions raised by the \sm ~into three
broad categories. One is the {\bf Problem of Mass}: do particle masses really
originate from a Higgs boson, and, if so, why are these masses not much
closer to the Planck mass $m_P\simeq 10^{19}$ GeV? This is the main
subject of the next two lectures. Another is the {\bf Problem of Unification}:
can all the particle interactions be unified in a simple gauge group,
and, if so, does it predict observable new phenomena such as baryon decay
and/or neutrino masses, and does it predict relations between parameters
of the \sm ~such as gauge couplings or fermion masses? This is the main
subject of the fourth lecture. Then there is the {\bf Problem of Flavour}: what
is the origin of the six flavours each of quarks and leptons, and what
explains their weak charged-current mixing and CP violation? This is the
main subject of Yossi Nir's lectures~\cite{Nir}. Finally, the quest for
the {\bf Theory
of Everything} seems most promising in the context of string theory,
particularly in its most recent incarnation of $M$ theory, as discussed
in the fifth lecture, and by Michael Green~\cite{Green}. In addition to
all the above
problems, this should also reconcile quantum mechanics with general
relativity, explain the origin of space-time and the number of
dimensions, make coffee, etc... .

\section{INTRODUCTION TO SUPERSYMMETRY}

\subsection{The Electroweak Vacuum}

We have discussed in Lecture 1 the fact that generating particle masses
requires breaking the electroweak gauge symmetry spontaneously:
\beq
m_{W,Z} \not= 0~ \Leftrightarrow ~~< 0 \vert X_{I,I_3} \vert 0 > ~\not= 0~
\label{twoone}
\eeq
for some spin-0 quantity $X$ with non-zero isospin $I$ and third
component $I_3$. The fact that experimentally $\rho \equiv m^2_W/
m^2_Z ~~\cos^2\theta_W \simeq 1$ is consistent with the \sm~ expectation
that $X$ has mainly $I = {1\over 2}$~\cite{RV}. This is also what is
required to
give non-zero fermion masses: $m_f \bar f_L f_R + h.c.$, since the
$f_{L,R}$ have $I = {1\over 2},0$. The question then remains: what is the
nature of $X$? In particular, is it elementary or composite?

The former is the option chosen in the \sm : $X = H : <0\vert H^0\vert
0 > \not= 0$. However, as discussed in more detail later, quantum
corrections to the squared mass of an elementary Higgs boson diverge
quadratically:
\beq
\delta m^2_H = 0 \left({\alpha\over\pi}\right) \Lambda^2
\label{twotwo}
\eeq
where $\Lambda$ is some cutoff, corresponding physically to the scale up
to which the \sm~ remains valid. We discuss later the possibility that
$\Lambda$ can be identified with the energy threshold for symmetry. This
should occur at $\lambda \lappeq$ 1 TeV, in order that the quantum
corrections (\ref{twotwo}) have the same magnitude as the physical Higgs
boson mass.

The alternative option is that $X$ is composite, namely a
fermion-antifermion condensate $<0\vert \bar FF \vert 0 > \not= 0$. This
idea is motivated by the existence of a quark-antiquark condensate 
$<0\vert \bar qq \vert 0 > \not= 0$ in QCD, and the r\^ole of Cooper pairs
$<0\vert e^-e^- \vert 0 > \not= 0$ in conventional superconductivity. Two
major possibilities for the condensate have been considered: a
top-antitop condensate $<0\vert \bar tt \vert 0 > \not= 0$ 
held together by a large Yukawa coupling $\lambda_{H\bar
tt}$~\cite{ttbar}, and
technicolour~\cite{TC}, in which new interactions 
that become strong at an energy
scale $\Lambda \sim$ 1 TeV bind new strongly-interacting technifermions:
$<0\vert \bar TT \vert 0 > \not= 0$. The $\bar tt$ condensate idea is
currently disfavoured, since simple implementations require $m_t >$ 200
GeV in contradiction with experiment, so we concentrate here on the
technicolour alternative.

The technicolour idea~\cite{TC} was initially modelled  on the known
dynamics of
QCD:
\beq
<0\vert\bar q_Lq_R + h.c.\vert 0> \not= 0 \rightarrow
<0\vert\bar F_LF_R + h.c.\vert 0> \not= 0
\label{twothree}
\eeq
which breaks isospin with $I = {1\over 2}$, if the electroweak multiplet
assignments of the $F_L$ and $F_R$ are the same as those of the $q_L$ and
$q_R$. The scale of this breaking will be appropriate if $\Lambda_{QCD}
\rightarrow \Lambda_{TC} \sim$ 1 TeV. Just as QCD contains (if $m_q=0$)
massless pions with the axial-current matrix element $<0\vert
A_\mu\vert\pi > = i p_\mu f_\pi$, one expects a similar coupling
\beq
<0\vert J_\mu\vert\pi_{TC}> = i p_\mu F_\pi
\label{twofour}
\eeq
for the technipion $\pi_{TC}$, which does the same Goldstone-eating job 
as (\ref{oneseven1}), (\ref{oneseven2}) if $F_\pi = v \simeq$ 250 GeV. If
there are two massless flavours of technifermions, one expects 3 massless
technipions to be eaten by the $W^\pm$ and $Z^0$, and the physical Higgs
boson is replaced by an effective massive composite scalar, analogous to
the scalar mesons of QCD and weighing 0(1) TeV. However, a single
technidoublet is not enough when one imposes the necessary cancellation
of anomalies and tries to give masses to conventional fermions~\cite{ETC}.
For these
reasons, the Standard Technicolour Model used to include a full
technigeneration: $[(N,E),(U,D)_{1,2,3}]_{1,\ldots ,N_{TC}}$, where the
indices denote colour and technicolour indices. For generality, one can
study models as functions of the numbers of techniflavours and
technicolours: $(N_{TF}, N_{TC})$. 

Their effects via one-loop quantum corrections can be parametrized in
terms of their contributions to electroweak observables via three
combinations of vacuum polarizations~\cite{PT,AB}. One example is:
\beq
T \equiv {\epsilon_1\over\alpha} \equiv {\Delta\rho\over\alpha} ~:~
\Delta\rho = {\pi_{ZZ}(0)\over m^2_Z} - {\pi_{WW}(0)\over m^2_W} -
2\tan\theta_W {\pi_{\gamma Z}(0)\over m^2_Z}
\label{twofive}
\eeq
which describes deviations from the tree-level relation
$\rho\equiv m^2_W/ m^2_Z\cos^2\theta = 1$ and measures
isospin-breaking effects. This is related to $\Delta r$ ~(\ref{onetwenty})
and receives contributions from Standard-Model particles:
\beq
T \ni {3\over 16\pi}~~{1\over
\sin^2\theta_W\cos^2\theta_W}~~\left({m^2_t\over m^2_W}\right) - {3\over
16\pi\cos^2\theta_W}~~\ln \left({m^2_H\over m^2_Z}\right) + \ldots
\label{twosix}
\eeq
The other relevant combinations of vacuum polarizations are~\cite{PT,AB}
\beq
S \equiv {4\sin^2\theta_W\over\alpha}~~\epsilon_3 \ni {1\over 12\pi}~~\ln
\left({m^2_H\over m^2_Z}\right) + \ldots
\label{twosevenone}
\eeq
and~\cite{PT,AB}
\beq
U \equiv - {4\sin^2\theta_W\over \alpha}~~ \epsilon_2
\label{twoeightone}
\eeq
The precision electroweak data may be used to constrain $(S,T,U)$ (or
$\epsilon_{1,2,3}$), and thereby possible extensions of the \sm ~with the
same $SU(2)\times U(1)$ gauge group and additional matter particles, such
as technicolour. Note, however, that this approach is not adequate for
precision analyses of theories with important vertex diagrams such as the
\sm~ or its minimal supersymmetric extension, to be discussed later. 
These have important vertex and box diagrams, as well as the
vacuum-polarization diagrams taken care of by $S,T,U
(\epsilon_{1,2,3}$. Some of
these be treated by introducing further parameters such as $\epsilon_b$
for the $Z\bar bb$ vertex. Even so, two-loop and other higher-order
effects are not treated exactly in this approach.

Figure 8 compiles the constraints on $\sin^2\theta_W$ and the overall
$Z$ weak coupling strength coming from the precision electroweak data
(top panel) and the resulting constraints on $S$ and $T$ (bottom
panel)~\cite{Matsumoto}.
We see that the lower-energy data are compatible with the high-energy
data at around the one-$\sigma$ level, and that the high-energy data
impose strong constraints on $S,T,U$ (equivalent to $\epsilon_{1,2,3}$). 
Figure 9 shows the available constraints in
the $(\epsilon_1,\epsilon_2)$ plane~\cite{ABC}. We see that the one-loop
corrections are certainly needed, since the data lie many $\sigma$ away from
the (improved) Born approximation. We also see that the data are quite
consistent with the \sm . A compilation of determinations of the
$\epsilon_i$ are shown in Fig. 10~\cite{ABC}, where we see a discrepancy
only in
$\epsilon_b$, but even this is only slightly more than one $\sigma$.
Finally, Fig. 11 compares the data constraint in the $(S,T)$ plane
 not only with the \sm ~but also with various technicolour
models~\cite{Matsumoto}. The models chosen all have one technidoublet, and
hence $N_{TF}
= 2$, and varying values of $N_{TC}$ = 2,3,4. We see that even the least
disfavoured model is at least four $\sigma$ away from the data, and
models with larger $N_{TC}$ (shown) and $N_{TF}$ (not shown) deviate
even further from experiment. 

\begin{figure}
\centerline{\includegraphics[height=3in]{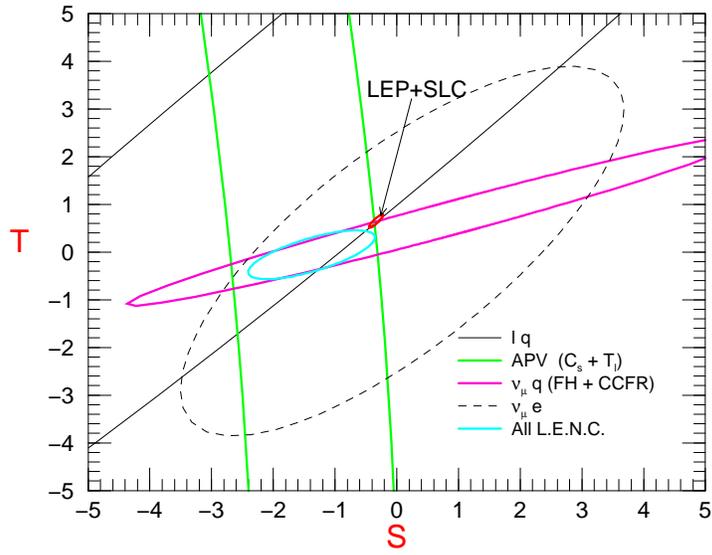}}
\caption[]{Constraints on $S$ and $T$ inferred from precision electroweak
measurements at low and high energies~\cite{Matsumoto}.}
\end{figure}

\begin{figure}
\centerline{\includegraphics[height=3in]{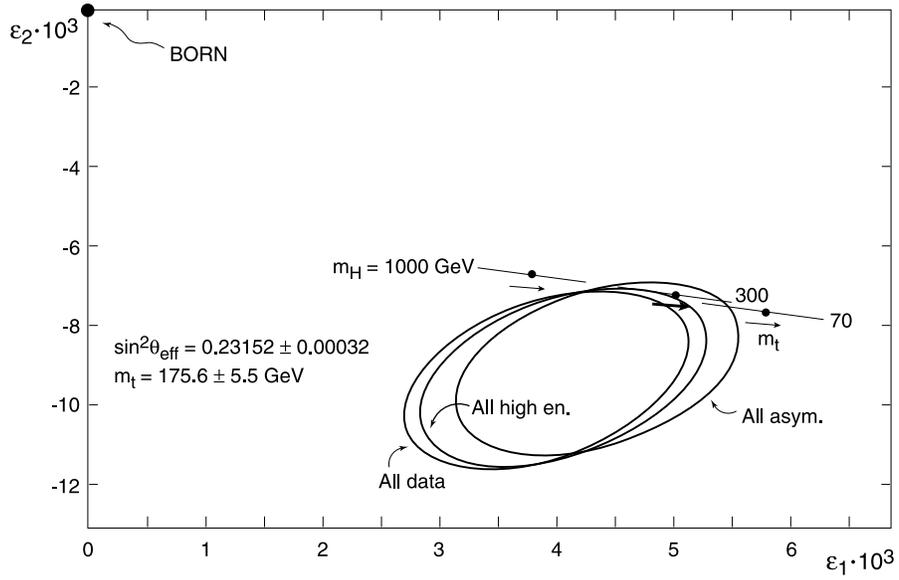}}
\caption[]{Constraints on $(\epsilon_1,\epsilon_2)$ from precision electroweak
measurements, which agree with the \sm ~predictions~\cite{ABC}. Note that
the data lie many
standard deviations away from the values expected (``Born") if quantum
corrections are neglected.}
\end{figure}

\begin{figure}
\centerline{\includegraphics[height=3in]{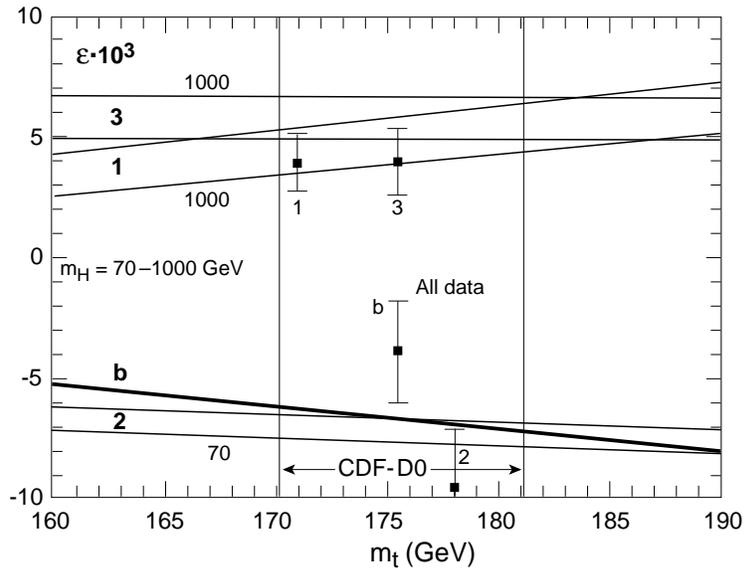}}
\caption[]{Values of the quantum corrections $\epsilon_i$ inferred from the
data~\cite{ABC}, compared with the \sm ~predictions.}
\end{figure}

\begin{figure}
\centerline{\includegraphics[height=3in]{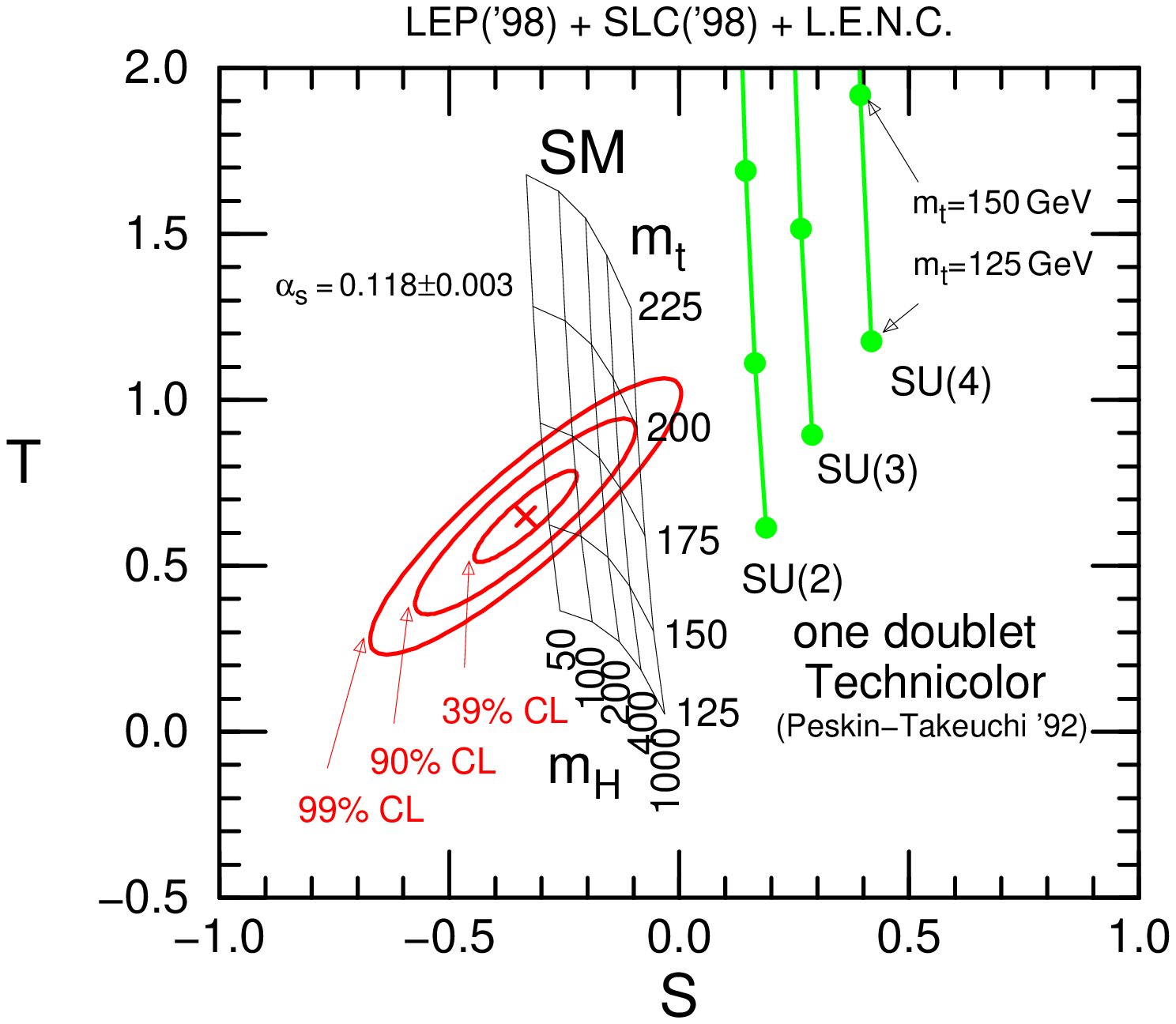}}
\caption[]{Values of $S$ and $T$ extracted from the data
(ellipses)~\cite{Matsumoto}, confronted
with the predictions of the \sm ~and of one-doublet technicolour models.}
\end{figure}

This large discrepancy has almost been the
death of technicolour models, but various suggestions have been made
that one could respect the experimental constraints if the technicolour
dynamics is somewhat different from that of QCD. Specifically, it has
been suggested that the technicolour coupling may not run as rapidly as
the strong coupling~\cite{walk}. Unfortunately, calculations in this
framework of
``walking technicolour" cannot be made as pecisely as in the
conventional technicolour models discussed above, rendering it difficult
to test or disprove. For the moment, no calculable technicolour model is
consistent with the precision electroweak data, so we turn to
supersymmetry.

\subsection{Introduction to supersymmetry}

Back in the 1960's there were many (forgettable) attempts to combine
internal symmetries such as flavour isospin or $SU(3)$ with relativistic
external symmetries such as Lorentz invariance. However, in 1967 Coleman
and Mandula~\cite{CM} proved that this could not be done using only
bosonic
charges. The way to avoid this no-go theorem was found in 1971, when
Gol'fand and Likhtman~\cite{GL} showed that one could extend the
Poincar\'e
algebra using  fermionic charges. In the same year, Neveu, Schwarz and
Ramond~\cite{NSR} invented supersymmetry in two dimensions when they
discovered how
to incorporate fermions in string models. Supersymmetric field theories
in four dimensions were discovered in 1973, by Volkov and Akulov~\cite{VA} 
in a
non-linear realization, and by Wess and Zumino~\cite{WZ} in the linear
realization
now used in most model-building. Soon afterwards, Wess, Zumino, Iliopoulos
and Ferrara~\cite{nonren} realized that supersymmetric models were free of
many of the
divergences found in other four-dimensional field theories. Then, in
1976, Freedman, van Nieuwenhuizen and Ferrara~\cite{FvF} and independently
Deser
and Zumino~\cite{DZ} showed how supersymmetry can be realized locally (by
analogy
with gauge theories) in the context of supergravity. 

Many of the ideas for using supersymmetry were motivated by the desire to
unify known bosons and fermions: for example, unifying mesons and baryons
motivated the early string work~\cite{NSR} and that of Wess and
Zumino~\cite{WZ}. It was
initially suggested that neutrino could be a Goldstone fermion in a
non-linear realization of supersymmetry~\cite{VA}, but it was soon pointed
out that
experimental data on $\nu$ interaction cross sections conflicted with
theorems on the low-energy behaviour in such theories~\cite{nolow}. The
fact that
supersymmetric theories had fewer (in some cases, no) divergences offered
 to some people who never liked infinite renormalizations hope that one
could construct a finite theory. Others were attracted by the idea that
supersymmetry might relate the odd-person-out Higgs
boson to fermionic matter and perhaps gauge bosons. At a more fundamental
level, the fact that local supersymmetry involves gravity suggested to
many the idea of unifying all the particles and their interactions in
some supergravity theory. However, this motivation did not provide a
clear clue as to the mass scale of supersymmetry breaking, so there was no
obvious reason why the sparticle masses should not be as heavy as $m_P
\simeq 10^{19}$ GeV.

Such a reason was eventually provided by the mass hierarchy
problem~\cite{hierarchy}: why 
is $m_W
\ll m_P$? The latter is the only candidate we have for a fundamental mass
scale in physics, where gravity is expected to become as strong as other
particle interactions, e.g., graviton exchange at LEP 10$^{19}$ would be
comparable to $\gamma$ and $Z^0$ exchange. The hierarchy problem can be
rephrased as: ``why is $G_F \gg G_N$?", since $G_F \sim 1/m^2_W$ and $G_N
= 1/m^2_P$. Alternatively, for the benefit of atomic, molecular and
condensed-matter physicists, not to mention chemists and biologists, one
can ask: why is the Coulomb potential in an atom so much larger than the
Newton potential? The former is $e^2/r : e^2$ = 0(1), whereas the latter
is $G_Nm_pm_e/r$, 
so the Newton potential is negligible just because conventional particle
masses $m_{p,e}$ are much lighter than $m_P$. 

You might think that one could just set $m_W \ll m_P$ by hand, and ignore
the problem. However, there is a threat from radiative
corrections~\cite{hierarchy}. Each
of the one-loop diagrams in Fig. 12 is individually quadratically
divergent, implying 
\beq
\delta m^2_{H,W} = {\cal O} \left({g^2\over 16\pi^2}\right)~~\int^\Lambda
d^4k {1\over k^2} = {\cal O} \left({\alpha\over\pi}\right) \Lambda^2
\label{twofivetwo}
\eeq

\begin{figure}
\centerline{\includegraphics[height=2in]{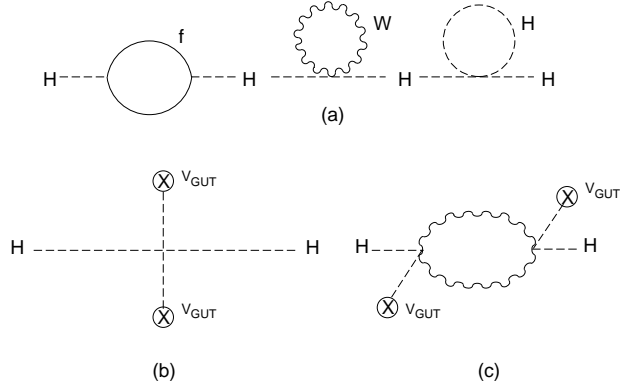}}
\caption[]{(a) One-loop quantum corrections to $m^2_H$ in the \sm. (b)
Tree-level and (c) one-loop corrections to $m^2_H$ in a GUT.}
\end{figure}

\noindent
where the cutoff $\Lambda$ in the integral represents the scale up to
which the \sm~ remains valid, and beyond which new physics sets in. If we
think $\Lambda\simeq m_P$ or the grand unification scale, the quantum
correction (\ref{twofivetwo}) is much larger than the physical value of
$m_{H,W}\sim$ 100 GeV. This is not a problem for renormalization theory:
there could be a large bare contribution with the opposite sign, and one
could fine-tune its value to many significant figures so that the
physical value $m^2_{H,W} \simeq \delta m^2_{H,W}$ (\ref{twofivetwo}).
However,
this seems unnatural, and would have to repeated order by order in
perturbation theory. In contrast, the one-loop corrections to a
fundamental fermion mass $m_f$ are proportional to $m_f$ itself, and only
logarithmically divergent:
\beq
\delta m_f = {\cal O}\left({g^2\over 16\pi^2}\right) m_f \int^\Lambda d^4k
{1\over k^4} = {\cal O} \left({\alpha\over\pi}\right) m_f \ln {\Lambda\over
m_f}
\label{twosixtwo}
\eeq
This correction is no larger numerically than the physical value, for any
$\Lambda \lappeq m_P$. This is because there is a chiral symmetry
reflected in the $m_f$
factor in (\ref{twosixtwo}) that keeps the quantum corrections naturally
(logarithmically)
small. The hope is to find a corresponding symmetry principle to make
small boson masses natural:  $\delta m^2_{H,W} \lappeq m^2_{H,W}$.

This is achieved by supersymmetry~\cite{FF}, exploiting the fact that the
boson and
fermion loop diagrams in Fig. 12a have opposite signs. If there are equal
numbers of fermions and bosons, and if they have equal couplings as in a
supersymmetric theory, the quadratic divergences (\ref{twofivetwo})
cancel:
\beq
\delta m^2_{H,W} = -\left({g^2_F\over 16\pi^2}\right)~~(\Lambda^2 +
M^2_F) + \left({g^2_B\over 16\pi^2}\right)~~(\Lambda^2 + M^2_B) =
{\cal O}\left({\alpha\over 4\pi}\right) ~~\vert m^2_B - m^2_f\vert
\label{twoseven}
\eeq
This  is no larger than the physical value: $\delta m^2_{H,W} \lappeq
m^2_{H,W}$, and hence naturally small\footnote{There is a logarithmic
multiplicative factor in the right-hand side of (\ref{twoseven}) that is
reflected in the discussion below of renormalization-group corrections to
supersymmetric particle masses.}, if
\beq
\vert m^2_B - m^2_F\vert \lappeq 1~{\rm TeV}^2
\label{twoeight}
\eeq
This naturalness argument~\cite{hierarchy} is the only available
theoretical motivation
for thinking that supersymmetry may manifest itself at an accessible
energy scale. 

However, this argument is qualitative, and a matter of
taste. It does not tell us whether sparticles should appear at 900 GeV, 1
TeV or 2 TeV, and some theorists reject it altogether. They say that,
in a renormalizable theory such as the \sm,
 one need not worry about the
fine-tuning of a bare parameter, since it is not physical. However, I
take  naturalness seriously as a physical argument: it is telling us
that a large hierarchy is intrinsically unstable, and supersymmetry is
the most plausible way of stabilizing it. Moreover, many logarithmic
divergences are absent in supersymmetry, which stabilizes the
possible GUT Higgs corrections to $m_H$ shown in Fig.~12b arising from the
loops shown in Fig.~12c, which is also important for
stabilizing the hierarchy $m_W \ll m_{GUT}$, as we see later.

\subsection{What is supersymmetry?}

After all this introduction and motivation, just what is
supersymmetry~\cite{FF}?
It is a symmetry that links bosons and fermions via spin-${1\over 2}$
charges $Q_\alpha$ (where $\alpha$ is a spinorial index). It seems to be
the last possible symmetry of the particle scattering matrix~\cite{HLS}.
As such,
many would argue that it must inevitably play a r\^ole in physics, and it
has in fact already appeared at a phenomenological level in
condensed-matter, atomic and nuclear physics. All previously-known
symmetries are generated by bosonic charges, which are, apart from the
momentum operator $P_\mu$ associated with Lorentz invariance, scalar
charges $Q$ that relate different particles of the same spin $J$: $Q\vert
J> = \vert J^\prime >$, $Q \in U(1), SU(2), SU(3), \ldots$. Indeed, as
already mentioned, Coleman and Mandula~\cite{CM} showed that it was
impossible to
mix such internal symmetries with Lorentz invariance using bosonic
charges. The essence of their proof is easy to grasp.

Consider $2\rightarrow 2$ scattering: $1+2\rightarrow 3+4$, and consider
the possibility that there is a conserved tensor charge $\Sigma_{\mu\nu}$
corresponding to some higher bosonic symmetry (there can be no other
charge with one vector index, besides $P_\mu$, and higher tensor charges
can be discussed analogously to $\Sigma_{\mu\nu}$). Its diagonal matrix
elements are required by Lorentz invariance to have the following tensor
decomposition:
\beq
<a\vert\Sigma_{\mu\nu}\vert a> = \alpha p^a_\mu p^a_\nu + \beta g_{\mu\nu}
\label{twonine}
\eeq
where $p^a_\mu$ is the four-momentum of the particle $a$ and $\alpha ,
\beta$ are unkown reduced matrix elements. For $\Sigma_{\mu\nu}$ to be
conserved in the scattering process, as long as $\alpha \not= 0$ 
\footnote{The case
$\alpha = 0$ corresponds to a scalar charge $\Sigma_{\mu\nu} = \hat\Sigma
g_{\mu\nu}$.} one must
require
\beq
p^1_\mu p^1_\nu + p^2_\mu p^2_\nu = p^3_\mu p^3_\nu + p^4_\mu p^4_\nu
\label{twoten}
\eeq
as well as $p^1_\mu + p^2_\mu = p^3_\mu + p^4_\mu$. It is easy to convince
oneself that the only possible simultaneous solutions to these
linear and quadratic conservation conditions correspond to purely forward
scattering. This conflicts~\cite{CM} with the basic principles of quantum
field theory as well as experiment.

This argument is fine as far as it goes, but it does not apply to any
spinorial charge $Q_\alpha$, since the diagonal matrix elements vanish:
$<a\vert Q_\alpha\vert a > = 0$.

Let us explore now what is the possible algebra of an algebra of such
spinorial charges $Q^i_\alpha : i = 1,2,\ldots , N$~\cite{HLS}. If they
are to be
symmetry generators, they must commute with the Hamiltonian:
\beq
[Q^i_\alpha , H] = 0
\label{twoeleven}
\eeq
Hence, their anticommutator (which is bosonic) must also commute with $H$:
\beq
[\{Q^i_\alpha , Q^j_\beta \},H] = 0
\label{twotwelve}
\eeq
By the Coleman-Mandula theorem~\cite{CM}, this anticommutator must be a
combination
of the conserved Lorentz vector charge $P_\mu$ and some scalar charge
$Z^{ij}$. The only possible form is in fact
\beq
\{Q^i_\alpha , Q^j_\beta \} = 2 \delta^{ij} (\gamma^\mu C)_{\alpha\beta}
P_\mu + Z^{ij}
\label{twothirteen}
\eeq
where we use four-component spinors, $C$ is the charge-conjugation matrix
and $Z^{ij}$ is antisymmetric in the supersymmetry indices $\{i,j\}$.
Thus, this so-called ``central charge" vanishes for the $N = 1$ case of
phenomenological relevance.

The basic building blocks of $N=1$ supersymmetric theories are
supermultiplets containing the following helicity states~\cite{FF}:
\beq
{\rm chiral}~:~~\left(\matrix{{1\over 2}  \cr 0}\right)~, ~~{\rm
gauge}~:~~\left(\matrix{1  \cr {1\over 2}}\right)~, ~~{\rm
graviton}~:~~\left(\matrix{2  \cr {3\over 2}}\right)~,
\label{twoforteen}
\eeq
which are used to describe matter and Higgses, gauge fields and gravity,
respectively. You may wonder why one does not use theories with extended
supersymmetry: $N \geq 2$. The building blocks for $N = 2$ are:
\beq
{\rm matter}~:~~\left(\matrix{{1\over 2}  \cr 0 \cr -{1\over 2}}\right)~,
~~{\rm gauge}~:~~\left(\matrix{1  \cr {1\over 2} \cr 0}\right)~, ~~{\rm
gravity}~:~~\left(\matrix{2  \cr {3\over 2}\cr 1}\right)~,
\label{twofifteen}
\eeq
and it is apparent that left- and right-handed particles (helicities $\mp$ 1/2)
must be in identical representations of the gauge group. This is
immediate for the matter supermultiplet in (\ref{twofifteen}), and must
also be the case for fermions in the gauge supermultiplet, since the
helicity $\mp 1$ must be in identical adjoint representations.
Hence an $N = 2$ theory cannot accommodate parity violation, and is not
suitable for phenomenology~\footnote{Moreover, there are
severe lower limits, in the context of unified theories, on the possible
renormalization scale down to which $N = 2$
supersymmetry may remain valid~\cite{AEL}.}.

The simplest $N=1$ supersymmetric field theory contains a free fermion
and a free boson~\cite{FF,Peskin}:
\beq
{\cal L} = \partial_\mu \phi^*~~\partial^\mu\phi + i~~\psi^+
\bar\sigma\cdot\partial\psi
\label{twosixteen}
\eeq
where we work with two-component spinors and denote 
$\sigma_\mu = (1,\underline{\sigma}), ~~\bar\sigma_\mu = (1,
-\underline{\sigma})$, where the $\underline{\sigma}$ are Pauli matrices.
The simple supersymmetry transformation laws are
\beq
\delta_\xi \phi = \sqrt{2} \xi^T C \psi~,~~ \delta_\xi\psi = \sqrt{2} i
\sigma\cdot\partial\phi C \xi^*
\label{twoseventeen}
\eeq
where $\xi$ is an infinitesimal spinor parameter and $C$ is the
conjugation matrix: $C = -i\sigma^2 = C^*$, $C^{-1} = C^T = -C$. It is
easy to check that under (\ref{twoseventeen}) the Lagrangian
(\ref{twosixteen}) changes by a total derivative $\partial_\mu (\ldots
)$, and hence the action $A = \int d^4x {\cal L}(x)$ is invariant. We can
also see in (\ref{twoseventeen}) a reflection of the supersymmetry
algebra (\ref{twothirteen}): after two supersymmetry transformations, the
fields $(\phi,\psi)$ are transformed by derivatives $(\partial\phi ,
\partial\psi)$, corresponding to the action of the momentum operator
$P_\mu = i\partial_\mu$. 

The example (\ref{twoseventeen}) can easily be
extended to include interactions~\cite{FF,Peskin}:
\beq
{\cal L} = \partial_\mu\phi^*\partial^\mu\phi + i\psi^\dagger
\bar\sigma\cdot\partial\psi + F^\dagger F + \left( F {\partial
W\over\partial\phi} - {1\over 2} \psi^T C\psi {
\partial^2W\over\partial\phi^2} + {\rm herm.conj.}\right)
\label{twoeighteen}
\eeq
with supersymmetry transformations:
\beq
\delta_\xi\phi = \sqrt{2} \xi^T C\psi~,~~\delta_\xi\psi = \sqrt{2}
i\sigma\cdot\partial\phi C\xi^* + \xi F~,~~
\delta_\xi F = -\sqrt{2} i\xi^\dagger \bar\sigma\cdot\partial\psi
\label{twonineteen}
\eeq
The field $F$ is called an auxiliary field: notice that it has no kinetic
term, and so may be eliminated by using an equation of motion:
\beq
F^\dagger = -{\partial W\over\partial\phi}
\label{twotwenty}
\eeq
Thus all the matter interactions are characterized by the analytic
function $W(\phi )$, which is called the superpotential. Renormalizability
of the
field theory requires the superpotential to be a cubic function: for $W =
\lambda\phi_1\phi_2\phi_3$, one obtains from (\ref{twoeighteen}) the
following particle interactions:
\beq
\lambda\left[ 
(\psi^T_1 C \psi_2)\phi_3 + 
(\psi^T_2 C \psi_3)\phi_1 + 
(\psi^T_3 C \psi_1)\phi_2\right] + \vert \lambda \phi_1\phi_2\vert^2 + 
\vert \lambda \phi_2\phi_3\vert^2 + 
\vert \lambda \phi_3\phi_1\vert^2 
\label{twotwentyone}
\eeq
where the last terms provide a quartic potential for the scalar fields
$\phi_i$ and are called in the jargon ``$F$ terms".

We shall not discuss here in detail the construction of the interactions
of a chiral supermultiplet with a gauge supermultiplet~\cite{FF}, limiting
ourselves to quoting the results. In addition to the gauge interactions
of the chiral fermions and their bosonic partners, there are gaugino
interactions
\beq
\sqrt{2} g \left[ (\psi^T_i C (T^a)^i_j V_a) \phi^{j*} + {\rm
herm.conj.}\right]
\label{twotwentytwo}
\eeq
where $ (T^a)^i_j$ is the gauge representation matrix for the chiral
fields. There is also another quartic potential term for the scalars:
\beq
V = {g^2\over 2} ~\sum_a~~\vert \phi^{i^*}~(T^a)^j_i~~\phi_j\vert^2
\label{twentythree}
\eeq
which are called in the jargon ``$D$ terms". Finally, we note for
completeness that the conventional gauge-boson kinetic term and the gauge
interactions of fermions in the adjoint representation of the gauge
group, such as the gauginos $\tilde V_a$, are automatically
supersymmetric.

\subsection{Minimal Supersymmetric Extension of the \sm}

Let us now return to  phenomenology. If one is to construct a minimal
supersymmetric model, the first natural question is: can one construct it
out of the \sm ~particles alone? It is easy to see that this is
impossible, because the known bosons and fermions have different
conserved quantum numbers~\cite{Fayet}. For example, gluons are in an
octet
(\underline{8}) representation of colour, whereas quarks are in triplet
(\underline{3}) representation of colour. Similarly, there are no known
weak-isotriplet fermions, as would be  
needed to partner the electroweak gauge bosons. The known leptons are
isodoublets like the Higgs boson, but they carry lepton number, unlike
the Higgs. For these reasons, new particles must be
postulated~\cite{Fayet} as
supersymmetric partners of known particles, as seen in the Table.

\begin{table}
\begin{center}
\begin{tabular}{lclc}\hline
&&&\\
Particle & Spin & Spartner & Spin \\ \hline
&&&\\
quark: $q$ & ${1\over 2}$ & squak: $\tilde q$ & 0 \\
&&&\\
lepton: $\ell$ & ${1\over 2}$ & slepton: $\tilde\ell$ & 0 \\
&&&\\
photon: $\gamma$ & 1 & photino: $\tilde\gamma$ & ${1\over 2}$ \\
&&&\\
$W$ & 1 & wino: $\tilde W$ & ${1\over 2}$ \\
&&&\\
$Z$ & 1 & zino: $\tilde Z$ & ${1\over 2}$ \\
&&&\\
Higgs: $H$ & 0 & higgsino: $\tilde H$ & ${1\over 2}$ \\ 
&&&\\ \hline
\end{tabular}
\end{center}
\caption[]{Particles in the \sm ~and their supersymmetric partners.}
\end{table}

The minimal supersymmetric extension of the \sm~ (MSSM)~\cite{MSSM} has
the same gauge
interactions as the \sm. In addition, there are couplings of the form
(\ref{twotwentyone}) derived from the following superpotential:
\beq
W = \lambda_d ~Q~D^C~H + \lambda_\ell ~L~E^C~H + \lambda_u~Q~U^C~\bar H +
\mu \bar HH
\label{twotwentyfour}
\eeq
Here, $Q[L]$ denote isodoublets of supermultiplets containing $(u,d)_L
[(\nu,\ell)_L],~D^C[U^C,E^C]$ are singlets containing the left-handed
conjugates $d^C_L[u^C_L, e^C_L]$ of the right-handed $d_R[u_R,e_R]$, and
the superpotential couplings $\lambda_d[\lambda_{u,\ell}]$ correspond to
the Yukawa couplings of the \sm~
that give masses to the $d[u,\ell^-]$, respectively:
\beq
m_d = \lambda_d <H>~,~~m_u = \lambda_u <\bar
H>~,~~m_\ell = \lambda_\ell <H>~.
\label{twotwentyfive}
\eeq
Each of these should be understood as a $3\times 3$ matrix in generation
space, which is to be diagonalized as in the \sm. 

In addition to the Standard-Model-like superpotential interactions shown in
(\ref{twotwentyfour}), the following superpotential couplings~\cite{Rviol} 
are also permitted by the gauge symmetries of the \sm:
\beq
W \ni \lambda LLE^C + \lambda^\prime Q D^C L + \lambda^{\prime\prime}
U^CD^CD^C
\label{twotwentyfive1}
\eeq
Each of these violate conservation of either lepton number $L$ or baryon
number $B$. The possible presence of such interactions attracted some
interest in 1997~\cite{AEGLM,Jerusalem} with the discovery of unexpectedly
many events at HERA
at large $x$ and $Q^2$~\cite{HERA}, but interest has now subsided. Their
potential
significance is only discussed intermittently in these lectures.

Note that (\ref{twotwentyfour}) requires two Higgs doublets $H, \bar H$
with opposite hypercharges in order to give masses to all the matter
fermions. In the \sm, one doublet $\phi$ and its complex conjugate
$\phi^\dagger$ would have sufficed. This does not work in the MSSM,
because the superpotential $W$ must be an analytic function of the
fields. Moreover, Higgs supermultiplets include Higgsino fermions that
generate triangle anomalies which must cancel among themselves, requiring
at least two Higgs doublets. These couple via the $\mu$ term in
(\ref{twotwentyfour}). Note also that the ratio of Higgs vacuum
expectation values
\beq
\tan\beta\equiv {<\bar H>\over <H>}
\label{twotwentysix}
\eeq
is undetermined and should be treated as a free parameter. Finally, we
comment that the superpotential and gauge couplings determine the MSSM's
quartic scalar couplings, providing important constraints on the Higgs
masses, as we see later.

Before discussing in more detail the phenomenology of the MSSM, it is
appropriate to mention two important but indirect experimental
indications that favour supersymmetry. One is the relatively light mass
of the Higgs boson inferred from the analysis of precision electroweak
data~\cite{LEPEWWG}, as seen in Fig. 5. As discussed in more detail in the
next
Lecture, the lightest MSSM Higgs boson must weigh $\lappeq$ 150
GeV~\cite{susyHiggs}, in
good agreement with the range favoured by the data. The other 
indication in favour of supersymmetry is the
measured value of $\sin^2\theta_W$~\cite{LEPEWWG}, as shown in Fig. 2. As
discussed in
more detail in Lecture 4, Grand Unified Theories predict $\sin^2\theta_W$
as a function of $\alpha_s(m_Z)$.
For the measured value of $\alpha_s(m_Z)$, GUTs without supersymmetry
predict 
$\sin^2\theta_W \sim$ 0.21 to 0.22~\cite{GQW}, whereas GUTs with
supersymmetry at the
TeV scale predict $\sin^2\theta_W \sim$ 0.23~\cite{DRW}, in much better
agreement with the data~\cite{EKN}.

These two experimental arguments buttress the theoretical argument given
earlier, which was based on the hierarchy problem. Put together, these
provide ample motivation for studying the phenomenology of the MSSM in
more detail, as we do in the next Lecture.

\section{PHENOMENOLOGY OF SUPERSYMMETRY}

\subsection{Soft Supersymmetry Breaking}

The first issue that must be addressed in the phenomenology of
supersymmetry is the sad fact
that no sparticles have ever been detected. This means
that sparticles do not weigh the same as their supersymmetric partners:
$m_{\tilde e} \not= m_e,~~m_{\tilde\gamma} \not= m_\gamma$, etc., and
hence that supersymmetry must be broken. We return in Lecture 5 to review
some theoretical ideas about the origin of supersymmetry breaking,
restricting ourselves here to a phenomenological
parametrization~\cite{DG}. Any
such parametrization  should retain the desirable features of
supersymmetry, particularly the absence of power-law divergences. This
``softness" requirement means that any supersymmetry-breaking
interactions ${\cal L}_{\rm susy X}$ should have quantum field dimension
$<4$ (recall that the quantum field dimension of a boson (derivative)
 (fermion) is $1(1)({3\over 2}$)), and
hence a positive power of some numerical mass parameters, so that $\int
d^4 x {\cal L}_{\rm susy X}$ is dimensionless. There are in fact further
restrictions on soft supersymmetry-breaking parameters~\cite{Grisaru}, and
a general
parametrization comprises scalar mass terms:
$m^2_{0_i}\vert\phi_i\vert^2$, gaugino masses: ${1\over 2} M_a~\tilde
V^T_a~C\tilde V_a$, and trilinear or bilinear scalar interactions
proportional to superpotential terms: $A_\lambda \lambda \phi^3$, 
$B_\mu \mu \phi^2$. Note some absences from this list, including masses
for fermions in chiral supermultiplets $m_\psi \psi^T C \psi$ and
non-analytic trilinear scalar couplings $\propto \phi^*\phi^2$.

We shall adopt for now the hypothesis (to be discussed in Lecture 5) that
the soft supersymmetry-breaking masses $m^2_{0_i}, M_a, A_\lambda ,
B_\mu$ originate at some high GUT or gravity scale, perhaps from some
supergravity or superstring mechanism. The physical values of the soft
supersymmetry-breaking parameters are then subject to logarithmic
renormalizations that may be calculated and resummed using the
renormalization-group techniques familiar from QCD~\cite{Inoue}, which
also figure in
the GUT calculations of $\sin^2\theta_W$ that are reviewed in Lecture 4.
Renormalizations by gauge interactions have the general structure
\beq
m^2_{0_i} \rightarrow m^2_{0_i} + C^a_i M^2_a~,~~ M_a \rightarrow
{\alpha_a\over\alpha_{GUT}}~M_a
\label{threeone}
\eeq
at the one-loop level, and higher-loop renormalizations are also well
understood~\cite{Jack}.

It is often assumed that the soft supersymmetry-breaking masses are
universal at the GUT or supergravity scale:
\beq
m^2_{0_i} \equiv m^2_0~,~~ M_a\equiv m_{1/2}~,~~A_\lambda\equiv A~,~~
B_\mu \equiv B
\label{threetwo}
\eeq
but this hypothesis is not very well motivated, since, in particular,
general supergravity models
give no theoretical hint why they should be universal. Some superstring
models give
hints of universality for the gaugino masses $m_{1/2}$, but
universality for the scalar masses $m^2_{0_i}$ is more questionable. Since a
high degree of universality is suggested (at least for the first two
generations) by flavour-changing neutral-current (FCNC)
constraints~\cite{EN}, this
provides some  impetus for models guaranteeing scalar-mass universality, such
as the gauge-mediated or messenger models~\cite{GR} discussed briefly in
Lecture 5. If
one assumes universality, the parameters $\mu , \tan\beta , m_0 ,
m_{1/2}, A$ suffice to characterize MSSM phenomenology.

Figure 13 shows the results of some typical renormalization-group
calculations assuming universal inputs~\cite{RGE}. We see that scalar
masses are
generally renormalized to larger values as the scale is reduced, but this
is not necessarily the case if there are large Yukawa interactions such as
those of the top quark, which may modify (\ref{threeone}) in the case of
Higgs masses. Such Yukawa effects involving the top quark must certainly be
taken into account, and could also be important for the bottom quark and the
$\tau$ lepton if
$\tan\beta$ is large. The potential significance of these Yukawa
interactions is that they tend to drive $m^2_H$ to smaller
values at smaller renormalization scales $\mu$~\cite{IR} 
\beq
\mu {d \over {\rm ln} \mu} m_h^2 = {1 \over (4 \pi)^2} \left( 3
\lambda_t^2
(m_h^2 + m_{\tilde q}^2 + m_t^2) + \dots \right)
\label{threethree}
\eeq
where $m_{\tilde q}$ is a squark mass.

\begin{figure}
\centerline{\includegraphics[height=3in]{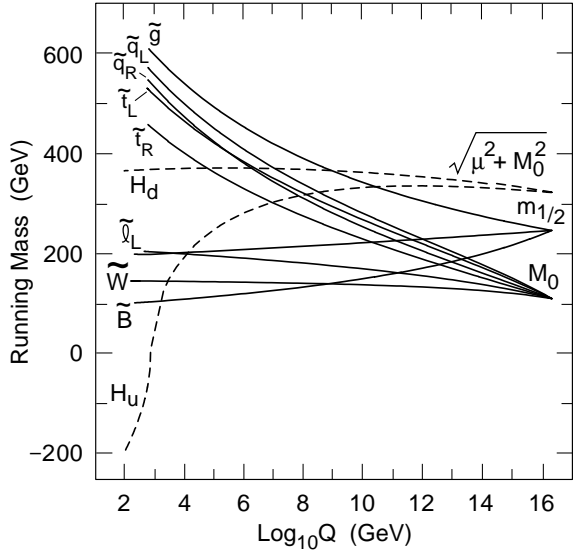}}
\caption[]{Renormalization-group evolution of soft supersymmetry-breaking mass
parameters~\cite{RGE}.}
\end{figure}

This makes it possible to generate electroweak symmetry breaking
dynamically, even if $m^2_H > 0$ at the input scale along with the other
scalar mass-squared parameters~\cite{IR}, as seen in Fig. 13. The
appropriate
renormalization scale for discussing the effective Higgs potential of the
MSSM is $Q\lappeq$ 1 TeV, and the electroweak gauge symmetry will be
broken if either or both of $m^2_{H_{1,2}}(Q) < 0$, as in the model
potential (\ref{onefive}). This is certainly possible for $m_t \sim$ 175
GeV as observed.

\subsection{Supersymmetric Higgs Bosons}

As was discussed in Lecture 2, one expects two complex Higgs doublets
$H_2\equiv (H^+_2 , H^0_2)~,~~H_1\equiv (H^+_1 , H^0_1)$ in the MSSM,
with a total of 8 real degrees of freedom. Of these, 3 are eaten via the
Higgs mechanism to become the longitudinal polarization states of the
$W^\pm$ and $Z^0$, leaving 5 physical Higgs bosons to be discovered by
experiment. Three of these are neutral: the lighter CP-even neutral $h$,
the heavier CP-even neutral $H$, the CP-odd neutral $A$, and charged
bosons $H^\pm$. The quartic potential is completely determined by the $D$
terms (\ref{twotwentytwo})
\beq
V_4 = {g^2 + g^{\prime 2}\over 8}~~\left( \vert H^0_1\vert^2 - \vert
H^0_2\vert^2 \right)
\label{threefour}
\eeq
for the neutral components, whilst the quadratic terms may be
parametrized at the tree level by
\beq
{1\over 2} = m^2_{H_1}~\vert H_1\vert^2 + m^2_{H_2}~\vert H_2\vert^2 +
(m^2_3~H_1H_2 + {\rm herm.conj.})
\label{threefive}
\eeq
where $m^2_3 = B_\mu\mu$. One combination of the three parameters
$(m^2_{H_1},m^2_{H_2},m^2_3)$ is fixed by the Higgs vacuum expectation $v
= \sqrt{v^2_1+v^2_2}$ = 246 GeV, and the other two combinations may be
rephrased as $(m_A,\tan\beta)$. These characterize all Higgs masses and
couplings in the MSSM at the tree level. Looking back at
(\ref{onefifteen}), we see that the gauge coupling strength of the
quartic interactions (\ref{threefour}) suggests a relatively low mass for
at least the lightest MSSM Higgs boson $h$, and this is indeed the case,
with $m_h \leq m_Z$ at the tree level:
\beq
m^2_h = m^2_Z~\cos^2 2\beta
\label{threesix}
\eeq
This raised considerable hope that the lightest MSSM Higgs boson could be
discovered at LEP, with its prospective reach to $m_H \sim$ 100 GeV.

However, radiative corrections to the Higgs masses are calculable in a
supersymmetric model (this was, in some sense, the whole point of
introducing supersymmetry!), and they turn out to be non-negligible for
$m_t \sim$ 175 GeV~\cite{susyHiggs}. Indeed, the leading one-loop
corrections to $m^2_h$ depend quartically on $m_t$:
\beq
\Delta m^2_h = {3m^4_t\over 4\pi^2v^2}~~\ln~~\left({m_{\tilde t_1} m_{\tilde t_2}\over
m^2_t}\right) + {3m^4_t \hat A^2_t\over 8\pi^2 v^2}~~\left[2h(m^2_{\tilde t_1}, m^2_{\tilde
t_2})+ \hat A^2_t ~~f(m^2_{\tilde t_1}, m^2_{\tilde t_2})\right] + \ldots
\label{threeseven}
\eeq
where $m_{\tilde t_{1,2}}$ are the physical masses of the two stop
squarks $\tilde t_{1,2}$ to be discussed in more detail shortly, $\hat A_t
\equiv A_t - \mu \cot\beta$, and
\beq
h(a,b) \equiv {1\over a-b}~\ln \left({a\over b}\right)~,~~f(a,b) =
{1\over (a-b)^2}~\left[2 - {a+b\over a-b}~\ln\left({a\over
b}\right)\right]
\label{threeeight}
\eeq
Non-leading one-loop corrections to the MSSM Higgs masses are also known,
as are corrections to coupling vertices, two-loop corrections and
renormalization-group resummations~\cite{moreradcorr}. For $m_{\tilde
t_{1m2}} \lappeq$ 1
TeV and a plausible range of $A_t$, one finds
\beq
m_h \lappeq 130~{\rm GeV}
\label{threenine}
\eeq
as seen in Fig. 14. There we see the sensitivity of $m_h$ to $(m_A,
\tan\beta)$, and we also see how $m_A, m_H$ and $m_{H^\pm}$ approach each
other for large $m_A$.

\begin{figure}
\centerline{\includegraphics[height=3in]{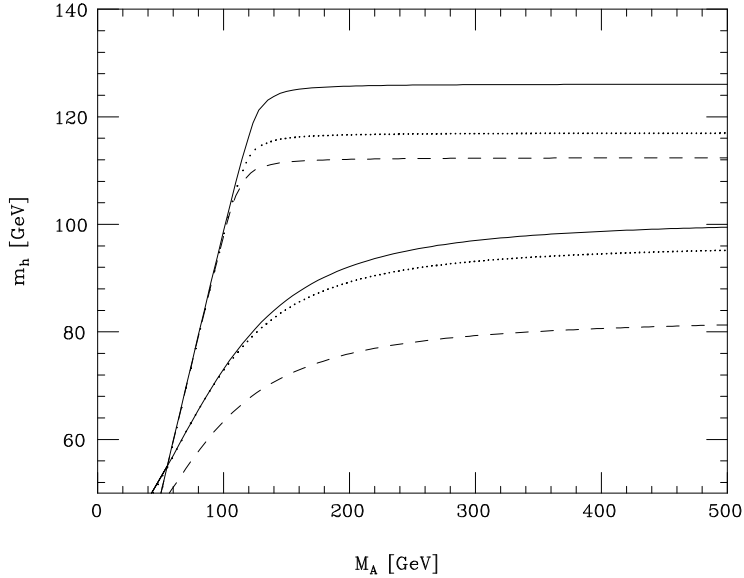}}
\caption[]{The lightest Higgs boson mass in the MSSM, for different values of
$\tan\beta$ and the CP-odd Higgs boson mass $M_A$~\cite{LEP2YB}.}
\end{figure}

The radiative corrections (\ref{threeseven}), (\ref{threeeight}) have
major implications for experiments and accelerators. They may push the
MSSM Higgs sector beyond the reach of LEP~2 and into the lap of the
LHC~\cite{LHC}.
They motivate the optimization of LHC detectors for the Higgs mass range
(\ref{threenine}). They may motivate the orientation of future $e^+e^-$
linear-collider construction so as to study such an MSSM Higgs boson in
more detail than is possible at the LHC~\cite{LC}.

The decay modes of the MSSM Higgs bosons have been carefully studied, as
seen in Fig. 15~\cite{LC}. Like the single Higgs boson of the \sm, the
lightest
MSSM Higgs boson $h$ prefers to decay into the heaviest particles
available, typically $h\rightarrow \bar bb$, and this has been the
primary focus of searches at LEP~2. However, there are ``blind spots" in
MSSM parameter space where this decay mode is suppressed by
cancellations, complicating the search at LEP~2. Ignoring this possible
complication, Fig. 16 shows the regions of MSSM parameter that may be
explored at LEP~2 for different centre-of-mass energies and
luminosities.
After LEP~2, the Fermilab Tevatron collider has a chance of observing $h
\rightarrow \bar bb$ and possibly other decays~\cite{TevII}, if it
accumulates sufficient luminosity. The
potential for LHC searches for MSSM  Higgs bosons is shown in Fig. 17
for one choice of the MSSM parameters~\cite{LHC}. We see that the entire
parameter
space is covered at maximum luninosity, though with considerable
reliance on the rare decay mode $h \rightarrow\gamma\gamma$.

\begin{figure}
\centerline{\includegraphics[height=3in]{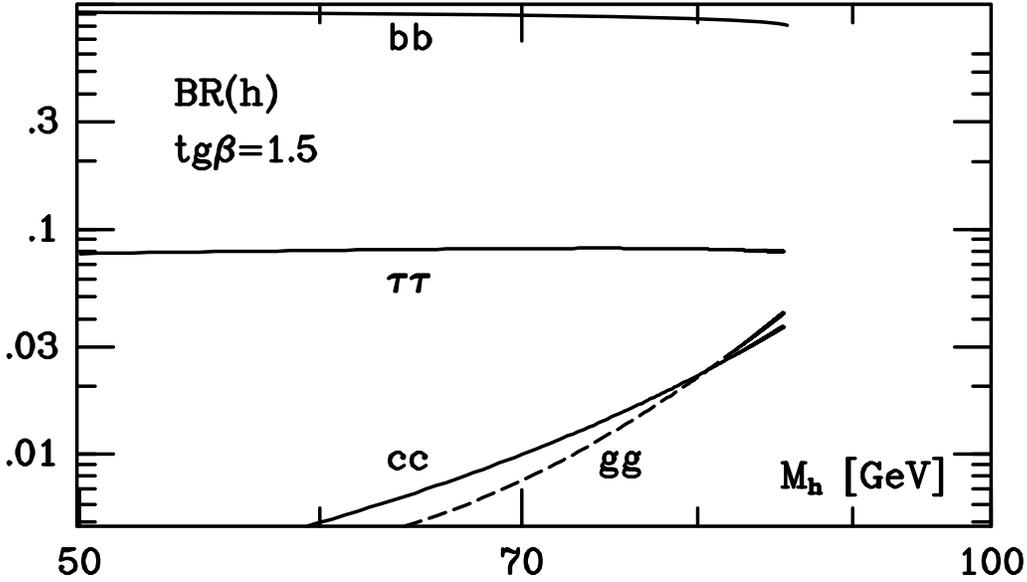}}
\caption[]{The expected decay modes of the lightest MSSM Higgs boson
$h$~\cite{LC}.}
\end{figure}

\begin{figure}
\centerline{\includegraphics[height=3in]{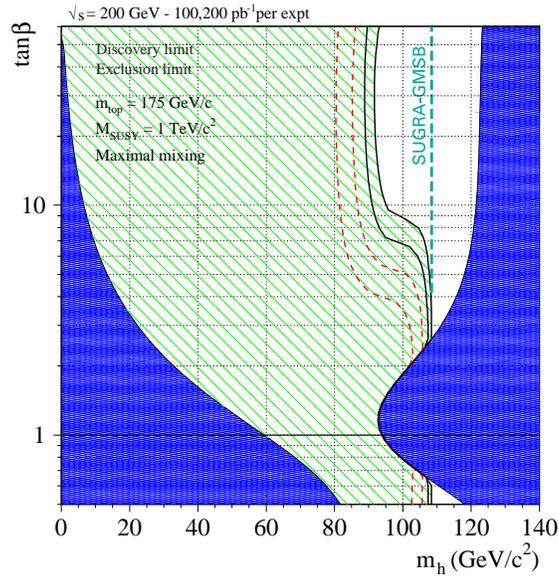}}
\caption[]{Prospective discovery and exclusion limits for MSSM Higgs searches
at LEP.}
\end{figure}

\begin{figure}
\centerline{\includegraphics[height=3in]{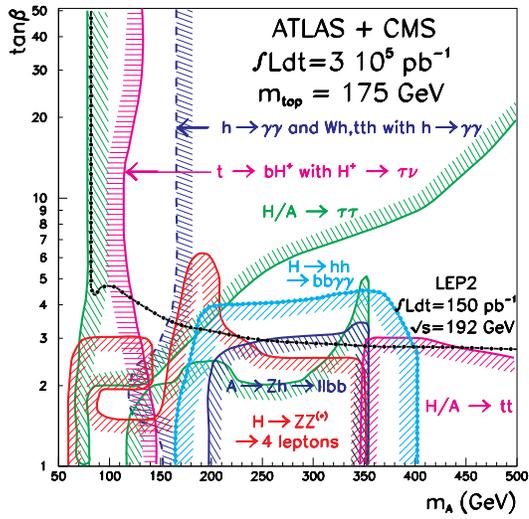}}
\caption[]{Prospective coverage of the MSSM Higgs sector at the
LHC~\cite{LHC}.}
\end{figure}

\subsection{Sparticle Masses and Mixing}

We now progress to a more complete discussion of sparticle masses and
mixing.

\noindent
{\bf Sfermions} : Each flavour of charged lepton or quark has both
left- and right-handed components $f_{L,R}$, and these have separate
spin-0 boson superpartners $\tilde f_{L,R}$. These have different
isospins $I = {1\over 2},~0$, but may mix as soon as the electroweak
gauge symmetry is broken. Thus, for each flavour we should consider a
$2\times 2$ mixing matrix for the 
$\tilde f_{L,R}$, which takes the following general form~\cite{ER}:
\beq
M^2_{\tilde f} \equiv \left( \matrix{m^2_{\tilde f_{LL}} & m^2_{\tilde
f_{LR}} \cr \cr m^2_{\tilde f_{LR}} & m^2_{\tilde f_{RR}}}\right)
\label{threeten}
\eeq
The diagonal terms may be written in the form
\beq
m^2_{\tilde f_{LL,RR}} = m^2_{\tilde f_{L,R}} + m^{D^2}_{\tilde
f_{L,R}} + m^2_f
\label{threeeleven}
\eeq
where $m_f$ is the mass of the corresponding fermion, $\tilde
m^2_{\tilde f_{L,R}}$ is the soft supersymmetry-breaking mass discussed
in the previous section, and $m^{D^2}_{\tilde f_{L,R}}$ is a
contribution due to the quartic $D$ terms in the effective potential:
\beq
m^{D^2}_{\tilde f_{L,R}} = m^2_Z~\cos 2\beta~~(I_3 + \sin^2\theta_WQ_{em})
\label{threetwelve}
\eeq
where the term $\propto I_3$ is non-zero only for the $\tilde f_L$.
Finally, the off-diagonal mixing term takes the general form
\beq
m^{2}_{\tilde f_{L,R}} = m_f \left(A_f +
\mu^{\tan\beta}_{\cot\beta}\right)~~{\rm for}~~f =
^{e,\mu,\tau,d,s,b}_{u,c,t}
\label{threethirteen}
\eeq
It is clear that $\tilde f_{L,R}$ mixing is likely to be important for
the $\tilde t$, and it may also be important for the $\tilde b_{L,R}$ and
$\tilde\tau_{L,R}$ if $\tan\beta$ is large. We also see from
(\ref{threeeleven}) that the diagonal entries for the $\tilde t_{L,R}$
would be different from those of the $\tilde u_{L,R}$ and $\tilde
c_{L,R}$, even if their soft supersymmetry-breaking masses were
universal, because of the $m^2_f$ contribution. In fact, we also expect
non-universal renormalization of $m^2_{\tilde t_{LL,RR}}$ (and also 
$m^2_{\tilde b_{LL,RR}}$ and $m^2_{\tilde \tau_{LL,RR}}$ if $\tan\beta$
is large), because of Yukawa effects analogous to those discussed in the
previous section for the renormalization of the soft Higgs masses.

For these reasons, the $\tilde t_{L,R}$ are not usually assumed to be
degenerate with the other squark flavours. Indeed, one of the $\tilde
t$ could well be the lightest squark, perhaps even lighter than the $t$
quark itself~\cite{ER}.  The mass limits~\cite{LEPC}
combined in Fig. 18 assume degenerate $(\tilde u, \tilde d, \tilde s,
\tilde c, \tilde b)_{L,R}$, even though this degeneracy should also be
broken by the flavour-universal $D$ terms (\ref{threetwelve}) and by
renormalization effects that are different for $\tilde f_{L,R}$. The search
for the stop mass eigenstates
$\tilde t_{1,2}$ requires a separate analysis. Figure
19 shows the experimental lower limits on $m_{\tilde t_1}$ from ALEPH and
$D\phi$ for different assumed values of the $\tilde t$ mixing angle
$\theta_{\tilde t}$~\cite{LEPC}, and assuming that $\tilde t \rightarrow c
\chi$ decay dominates, where $\chi$ is the lightest neutralino.

\begin{figure}
\centerline{\includegraphics[height=3in]{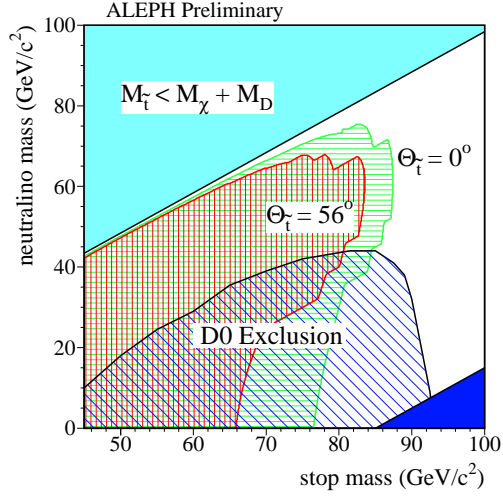}}
\caption{Excluded domains in the search for light stops and neutralinos, as a
function of the stop mixing angle $\theta_{\tilde t}$~\cite{LEPC}.}
\end{figure}

\begin{figure}
\centerline{\includegraphics[height=3in]{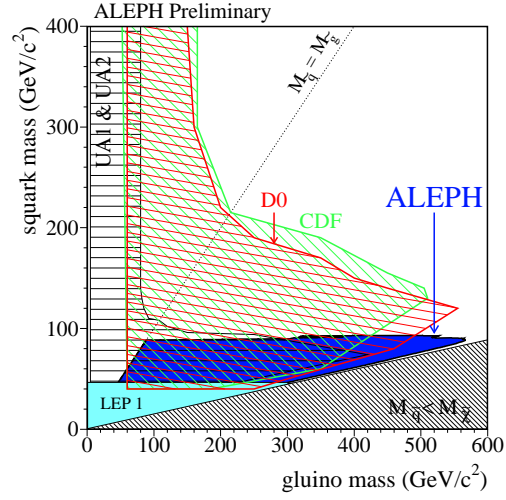}}
\caption[]{Excluded regions in gluino and squark masses~\cite{LEPC}.}
\end{figure}

\noindent
{\bf Charginos}: These are the supersymmetric partners of the
$W^\pm$ and $H^\pm$, which mix through a $2\times 2$ matrix
\beq
-{1\over 2} ~(\tilde W^-, \tilde H^-)~~M_C ~~\left(\matrix{\tilde
W^+\cr\tilde H^+}\right)~~+~~{\rm herm.conj.}
\label{threeforteen}
\eeq
where
\beq
M_C \equiv \left(\matrix{M_2 & \sqrt{2} m_W\sin\beta \cr \sqrt{2}
m_W\cos\beta & \mu}\right)
\label{threefifteen}
\eeq
Here $M_2$ is the unmixed $SU(2)$ gaugino mass and $\mu$ is the Higgs
mixing parameter introduced in (\ref{twotwentyfour}). Figure 20
displays (among other lines to be discussed later) the contour
$m_{\chi^\pm}$ = 91 GeV for the lighter of the two chargino mass
eigenstates~\cite{EFGOS}. Some recent experimental lower limits on
$m_{\chi^\pm}$ as
functions of the other MSSM parameters are shown in Fig.
21~\cite{LEPC}.

\begin{figure}
\centerline{\includegraphics[height=5in]{EllisScotfig20ab.eps}}
\caption[]{The $(\mu , M_2)$ plane characterizing charginos and neutralinos,
for (a) $\mu < 0$ and (b) $\mu > 0$, including contours of $m_\chi$ and
$m_{\chi^\pm}$, and of neutralino purity~\cite{EFGOS}.}
\end{figure}

\begin{figure}
\centerline{\includegraphics[height=5in]{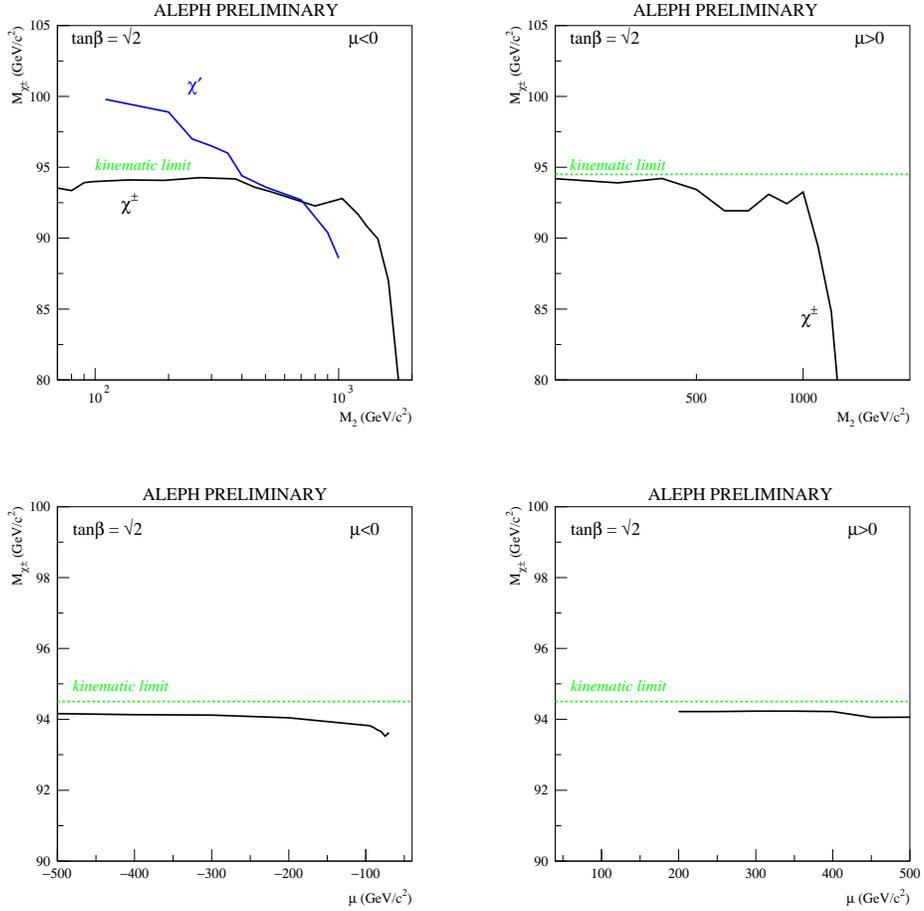}}
\caption[]{Lower limits on the chargino mass inferred in the higgsino region
(top panels) and the gaugino region (bottom panels)~\cite{LEPC}.}
\end{figure}

\noindent
{\bf Neutralinos}: These are characterized by a $4\times 4$ mass
mixing matrix~\cite{EHNOS}, which takes the following form in the $(\tilde
W^3, \tilde
B, \tilde H^0_2, \tilde H^0_1)$ basis :
\beq
m_N = \left( \matrix{
M_2 & 0 & {-g_2v_2\over\sqrt{2}} & {g_2v_1\over\sqrt{2}}\cr\cr
0 & M_1 & {g^\prime v_2\over\sqrt{2}} & {-g^\prime v_1\over\sqrt{2}}\cr\cr
{-g_2 v_2\over\sqrt{2}} & {g^\prime v_2\over\sqrt{2}} & 0 & \mu \cr\cr
{g_2v_1\over\sqrt{2}} & {-g^\prime v_1\over\sqrt{2}} & \mu & 0}\right)
\label{threesixteen}
\eeq
Note that this has a structure similar to $M_C$ (\ref{threefifteen}), but 
with its entries replaced by $2\times 2$ submatrices. As has already been
mentioned, one conventionally assumes that the $SU(2)$ and $U(1)$ gaugino
masses $M_{1,2}$ are universal at the GUT or supergravity scale, so that
\beq
M_1 \simeq M_2~~{\alpha_1\over\alpha_2}
\label{threeseventeen}
\eeq
so the relevant parameters of (\ref{threesixteen}) are generally taken to
be $M_2 = (\alpha_2/ \alpha_{GUT}) m_{1/2}$, $\mu$ and
$\tan\beta$.

Figure 20 also displays contours of the mass of the lightest neutralino
$\chi$, as well as contours of its gaugino and Higgsino
contents~\cite{EFGOS}. In the
limit $M_2\rightarrow 0$, $\chi$ would be approximately a photino and it
would be approximately a Higgsino in the limit $\mu\rightarrow 0$.
Unfortunately, these idealized limits are excluded by unsuccessful LEP
and other searches for neutralinos and charginos, as we now discuss in
more detail.

\subsection{The Lightest Supersymmetric Particle}

This is expected to be stable in the MSSM,  and hence should be present in
the Universe today as a cosmological relic from the Big
Bang~\cite{Goldberg,EHNOS}. Its stability
arises because there is a multiplicatively-conserved quantum number called
$R$ parity, that takes the values +1 for all conventional particles and -1
for all sparticles~\cite{Fayet}. The conservation of $R$ parity can be
related to that of
baryon number $B$ and lepton number $L$, since
\beq
R = (-1)^{3B+L+2S}
\label{threeeighteen}
\eeq
where $S$ is the spin. Note that $R$ parity could be violated either
spontaneously if $<0\vert\tilde\nu\vert 0> \not= 0$ or
explicitly if one
of the supplementary couplings (\ref{twotwentyfive1}) is
present. There
could also be a coupling $HL$, but this can be defined away be choosing a
field basis such that $\bar H$ is defined as the superfield with a
bilinear coupling to $H$. Note that $R$ parity is not violated by the
simplest models for neutrino masses, which have $\Delta L = 0, \pm 2$,
nor by the simple GUTs discussed in the next Lecture, which violate
combinations of $B$ and $L$ that leave $R$ invariant.
There are three important consequences of $R$ conservation: 
\begin{enumerate}
\item sparticles are always produced in pairs, e.g., $\bar
pp\rightarrow\tilde q \tilde g X$, $e^+e^-\rightarrow \tilde\mu +
\tilde\mu^-$, 
\item heavier sparticles decay to lighter ones, e.g., $\tilde q \rightarrow
q\tilde g, \tilde\mu\rightarrow\mu\tilde\gamma$, and 
\item the lightest sparticles is stable, 
\end{enumerate}
because it has no legal decay
mode.

This feature constrains strongly the possible nature of the lightest
supersymmetric sparticle. If it had either electric charge or strong
interactions, it would surely have dissipated its energy and condensed
into galactic disks along with conventional matter. There it would surely
have bound electromagnetically or via the strong interactions to
conventional nuclei, forming anomalous heavy isotopes that should have
been detected. There are upper limits on the possible abundances of such
bound relics, as compared to conventional nucleons~\cite{isotopes}:
\beq
{n({\rm relic})\over n(p)} \lappeq 10^{-15}~~{\rm to}~~10^{-29}
\label{threenineteen}
\eeq
for 1 GeV $\lappeq m_{\rm relic} \lappeq$ 1 TeV. These are far below the
calculated abundances of such stable relics:
\beq
{n({\rm relic})\over n(p)} \gappeq 10^{-6}~~(10^{-10})
\label{threetwenty}
\eeq
for relic particles with electromagnetic (strong) interactions. We may
conclude~\cite{EHNOS} that any supersymmetric relic is probably
electromagnetically neutral with only weak interactions, and could in
particular not be a gluino. Whether the lightest hadron containing
a gluino is charged or neutral, it would surely bind to some nuclei.
Even if one pleads for some level of fractionation, it is difficult
to see how such gluino nuclei could avoid the stringent bounds
established for anomalous isotopes of many species~\cite{isotopes}.

Plausible scandidates of different spins are the sneutrinos $\tilde\nu$
of spin 0, the lightest neutralino $\chi$ of spin 1/2, and the
gravitino $\tilde G$ of spin 3/2. The sneutrinos have been
ruled out by the combination of LEP experiments and direct searches for
cosmological relics. Neutrino counting (\ref{oneforteenone}) requires
$m_{\tilde\nu}\gappeq$ 43 GeV~\cite{EFOS}, in which case the direct relic
searches in
underground low-background experiments require  
$m_{\tilde\nu}\gappeq$ 1 TeV~\cite{uground}. The gravitino cannot be ruled
out, and its
popularity has revived somewhat with the renaissance of gauge-mediated
(messenger) models~\cite{GR}, as described in Lecture 5. For the rest of
this Lecture, however, we condentrate on the neutralino possibility.

A very attractive feature of the neutralino candidature for the lightest
supersymmetric particle is that it has a relic density of interest to
astrophysicists and cosmologists: $\Omega_\chi h^2 = {\cal O}(0.1)$ over
generic domains of the MSSM parameter space~\cite{EHNOS}. This feature is
seen clearly in
Fig.~22, where $0.1 < \Omega_\chi h^2 < 0.3$ is possible in a large area of
the $(\mu,M_2)$ plane for suitable choices of the other MSSM
parameters~\cite{EFGOS}.
In this domain, the lightest neutralino $\chi$ could constitute the cold
dark matter favoured by theories of cosmological structure
formation~\cite{structure}.

\begin{figure}
\centerline{\includegraphics[height=5in]{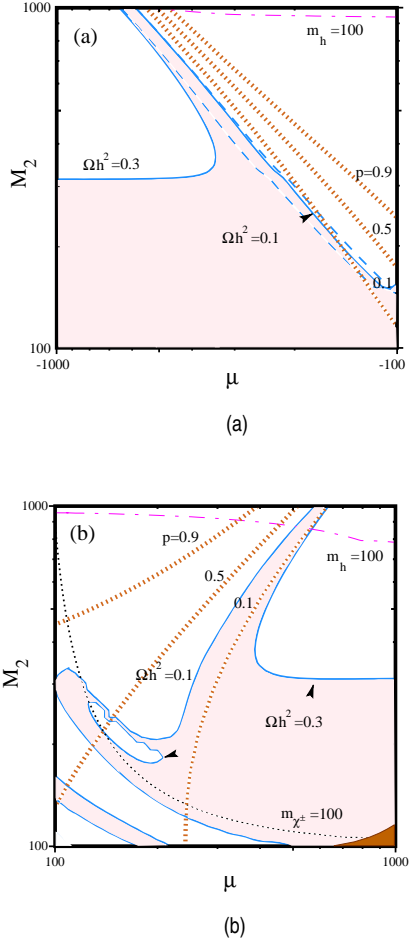}}
\caption[]{Contours of $\Omega h^2, m_h, m_{\chi^\pm}$ and higgsino purity in
the $(\mu , M_2)$ plane, for (a) $\mu < 0$ and (b) $\mu >
0$~\cite{EFGOS}.}
\end{figure}

We have already seen in Fig. 21 some of the experimental limits on
chargino and neutralino production, that may be used to set interesting
limits on $m_\chi$. One example is shown in Fig.~23, where one
particular choice of $m_0$ is assumed~\cite{LEPC}. (This parameter is
relevant because $\tilde \nu$ exchange contributes to
$\sigma(e^+e^-\rightarrow\chi^+\chi^-)$, and the $\tilde \nu$ and $\tilde
e$ masses influence
$\chi^\pm$ decay patterns~\cite{EFOS}.) It is interesting to note in 
Fig.~23 that
LEP~1 data (e.g., neutrino counting in $Z^0$ decays (\ref{oneforteenone}))
did not by themselves provide an absolute lower limit on $m_\chi$: this
became possible only by combining them with higher-energy LEP
data~\cite{EFOS}.

\begin{figure}
\centerline{\includegraphics[height=3in]{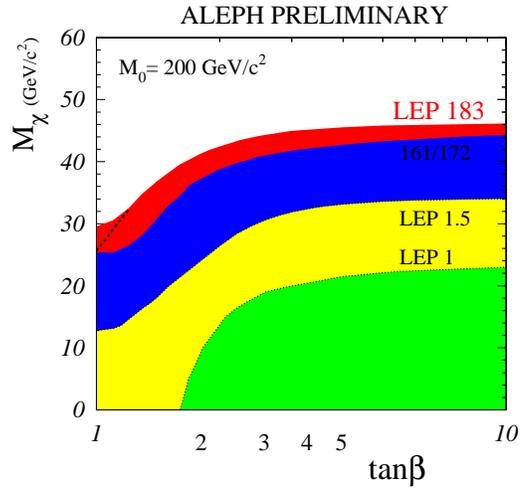}}
\caption[]{Experimental lower limit on the lightest neutralino mass, as a
function of $\tan\beta$, assuming $m_0$~=~200~GeV~\cite{LEPC}.}
\end{figure}

The lower limit on $m_\chi$ can be strengthened by combining the direct
chargino/neutralino searches with other experimental and theoretical
constraints~\cite{EFOS,EFGOS}, as illustrated in Fig. 24. The dotted lines
labelled LEP
are the analogues of Fig. 23, but with $m_0$ allowed to float
freely. The dotted lines marked $H, C$ incorporate the experimental lower
limit on $m_h$ and the cosmological relic-density constraint $\Omega_\chi
h^2 \leq$ 0.3, respectively. The solid lines marked $UHM$ further assume
universal scalar masses for the Higgs multiplets. The lines marked cosmo,
$DM$ combine this assumption with the relic-density assumptions
$\Omega_\chi h^2 < $ 0.3, $>$0.1, respectively. Figure 24 documents a
lower limit $m_\chi >$ 40 GeV~\cite{EFOS2}, which can be strengthened
using more
recent LEP~2 data to about 45 GeV~\cite{EFGOS}. We expect that
higher-energy runs of
LEP will extend this sensitivity to $m_\chi \sim$ 50 GeV. We also see in
Fig. 24 that this type of combined analysis of the MSSM parameter space
imposes an absolute lower limit on $\tan\beta$. Data from LEP that have
been published so far indicate that $\tan\beta\gappeq$ 1.8~\cite{EFGOS},
and future
LEP runs will be sensitive up to $\tan\beta \sim 3$, principally via
Higgs searches.

\begin{figure}
\centerline{\includegraphics[height=3in]{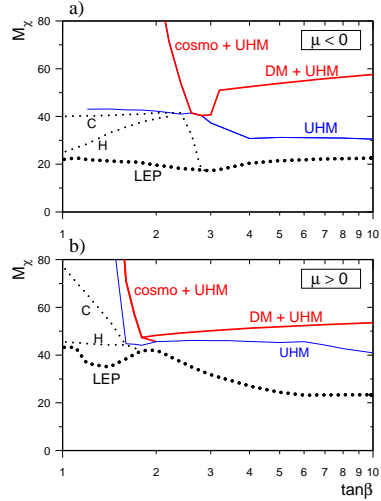}}
\caption[]{Lower limits on $m_\chi$ from LEP (thick dots), Higgs searches
(H), cosmological limits on $\Omega h^2$ (c), which are strengthened by
assuming universal scalar masses for Higgs bosons (UHM), also in combination
with the constraints $\Omega h^2 < 0.3$ (cosmo + UHM) and $\Omega h^2 > 0.1$
(DM + UHM)~\cite{EFOS}.}
\end{figure}

We close with a few comments on the prospects for sparticle searches at
the LHC. These should be able to extend the squark and gluino searches up
to masses on 2.5 GeV, as seen in Fig. 25~\cite{Abdullin}. By looking in
several
different channels: missing energy $E^{miss}_T$ with 0, 1, 2, etc.
leptons, it should be possible to explore several times over the domain of
parameter space of interest to cosmologists where $\Omega_\chi h^2
\lappeq 0.4$, as also seen in Fig. 25~\footnote{This statement may
require some re-examination in the light of co-annihilation effects on
the relic $\chi$ density~\cite{EFO}.}.
Moreover, it should be possible
to reconstruct several different sparticles via the cascade decays of
squarks and gluinos, and even make detailed mass
measurements that could
test supergravity mass relations~\cite{Hinchliffe}.  

\begin{figure}
\centerline{\includegraphics[height=4in]{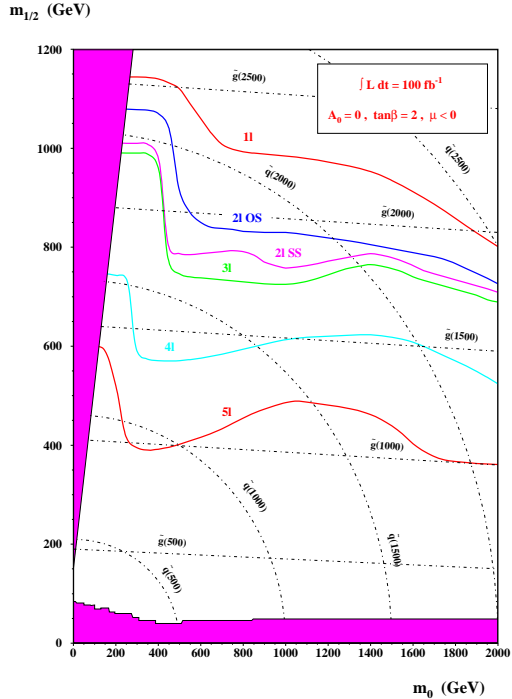}}
\caption[]{Regions of the $(m_0, m_{1/2})$ plane that can be explored at the
LHC~\cite{Abdullin}.}
\end{figure}

\subsection{The ``Anomaly" that Went Away$\ldots$}

For some time, measurements of $R_b = \Gamma (Z^0\rightarrow \bar
bb)/\Gamma (Z \rightarrow$ all hadrons) seemed to be in significant
disagreement with the \sm, generating considerable interest. It was
suggested that the discrepancy might be explicable by one-loop
supersymmetric radiative corrections, due either to Higgs exchange if
$m_A$ were small and $\tan\beta$  large, or to chargino and stop
exchange if both $m_{\chi^\pm}$ and $m_{\tilde t}$ were small, as well
as $\tan\beta$~\cite{susyRb}. The Higgs former scenario was early
effectively excluded
by early Higgs searches at LEP, but the $\chi^\pm / \tilde t$ scenario
fitted well with theoretical prejudices and survived somewhat longer. It
was particularly interesting, because it suggested that either a chargino
or a stop might be light enough to be produced at LEP~2 or at the
Fermilab Tevatron collider.

As time has progressed, the $R_b$ anomaly has steadily decreased in
significance, and is now barely a one-$\sigma$ discrepancy, as seen in
Fig. 26~\cite{LEPEWWG}. In parallel, both LEP~2 and the Tevatron have
explored considerable domains of MSSM parameter space, excluding
significant domains of $m_{\chi^\pm} $ and $m_{\tilde t}$. Might there
still be a significant supersymmetric contribution to $R_b$, comparable
to the experimental error $\Delta R_b\sim$ 0.0010 ? 
Even before the latest exclusion domains from LEP {\sl et al}.,
we~\cite{ELN} found that, of
about 500,000 possible choices of the basic MSSM model parameters, only
210 of those that respected the experimental constraints in early 1997
(including $b\rightarrow s\gamma$ and the cosmological relic density)
could yield $\Delta R_b \geq$ 0.0010. This already made the case for a
significant supersymmetric contribution  to $R_b$ appear somewhat
implausible. (Though everybody would have been happy if one of these
``unusual" models was close to reality!) Another possible strike against
these models was that they required a departure from the universality
assumptions favoured in supergravity models, as seen in Fig. 27, where
the ``globular cluster" of ``interesting" models with $\Delta R_b >$
0.0010 is outside the zone of parameter space accessible in such
universal supergravity models, which can only yield~\cite{ELN}
\beq
\Delta R_b < 0.0003
\label{threenine1}
\eeq

\begin{figure}
\centerline{\includegraphics[height=3in]{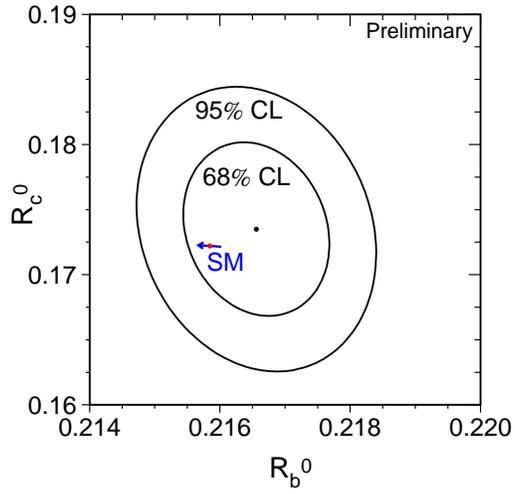}}
\caption[]{Experimental constraints on $Z\rightarrow\bar bb$ and $\bar cc$
decays compared with the \sm ~prediction~\cite{LEPEWWG}.}
\end{figure}

\begin{figure}
\centerline{\includegraphics[height=3in]{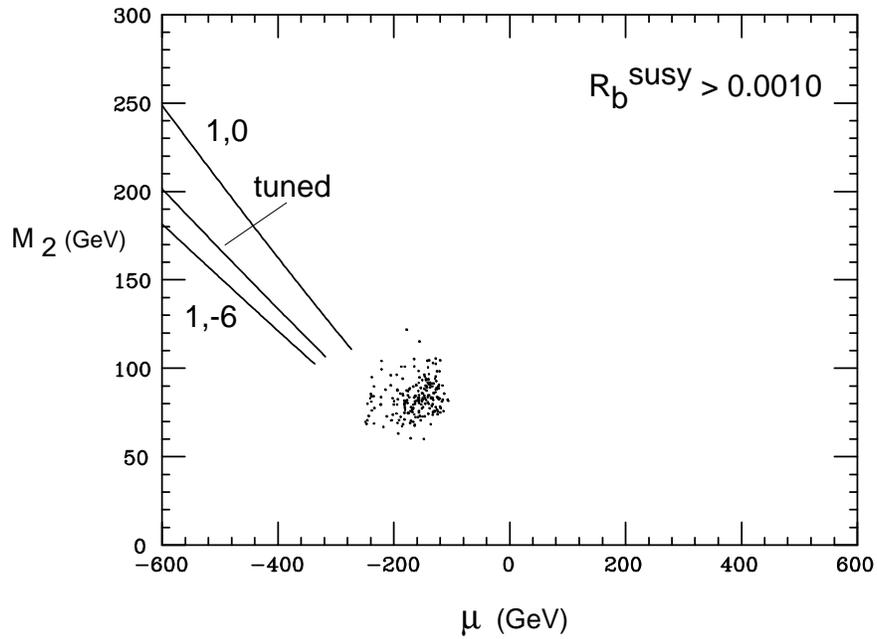}}
\caption[]{Region of the $(\mu,M_2)$ plane where MSSM contributions to $R_b$
larger than 0.0010 are possible which cannot be attained in models with
universal scalar masses (solid lines)~\cite{ELN}.}
\end{figure}

Moreover, the ``interesting" models all had $m_{\tilde t_1} <$ 100 GeV,
and 90\%  of them have now been excluded by the more sensitive LEP~2
searches shown in Fig. 19. The conclusion must be that plausible
parameter choices for the MSSM do not yield a significant contribution to
$R_b$, and hence that it is legitimate to use the measurements in a global
fit
to the precision electroweak data in the \sm, as was assumed in Lecture 1.

\section{GRAND UNIFICATION}

\subsection{Basic Strategy}

The philosophy of grand unification~\cite{GG,PS} is to seek a simple gauge
group that
includes the untidy $SU(3)$, $SU(2)$ and $U(1)$ gauge groups of QCD and
the electroweak sector of the \sm. The hope is that this grand
unification can be achieved while neglecting gravity, at least as a first
approximation. If the grand unification scale turns out to be
significantly less than the Planck mass, this is not obviously a false
hope. We discuss later in this Lecture and the next the extent to which
this hope is indeed realistic: for the moment we just note that the grand
unification scale is indeed expected to be exponentially large:
\beq
{m_{GUT}\over m_W} = \exp \left( {\cal
O}\left({1\over\alpha_{em}}\right)\right)
\label{fourone}
\eeq
and typical estimates will be that $m_{GUT} = 0(10^{16}$ GeV). 
Such a calculation involves an extrapolation of known physics by many
orders of magnitude further than, e.g., the extrapolation that Newton
made from the apple to the Solar System. However, it is not excluded by
our current knowledge of the \sm. For example, we see in Fig. 6 that
the estimate (\ref{onetwentynine}) of the Higgs mass is consistent with
the \sm ~remaining valid all the way to the Planck scale $m_P \simeq
10^{19}$ GeV, and even far beyond.

If the grand unification scale is indeed so large, most tests of it are
likely to be indirect, and we meet some later, such as relations between
\sm ~gauge couplings and between particle masses. Any new interactions,
such as those that might cause protons to decay or give masses to
neutrinos, are likely to be very strongly suppressed.

The first apparent obstacle to the philosophy of grand unification is the
fact that the strong coupling $\alpha_3 = g^2_3/ 4\pi$ is indeed
much stronger than the electroweak couplings at present-day energies:
$\alpha_3 \gg \alpha_2, \alpha_1$. However, you have seen here in the
lectures by Michelangelo Mangano~\cite{Mangano}  that the strong coupling
is
asymptotically free:
\beq
\alpha_3(Q) \simeq {12\pi\over (33-2N_q) \ln (Q^2/\Lambda^2_3)} + \ldots
\label{fourthree}
\eeq
where $N_q$ is the number of quarks, $\Lambda_3 \simeq$ few hundred MeV
is an intrinsic scale of the strong interactions, and the dots in
(\ref{fourthree}) represent higher-loop corrections to the leading
one-loop behaviour shown. The other \sm~ gauge couplings also exhibit
logarithmic violations analogous to (\ref{fourthree}). For example, the
effective value of $\alpha_{em}(m_Z) \sim 1/ 128$, with  
estimated ranges displayed in Fig. 5. The renormalization-group
evolution for the $SU(2)$ gauge coupling is
\beq
\alpha_2(Q) \simeq {12\pi\over (22-2N_q - N_{H/2}) \ln
(Q^2/\Lambda^2_2)} + \ldots
\label{fourfour}
\eeq
where we have assumed equal numbers of quarks and leptons, and $N_H$ is
the number of Higgs doublets. Taking the inverses of (\ref{fourthree})
and (\ref{fourfour}), and then taking their difference, we find
\beq
{1\over\alpha_3(Q)} - {1\over \alpha_2(Q)} = \left({11+N_{H/2}\over
12\pi}\right) \ln \left({Q^2\over m^2_X}\right) + \ldots
\label{fourfive}
\eeq
We have absorbed the scales $\Lambda_3$ and $\Lambda_2$ into a single
grand unification scale $M_X$ where $\alpha_3 = \alpha_2$.

Evaluating (\ref{fourfive}) when $Q = {\cal O}(M_W)$, where $\alpha_3 \gg
\alpha_2 = 0(\alpha_{em})$, we derive the characteristic feature
(\ref{fourone}) that the grand unification scale is exponentially large.
As we see in more detail later, in most GUTs there are new interactions 
mediated by bosons weighing ${\cal O}(m_X)$ that cause protons to decay with
a lifetime $\alpha m^4_X$. In order for the proton lifetime to exceed the
experimental limit, we need $m_X \gappeq 10^{14}$ GeV and hence
$\alpha_{em} \lappeq 1/120$ in (\ref{fourone})~\cite{ENnature}. On the
other hand,
if the neglect of gravity is to be consistent, we need $m_X \lappeq
10^{19}$ GeV and hence $\alpha_{em} \gappeq 1/ 170$ in
(\ref{fourone})~\cite{ENnature}. The fact that the measured value of the
fine-structure
constant $\alpha_{em}\simeq 1/ 137.035 999  
59(38)13)$ lies in this allowed range
may be another hint favouring the GUT philosophy.

Further empirical evidence for grand unification is provided by the
previously-advertized
prediction it makes for the neutral electroweak mixing angle~\cite{GQW}.
Calculating
the renormalization of the electroweak couplings, one finds:
\beq
\sin^2\theta_W = {\alpha_{em}(m_W)\over\alpha_2(m_W)} \simeq {3\over
8}~~\left[ 1 - {\alpha_{em}\over 4\pi}~~{110\over 9} \ln {m^2_X\over
m^2_W}\right]
\label{foursix}
\eeq
which can be evaluated to yield $\sin^2\theta_W \sim$ 0.210 to 0.220,
if there are only \sm ~particles with masses $\lappeq m_X$~\cite{GQW}.
This is to be
compared with the experimental value $\sin^2\theta_W = 0.23155 \pm
0.00019$ shown in Fig. 2. Considering that $\sin^2\theta_W$ could
{\it a~priori}  have had any value between 0 and 1, this is an
impressive qualitative success. The small discrepancy can be removed by
adding some extra particles, such as the supersymmetric particles in the
MSSM.

Another qualitative success is the prediction of the $b$ quark
mass~\cite{CEG,BEGN}. In many
GUTs, such as the minimal $SU(5)$ model discussed shortly, the $b$ quark
and the $\tau$ lepton have equal Yukawa couplings when renormalized at
the GUT sale. The renormalization group then tells us that
\beq
{m_b\over m_\tau} \simeq \left[\ln \left({m^2_b\over m^2_X}
\right)\right]^{12\over 33-2N_q}
\label{fourseven}
\eeq
Using $m_\tau =$ 1.78 GeV, we predict that $m_b\simeq$ 5 GeV, in
agreement with experiment\footnote{This prediction was made~\cite{CEG} 
shortly
before the $b$ quark was discovered. When we received the proofs of
this article, I gleefully wrote by hand in the margin our then
prediction, which was already in the text, as $2~~to~~5$. This was
misread by the typesetter to become 2605: a spectacular disaster!}.
Happily, this prediction remains
successful if the effects of supersymmetric particles are included in the
renormalization-group calculations~\cite{susymb}.

To examine the GUT predictions for $\sin^2\theta_W$, etc. in more detail,
one needs to study the renormalization-group equations beyond the leading
one-loop order. Through two loops, one finds that
\beq
Q~~{\partial\alpha_i(Q)\over\partial Q} = -{1\over 2\pi}~~\left( b_i +
{b_{ij}\over 4\pi}~~\alpha_j(Q)\right)~~\left[\alpha_i(Q)\right]^2
\label{fourseven1}
\eeq
where the $b_i$ receive the one-loop contributions
\beq
b_i = \left(\matrix{ 0 \cr -{22\over 3} \cr -11}\right) + N_g
\left(\matrix{{4\over 3} \cr\cr {4\over 3} \cr\cr {4\over 3}}\right) + N_H
\left(\matrix{{1\over 10} \cr\cr {1\over 6} \cr \cr 0}\right)
\label{foureight}
\eeq
from gauge bosons, $N_g$ matter generations and $N_H$ Higgs doublets,
respectively, and at two loops
\beq
b_{ij} = \left(\matrix{0&0&0\cr\cr 0&-{136\over 3} & 0 \cr\cr
0&0&-102}\right) + N_g
\left(\matrix{{19\over 15} & {3\over 5} & {44\over 15} \cr\cr {1\over 5} & {49\over 3} & 4
\cr\cr {4\over 30} & {3\over 2} & {76\over 3}}\right) +  N_H \left(
\matrix{{9\over 50} &
{9\over 10} & 0 \cr\cr {3\over 10} & {13\over 6} & 0 \cr\cr 0 & 0 & 0}\right)
\label{fournine}
\eeq
These coefficients are all independent of any specific GUT model,
depending only on the light particles contributing to the
renormalization. Including supersymmetric particles as in the MSSM, one
finds~\cite{DRW}
\beq
b_i = \left(\matrix{0 \cr\cr -6 \cr\cr -9}\right) + N_g
\left(\matrix{2\cr\cr 2 \cr\cr 2}\right) + N_H \left(\matrix{{3\over 10}
\cr\cr {1\over 2}\cr\cr 0}\right)
\label{fourten}
\eeq
and
\beq
b_{ij} = \left(\matrix{0&0&0\cr\cr 0&-24 & 0 \cr\cr 0&0&-54}\right) + N_g
\left(\matrix{{38\over 15} & {6\over 5} & {88\over 15} \cr\cr {2\over 5} & 14 & 8
\cr\cr {11\over 5} & 3 & {68\over 3}}\right) +  N_H \left( \matrix{{9\over 50} &
{9\over 10} & 0 \cr\cr {3\over 10} & {7\over 2} & 0 \cr\cr 0 & 0 & 0}\right)
\label{foureleven}
\eeq
again independent of any specific supersymmetric GUT.

One can use these two-loop equations to make detailed calculations of 
$\sin^2\theta_W$ in different GUTs. These
confirm that non-supersymmetric models are not consistent with the 
determinations of the
gauge couplings from LEP and elsewhere~\cite{EKN}. 
Previously, we argued that these models predicted a
wrong value for $\sin^2\theta_W$, given the experimental 
value of $\alpha_3$. In Fig. 28a we
see the converse, namely that extrapolating the experimental 
determinations of the
$\alpha_i$ using the non-supersymmetric 
renormalization-group equations (\ref{foureight}),
(\ref{fournine}) does not lead to a common value at 
any renormalization scale. In contrast,
we see in Fig. 28b that extrapolation using the 
supersymmetric renormalization-group
equations (\ref{fourten}), (\ref{foureleven}) {\bf does} lead to possible
unification
at $m_{GUT} \sim 10^{16}$ GeV~\cite{ADF}.

\begin{figure}
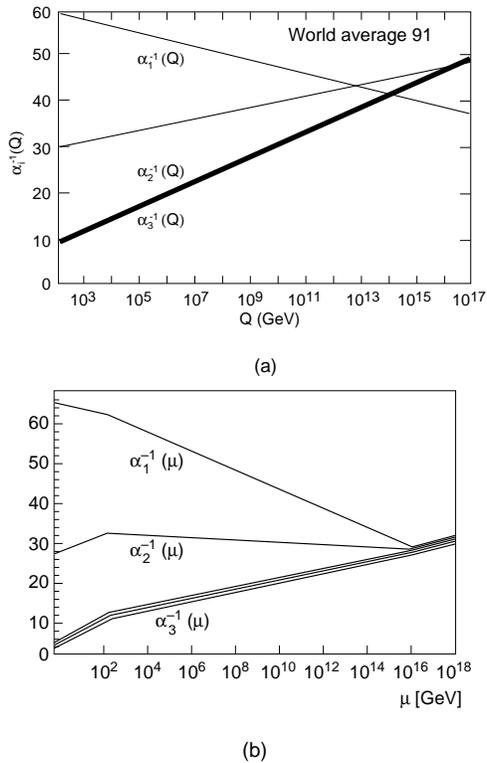

\centerline{\includegraphics[height=2in]{EllisScotfig28a.eps}}
\centerline{\includegraphics[height=2in]{EllisScotfig28b.eps}}
\caption[]{The measurements of the gauge coupling strengths at LEP 
(a) do not 
evolve to a unified value if there is no supersymmetry but do (b) if
supersymmetry is included~\cite{ADF}.} 
\end{figure}

Turning this success around, and assuming 
$\alpha_3 = \alpha_2 = \alpha_1$ at $m_{GUT}$ with
no threshold corrections at this scale, one may estimate that~\cite{EKN2}
\bea
\sin^2\theta_W(M_Z)\bigg\vert_{\overline{\rm MS}} &=& 
0.2029 + {7 \alpha_{em}\over 15
\alpha_3}+ {\alpha_{em}\over 20\pi}
\left[ -3 \ln \left({m_t\over m_Z}\right) + {28\over 3}
\ln \left({m_{\tilde g}\over m_Z}\right) \right.\nonumber \\ \nonumber \\
&&\left. - {32\over 3} \ln \left({m_{\tilde
W}\over m_Z}\right) - \ln \left({m_A\over m_Z}\right) - 
4\ln \left({\mu\over m_Z}\right) +
\ldots
\right]
\label{fourtwelve}
\eea
Setting all the sparticle masses to 1 TeV reproduces approximately the value of
$\sin^2\theta_W$ observed experimentally. 
Can one invert this successful argument to
estimate the supersymmetric particle mass scale? 
One can show~\cite{lump} that the sparticle mass
thresholds in (\ref{fourtwelve}) can be lumped into the parameter
\beq
T_{susy} \equiv \vert\mu\vert \left({m^2_W\over m_{\tilde
g}}\right)^{14/19}~~\left({m^2_A\over \mu^2}\right)^{3/38} ~~
\left({m^2_{\tilde W}\over \mu^2}\right)^{2/19}~~\prod^3_{i=1}~~
\left( {m^3_{\tilde \ell_{Li}} m^7_{\tilde q_i}\over m^2_{\tilde\ell_{R_i}} m^5_{\tilde u_i}
m^3_{\tilde d_i}}\right)^{1/19}
\label{fourthirteen}
\eeq
If one assumes sparticle mass universality at the GUT scale,
then~\cite{lump}
\beq
T_{susy} \simeq \vert\mu\vert \left({\alpha_2\over 
\alpha_3}\right)^{3/2} \simeq {\mu\over 7}
\label{fourforteen}
\eeq
approximately. The measured value of 
$\sin^2\theta_W$ is consistent with $T_{susy} \sim$ 100
GeV to 1 TeV, roughly as expected from the 
hierarchy argument. However, the uncertainties are
such that one cannot use this consistency to 
constrain $T_{susy}$ very tightly~\cite{Zich}. In
particular, even if one accepts the universality hypothesis, 
there could be important
model-dependent threshold corrections around the GUT
scale~\cite{EKN2,GUTthresh}. We are at the limit of what one
can say without studying specific models, so let us now do so.

\subsection{GUT Models}

Before embarking on their study, however, we first clarify some necessary
technical points. As well as looking for a simple unifying group $G
\supset SU(3)\times SU(2)\times U(1)$, we shall be looking for unifying
representations $R$ that contain both quarks and leptons. Since gauge
interactions conserve helicity, any particles with the same helicity are
fair game to appear in any GUT representation $R$, and it is convenient
to work with states of just one helicity, say left-handed. The
left-handed particle content of the \sm~ is as follows. In each
generation, there is a quark doublet $(u,d)_L$ which transforms as (3,2)
of $SU(3)\times SU(2)_L$. Instead of working with the right-handed
singlets $u_R, d_R$ that have (3,1) representations, it is convenient to
work with their antiparticles, which are left-handed: the $u^c_L$ and
$d^c_L$ transform as $(\bar 3, 1)$ of $SU(3)\times SU(2)_L$. Similarly
each generation contains a lepton doublet $(\nu, \ell^-)_L$ transforming
as (1,2), and the right-handed charged lepton $\ell_R$ is replaced by its
conjugate $\ell^C_L$, which transforms as a (1,1) of $SU(3)\times
SU(2)_L$. We should also keep track of the hypercharges $Y = Q - I_3$.
One of the major puzzles of the \sm~ is why
\beq
\sum_{q,\ell} Q_i = 3Q_u + 3Q_d + Q_e = 0
\label{fourtwo}
\eeq
In the \sm, the hypercharge assignments are {\bf a~priori}
independent of the $SU(3)\times SU(2)_L$ assignments, although 
constrained by the fact that quantum
consistency requires the resulting triangle anomalies to cancel. In a
simple GUT, the relation (\ref{fourtwo}) is automatic: whenever $Q$ is a
generator of a simple gauge group, $\sum_RQ = 0$ for particles in any
representation $R$ (consider, e.g., the values of $I_3$ in any
representation of $SU(2)$).

The basic rules of GUT model-building are that 
one must look for (a) a gauge group of rank 4
or more -- to accommodate the \sm~ $SU(3)\times 
SU(2)\times U(1)$ gauge group -- which (b)
admits complex representations -- to accommodate 
the known matter fermions. The rank of a
gauge group is the number of generators that 
can be diagonalized simultaneously, i.e., the
number of quantum numbers that it admits. For 
example, $SU(2)$ and $U(1)_{em}$ both have
rank 1 corresponding to $I_3$ and $Q_{em}$, 
respectively, and $SU(3)$ has rank 2 
corresponding to $T_3$ and $Y$. Complex 
representations are required to allow the violation
of charge conjugation $C$, as required by the \sm, which has
\beq
(\nu , e)_L \in (1,2)~, ~~ (u,d)_L \in (3,2)~,~~ e^c_L \in (1,1)~,~~ u^c_L~,~~ d^c_L \in
(\bar 3,1)
\label{fourfifteen}
\eeq
as discussed above.

The following is the mathematical catalogue~\cite{GG} of rank-4 gauge
groups which are either simple
or the direct products of identical simple gauge groups:
\beq
Sp(8)~,~~ SO(8)~,~~ SO(9)~,~~F_4~,~~ SU(3)\times SU(3)~,~~SU(5)
\label{foursixteen}
\eeq
Among these, only $SU(3)\times SU(3)$ and $SU(5)$ have 
complex representations. Moreover, if
one tried to use $SU(3)\times SU(3)$, one would need 
to embed the electroweak gauge group in
the second $SU(3)$ factor. This would be possible 
only if $\sum_q Q_q = 0 = \sum_\ell
Q_\ell$, which is not the case for the known 
quarks and leptons. Therefore, attention has
focussed on
$SU(5)$~\cite{GG} as the only possible rank-4 GUT group.

The useful representations of $SU(5)$ are the complex 
vector \underline{5} representation
denoted by $F_\alpha$, its conjugate $\underline{\bar 5}$ 
denoted by $\bar F^\alpha$, the
complex two-index antisymmetric tensor \underline{10} 
representation $T_{[\alpha\beta]}$,
and the adjoint \underline{24} representation 
$A^\alpha_\beta$. The latter is used to
accommodate the gauge bosons of $SU(5)$:
\beq
\left( \matrix{
&&& \vdots & \bar X \;\;\bar Y \cr
&g_{1,\ldots ,8} && \vdots & \bar X \;\;\bar Y \cr
&&& \vdots & \bar X \;\;\bar Y \cr
\multispan5 \dotfill \cr
X & X & X & \vdots & \cr
&&& \vdots& W_{1,2,3} \cr
Y & Y & Y & \vdots &}
\right)
\label{fourseventeen}
\eeq
where the $g_{1,\ldots,8}$ are the gluons of $SU(3)$, 
the $W_{1,2,3}$ are the $SU(2)$ weak 
bosons, the $U(1)$ hypercharge boson is proportional 
to the traceless diagonal generator
$(1,1,1,-3/2, -3/ 2)$, and the $(X,Y)$ are (3,2) of new gauge bosons that we
discuss in the next section.

The quarks and leptons of each generation are 
accommodated in $\underline{\bar 5}$ and 
 $\underline{10}$ representations of $SU(5)$:
\beq
\bar F = \left(\matrix{d^c_R\cr\cr d^c_Y \cr\cr d^c_B \cr \dotfill \cr -e^- \cr
\nu_e}\right)_L~,~~~T = 
\left(
\matrix{0  & u^c_B & -u^c_Y & \vdots & -u_R & -d_R \cr
-u^c_B & 0 & u^c_R & \vdots & -u_Y & -d_Y \cr
u^c_Y & -u^c_R & 0 & \vdots & -u_B & -d_B \cr 
\multispan6 \dotfill \cr
u_R & u_Y & u_B & \vdots & 0 & -e^c \cr
d_R & d_Y & d_B & \vdots & e^c & 0} \right)_L
\label{foureighteen}
\eeq
The particle assignments are unique up to the 
effects of mixing between generations, which we do
not discuss in detail here~\cite{SU5mix}. The uniqueness
is because
\beq
\underline{\bar 5} = (\bar 3,1) + (1,2) ~, ~~ 
\underline{10} = (3,2) + (\bar 3,1) + (1,2)
\label{fournineteen}
\eeq
in terms of $SU(3)\times SU(2)$ representations. 
Therefore, the $(\nu , e)_L$ doublet in
(\ref{fourfifteen}) can only be assigned to the 
$\underline{\bar 5}$, and since $\Sigma
Q_{em} = 0$ in any GUT representation, the $(\bar 3,1)$ 
in the $\underline{\bar 5}$ must
be assigned to the $d^c$ in (\ref{fourfifteen}). 
The remaining $(u,d)_L \in (3,2), u^c \in
(\bar 3,1)$ and $e^c \in (1,1)$ in (\ref{fourfifteen}) 
fit elegantly into the
\underline{10}, as seen in (\ref{foureighteen}) and
(\ref{fournineteen})~\footnote{Different particle assignments are
possible in the flipped $SU(5)$ model inspired and derived from
string~\cite{AEHN}, because it contains an external $U(1)$ factor
not icnluded in the simple $SU(5)$ group.}.

The remaining steps in constructing an $SU(5)$ GUT 
are the choices of representations for
Higgs bosons, first to break $SU(5)\rightarrow 
SU(3)\times SU(2)\times U(1)$ and
subsequently to break the electroweak $SU(2)\times 
U(1)_Y\rightarrow U(1)_{em}$. The
simplest choice for the first stage is an 
adjoint \underline{24} of Higgs bosons $\Phi$:
\beq
<0\vert\Phi\vert 0 > = \left( \matrix{
1 & 0 & 0 &\vdots & 0 & 0 \cr
0 & 1 & 0 & \vdots & 0 & 0 \cr
0 & 0 & 1 & \vdots & 0 & 0 \cr
\multispan6 \dotfill \cr
0 & 0 & 0 & \vdots & -{3\over 2} & 0 \cr
0 & 0 & 0 & \vdots & 0 & -{3\over 2} }
\right) \times {\cal O} (m_{GUT})
\label{fourtwenty}
\eeq
It is easy to see that this v.e.v. preserves 
colour $SU(3)$ acting on the first three rows
and columns, weak $SU(2)$ acting on the last two 
rows and columns, and the hypercharge
$U(1)$ along the diagonal. The subsequent breaking 
of  $SU(2)\times U(1)_Y\rightarrow
U(1)_{em}$ is most economically accomplished by a 
\underline{5} representation of Higgs
bosons $H$:
\beq
< 0 \vert\phi\vert 0 > = (0,0,0,0,1) \times 0 (m_W)
\label{fourtwentyone}
\eeq
It is clear that this has an $SU(4)$ symmetry which yields~\cite{CEG} the
relation $m_b = m_\tau$
that leads, after renormalization (\ref{fourseven}), to a 
successful prediction for $m_b$
in terms of $m_\tau$. However, the same trick does not 
work for the first two generations,
indicating a need for epicycles in this simplest GUT model~\cite{EG}.

Making the minimal $SU(5)$ GUT supersymmetric, as motivated 
by the naturalness of the
gauge hierarchy, is not difficult~\cite{DG}. One must replace the above
GUT multiplets by supermultiplets:
$\underline{\bar 5} \bar F$ and \underline{10} $T$ for the matter particles,
\underline{24} $\Phi$ for the GUT Higgs fields 
that break $SU(5) \rightarrow SU(3)\times
SU(2) \times U(1)$. The only complication is that one needs \underline{5} and
$\underline{\bar 5}$ Higgs representations $H$ 
and $\bar H$ to break $SU(2)\times U(1)_Y
\rightarrow U(1)_{em}$, just as two doublets were 
needed in the MSSM. The Higgs potential
is specified by the appropriate choice of superpotential~\cite{DG}:
\beq
W = (\mu + {3\lambda\over 2} M) + \lambda \bar H \Phi H + f(\Phi )
\label{fourtwentytwo}
\eeq
where $f(\Phi )$ is chosen so that $\partial f/ \partial\Phi = 0$ when
\beq
<0\vert\Phi\vert 0 > = M \left( \matrix{
1 & 0 & 0 &\vdots & 0 & 0 \cr
0 & 1 & 0 & \vdots & 0 & 0 \cr
0 & 0 & 1 & \vdots & 0 & 0 \cr
\multispan6 \dotfill \cr
0 & 0 & 0 & \vdots & -{3\over 2} & 0 \cr
0 & 0 & 0 & \vdots & 0 & -{3\over 2} }
\right) 
\label{fourtwentythree}
\eeq
Inserting this into the second term of (\ref{fourtwentytwo}), 
one finds terms $\lambda M
\bar H_3 H_3,~~-3/ 2 \lambda M \bar H_2 H_2$ for the 
colour-triplet and weak-doublet
components of $\bar H$ and $H$, respectively. Combined 
with the bizarre coefficient of the
first term, these lead to
\beq
W \ni (\mu + {5\lambda\over 2} M) \bar H_3 H_3 + \mu \bar H_2 H_2
\label{fourtwentyfour}
\eeq
Thus we have heavy Higgs triplets (as needed for baryon stability, 
see the next section)
and light Higgs doublets. This requires fine tuning the 
coefficient of the first term in
$W$ (\ref{fourtwentytwo}) to about 1 part in $10^{13}$! 
The advantage of supersymmetry
is that its no-renormalization theorems~\cite{nonren} guarantee that this
fine tuning is ``natural", in the sense 
that quantum corrections like those in Fig.~12c do not
destroy it, unlike the situation without supersymmetry. 
On the other hand, supersymmetry
alone does not explain the origin of the hierarchy.

\subsection{Baryon Decay}

Baryon instability is to be expected on general grounds, 
since there is no exact gauge
symmetry to guarantee that baryon number $B$ is conserved. 
Indeed, baryon decay is a
generic prediction of GUTs, which we illustrate with the 
simplest $SU(5)$ model, that is
anyway embedded in larger and more complicated GUTs. 
We see in (\ref{fourseventeen}) that
there are two species of gauge bosons in $SU(5)$ 
that couple the colour $SU(3)$ indices
(1,2,3) to the electroweak $SU(2)$ indices (4,5), 
called $X$ and $Y$. As we can see from the
matter representations (\ref{foureighteen}), 
these may enable two quarks or a quark and
lepton to annihilate, as seen in Fig. 29a. 
Combining these possibilities leads to
interactions with $\Delta B  =
\Delta L = 1$. The forms of effective 
four-fermion interactions mediated by the exchanges of
massive $Z$ and $Y$ bosons, respectively, are~\cite{BEGN}:
\bea
&&\left(\epsilon_{ijk} u_{R_k} \gamma_\mu u_{L_j}\right)~~{g^2_X\over 8 m^2_X} ~~ \left(2 e_R
~\gamma^\mu ~d_{L_i} + e_L~\gamma^\mu~d_{R_i} \right)~, \nonumber \\
&&\left(\epsilon_{ijk} u_{R_k} \gamma_\mu d_{L_j}\right)~~{g^2_Y\over 8 m^2_X} ~~ \left(\nu_L
~\gamma^\mu ~d_{R_i}\right)~.
\label{fourtwentyfive}
\eea
up to generation mixing factors.

\begin{figure}
\centerline{\includegraphics[height=1.5in]{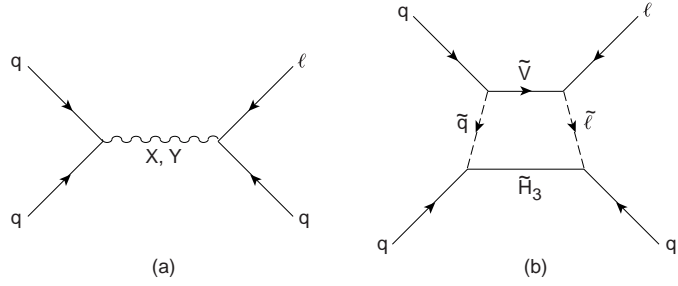}}
\caption[]{Diagrams contributing to baryon decay (a) in minimal $SU(5)$
and (b) in minimal supersymmetric $SU(5)$.}
\end{figure}

\noindent
Since the gauge couplings $g_X = g_Y = g_{3,2,1}$ 
in an $SU(5)$ GUT, and $m_X \simeq m_Y$,
we expect that
\beq
G_X \equiv {g^2_X\over 8m^2_X}\simeq G_Y \equiv {g^2_Y\over 8m^2_Y}
\label{fourtwentysix}
\eeq
It is clear from (\ref{fourtwentyfive}) that the baryon 
decay amplitude $A\propto G_X$, and
hence the baryon $B\rightarrow \ell +$ meson decay rate
\beq
\Gamma_B = c G^2_X m^5_p
\label{fourtwentyseven}
\eeq
where the factor of $m^5_p$ comes from dimensional 
analysis, and $c$ is a coefficient that
depends on the GUT model and the non-perturbative 
properties of the baryon and meson.

The decay rate (\ref{fourtwentyseven}) corresponds to a proton lifetime
\beq
\tau_p = {1\over c} ~{m^4_X\over m^5_p}
\label{fourtwentyeight}
\eeq
It is clear from (\ref{fourtwentyeight}) that 
the proton lifetime is very sensitive to
$m_X$, which must therefore be calculated very 
precisely. In minimal $SU(5)$, the best
estimate was~\cite{Marciano}
\beq
m_X \simeq (1~{\rm to}~2) \times 10^{15} \times \Lambda_{QCD}
\label{fourtwentynine}
\eeq
where $\Lambda_{QCD}$ is the characteristic 
QCD scale in the $\overline{\rm MS}$
prescription with four active flavours. 
Making an analysis of the generation mixing factors~\cite{SU5mix}, one
finds that the preferred proton (and bound neutron) decay
modes in minimal
$SU(5)$ are
\bea
&&p \rightarrow e^+\pi^0~,~~e^+\omega~,~~\bar\nu \pi^+~,
~~\mu^+K^0~,~~\ldots \nonumber \\
&& n \rightarrow e^+\pi^-~,~~ e^+\rho^-~,~~\bar\nu \pi^0~,~~\ldots
\label{fourthirty}
\eea
and the best numerical estimate of the lifetime is~\cite{Marciano}
\beq
\tau (p\rightarrow e^+\pi^0) \simeq 2\times 10^{31\pm 1} \times
\left({\Lambda_{QCD}\over 400~{\rm MeV}}\right)^4~~y
\label{fourthirtyone}
\eeq
This is in {\it prima facie} conflict with the latest experimental lower limit
\beq
\tau (p \rightarrow e^+\pi^0) > 1.6 \times 10^{33}~y
\label{fourthirtytwo}
\eeq
from super-Kamiokande~\cite{SKpdk}. However, this failure of minimal
$SU(5)$ is not as conclusive as
the failure of its prediction for $\sin^2\theta_W$.

We saw earlier that supersymmetric GUTs, including $SU(5)$, fare better with
$\sin^2\theta_W$. They also predict a larger GUT scale~\cite{DRW}:
\beq
m_X \simeq 2\times 10^{16}~{\rm GeV}
\label{fourthirtythree}
\eeq
so that $\tau (p\rightarrow e^+\pi^0)$ is considerably 
longer than the experimental lower
limit. However, this is not the dominant proton 
decay mode in supersymmetric $SU(5)$~\cite{susySU5pdk}. In
this model, there are important $\Delta B = \Delta L = 1$ 
interactions mediated by the
exchange of colour-triplet Higgsinos $\tilde H_3$, 
dressed by gaugino exchange as seen in
Fig. 29b~\cite{WSY}:
\beq
G_X\rightarrow {\cal O}~\left({\lambda^2g^2\over 
16\pi^2}\right)~{1\over m_{\tilde
H_3}\tilde m}
\label{fourthirtyfour}
\eeq
where $\lambda$ is a Yukawa coupling. 
Taking into account colour factors and the increase
in $\lambda$ for more massive particles, 
it was found~\cite{susySU5pdk} that decays into neutrinos and
strange particles should dominate:
\beq
p\rightarrow \bar\nu K^+~,~~n\rightarrow\bar\nu K^0~,~~\ldots
\label{fourthirtyfive}
\eeq
Because there is only one factor of a heavy mass 
$m_{\tilde H_3}$ in the denominator of
(\ref{fourthirtyfour}), these decay modes are 
expected to dominate over $p\rightarrow
e^+\pi^0$, etc.,  in minimal supersymmetric $SU(5)$. 
Calculating carefully the other factors
in (\ref{fourthirtyfour})~\cite{EGR}, it seems that the modes
(\ref{fourthirtyfive}) may be close to
detectability in this model. The current experimental 
limit is $\tau(p\rightarrow \bar\nu
K^+) > 10^{32} y$, and super-Kamiokande may 
soon be able to improve this
significantly. 

There are non-minimal supersymmetric GUT models 
such as flipped $SU(5)$~\cite{AEHN} in which the $\tilde
H_3$- exchange mechanism 
(\ref{fourthirtyfour}) is suppressed. 
In such models, $p\rightarrow e^+\pi^-$ may again be
the preferred decay mode~\cite{aspects}. However, this is not necessarily
the case, as colour-triplet
Higgs boson exchange may be important, in 
which case $p\rightarrow \mu^+K^0$ could be
dominant~\cite{CER}, or there may be non-intuitive generation mixing in
the couplings of the $X$ and
$Y$ bosons, offering the possibility 
$p\rightarrow \mu^+\pi^0$, etc. . Therefore, the
continuing search for proton decay should be open-minded about the
possible decay modes.

\subsection{Neutrino Masses and Oscillations}

The experimental upper limits on neutrino 
masses are far below the corresponding lepton
masses~\cite{CDFD0}. From studies of the end-point of Tritium $\beta$
decay,
we have
\beq
m_{\nu_e} \lappeq 3.5~{\rm eV}
\label{fourtyirthsix}
\eeq
to be compared with $m_e = 0.511$ MeV. 
From studies of $\pi\rightarrow \mu\nu_\mu$ decays,
we have
\beq
m_{\nu_\mu} < 160~{\rm keV}
\label{fourthirtyseven}
\eeq
to be compared with $m_\mu$ = 105 MeV, and 
from studies~\cite{mnutau} of $\tau\rightarrow$ pions +
$\nu_\tau$ we have
\beq
m_{\nu_\tau} < 18~{\rm MeV}
\label{fourthirtyeight}
\eeq
to be compared with $m_\tau$ = 1.78 GeV. On the other
hand, there is no good symmetry reason to expect 
the neutrino masses to vanish. We expect
masses to vanish only if there is a corresponding 
exact gauge symmetry, cf., $m_\gamma$ =
0 in QED with an unbroken $U(1)$ gauge symmetry.

Although there is no candidate gauge symmetry to 
ensure $m_\nu = 0$, this is a prediction
of the \sm. We recall that the neutrino couplings 
to charged leptons take the form
\beq
J_\mu = \bar e\gamma_\mu (1-\gamma_5) \nu_e + \bar\mu \gamma_\mu (1-\gamma_5)\nu_\mu +
\bar\tau\gamma_\mu (1-\gamma_5)\nu_\tau
\label{fourthirtynine}
\eeq
and that only left-handed neutrinos have 
ever been detected. In the cases of charged
leptons and quarks, their masses arise in 
the \sm~ from couplings between left- and
right-handed components via a Higgs field:
\beq
g_{H\bar ff}~H_{\Delta I = {1\over 2},\Delta L = 0}~~
\bar f_R f_L + h.c. \rightarrow
m_f = g_{H\bar ff} < 0\vert H_{\Delta I = {1\over 2},\Delta L = 0} \vert 0 >
\label{fourforty}
\eeq
Such a left-right coupling is conventionally 
called a Dirac mass. The following questions
arise for neutrinos: if there is no $\nu_R$, 
can one have $m_\nu \not= 0$? and if there is
a $\nu_R$ why are the neutrino masses so small?

The answer to the first question is positive, 
because it is possible to generate neutrino
masses via the Majorana mechanism that involves the $\nu_L$ alone. 
This is possible
because an $(\overline{f_R})$ field is in fact left-handed:
$(\overline{f_R}) = (f^c)_L = f^T_L C$, where the 
superscript $T$ denotes a transpose, and
$C$ is a $2\times 2$ conjugation matrix. We can therefore imagine replacing
\beq
(\overline{f_R}) f_L \rightarrow f^T_L~C~ f_L
\label{fourfortyone}
\eeq
which we denote by $f_L \cdot f_L$. In the 
cases of quarks and charged leptons, one
cannot generate masses in this way, because 
$q_L \cdot q_L$ has $\Delta Q_{em}$, $\Delta
($colour)$ \not= 0$ and $\ell_L\cdot \ell_L$ 
has $\Delta Q_{em}\not= 0$. However, the
coupling
$\nu_L\cdot\nu_L$ is not forbidden by such 
exact gauge symmetries from leading to a
neutrino mass:
\beq
m^M~\nu_L^T~C~\nu_L = m^M (\overline{\nu^c})_L\nu_L = m^M~\nu_L\cdot\nu_L
\label{fourfortytwo}
\eeq
Such a combination has non-zero net lepton 
number $\Delta L = 2$ and weak isospin $\Delta I
= 1$. There is no corresponding Higgs field in 
the \sm~ or in the minimal $SU(5)$ GUT, but
there is no obvious reason to forbid one. 
If one were present, one could generate a
Majorana neutrino mass via the renormalizable coupling
\beq
\tilde g_{H\bar\nu\nu} ~~H_{\Delta I=1, 
\Delta L = L}~~\nu_L\cdot\nu_L \Rightarrow m^M =
\tilde g_{H\bar\nu\nu} < 0 \vert H_{\Delta I = 1,\Delta L = 2}\vert 0 >
\label{fourfortythree}
\eeq
However, one could also generate a Majorana mass 
without such an additional Higgs field,
via a non-renormalizable coupling to the 
conventional $\Delta I = {1\over 2}$ \sm ~Higgs
field~\cite{BEG}:
\beq
{1\over M}~~\left(H_{\Delta I = {1\over 2}} 
\nu_L\right) \cdot \left( H_{\Delta I =
{1\over 2}} \nu_L\right) \Rightarrow m^M = 
{1\over M} <0\vert H_{\Delta I = {1\over 2}}
\vert 0>^2
\label{fourfortyfour}
\eeq
where $M$ is some (presumably heavy: $M \gg m_W)$ 
mass scale. The simplest possibility of
generating a non-renormalizable interaction of 
the form (\ref{fourfortyfour}) would be via
the exchange of a heavy field $N$ that is a 
singlet of $SU(3)\times SU(2)\times U(1)$ or
$SU(5)$:
\beq
{1\over M} \rightarrow {\lambda^2\over M^2_N}
\label{fourfortyfive}
\eeq
where one postulates a renormalizable coupling $\lambda H_{\Delta I=1/2}
\nu_L\cdot N$. Such a heavy singlet field 
appears automatically in extensions of the $SU(5)$
GUT, such as $SO(10)$, but does not actually {\it require}
the existence of any new GUT gauge bosons.

We now have all the elements we need for 
the see-saw mass matrix~\cite{GRY} favoured by GUT
model-builders:
\beq
(\nu_L , N) \cdot \left(\matrix{m^M & m^D\cr 
m^D & M^M}\right)~~\left(\matrix{\nu_L \cr
N}\right)
\label{fourfortysix}
\eeq
where the $\nu_L\cdot\nu_L$ Majorana mass 
$m^M$ might arise from a $\Delta I = 1$ Higgs
with coupling $\tilde g_{H\bar\nu\nu}$, 
(\ref{fourfortythree}), the $\nu_L\cdot N$ Dirac
mass $m^D$ could arise from a conventional 
Yukawa coupling $\lambda$ (\ref{fourfortyfive})
and should be of the same order as a conventional 
quark or lepton mass, and $M^M$ could {\it
a priori} be ${\cal O}(M_{GUT})$. Diagonalizing 
(\ref{fourfortysix}) and assuming that $m^M =
0$ or that $<0\vert H_{\Delta I = 1}\vert 0> = 
{\cal O}(m^2_W/ m_{GUT})$, as generically
expected in GUTs, it is easy to diagonalize 
(\ref{fourfortysix}) and obtain the mass
eigenstates
\bea
\nu_L + 0\left({m_W\over m_X}\right) N &:& 
m = {\cal O}\left({m^2_W\over m_{GUT}}\right)
\nonumber \\
N + 0\left({m_W\over m_X}\right) \nu_L &:& M = {\cal O} (M_{GUT})
\label{fourfortyseven}
\eea
So far, we have not touched on the generation 
structure of the neutrino masses. It is
often suggested that $m^M$ is negligible, 
$M^M$ is (approximately) generation-independent,
and $m^D\propto m_{2/ 3}$ (the $u$-quark mass matrix). If so, one sees that
\beq
m_{\nu_i} \sim {m^2_{2/ 3_i}\over M_{GUT}}
\label{fourfortyeight}
\eeq
and one might expect that
\beq
m_{\nu_e} \ll m_{\nu_\mu} \ll m_{\nu_\tau}
\label{fourfortynine}
\eeq
with mixing related to the Cabibbo-Kobayashi-Maskawa matrix.

As you know~\cite{PL}, evidence has recently been presented for
atmospheric neutrino oscillations~\cite{SK}
between $\nu_\mu$ and $\nu_\tau$ with 
$\Delta m^2_A \sim (10^{-2}$ to $10^{-3}$) eV$^2$
and a large mixing angle: $\sin^2\theta_{\mu\tau} 
\gappeq$ 0.8. This is in addition to the
previous evidence~\cite{solar} for solar neutrino oscillations with
$\Delta m^2_S \simeq 10^{-5}$ eV$^2$ 
and $\sin^2\theta \sim 10^{-3}$ or $\sim$ 1 (Mikheev-Smirnov-Wolfenstein
or MSW~\cite{MSW}
oscillations) or $\Delta m^2_S \sim 10^{-10}$ eV$^2$  
and $\sin^2\theta \sim 1$ (vacuum
oscillations), as seen in Fig. 30. 

Various 
theoretical groups~\cite{newnu} have restudied the previous
see-saw prejudices in the light of the new data. 
In a hierarchical pattern of neutrino
masses, one would expect
\beq
m_{\nu_3} \sim \sqrt{\Delta m^2_A} > m_{\nu_2} 
\sim \sqrt{\Delta m^2_S} > m_{\nu_3}
\label{fourfifty}
\eeq

\begin{figure}
\centerline{\includegraphics[height=3in]{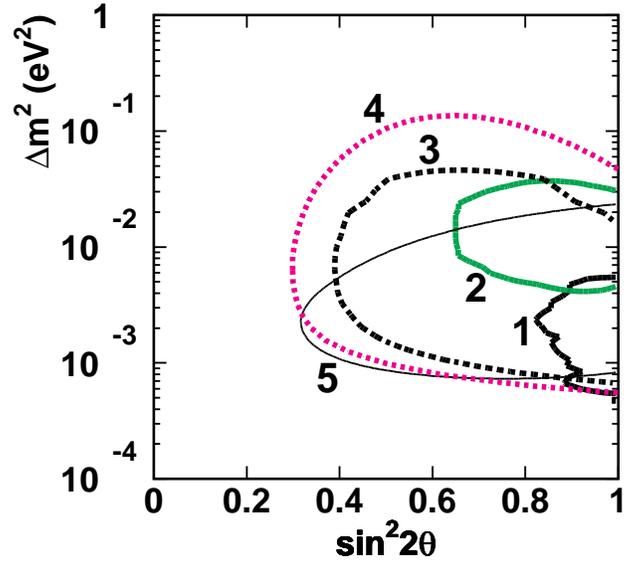}}
\caption[]{Constraints on $\nu_\mu - \nu_\tau$ mixing from (1,2)
contained events in super-Kamiokande and Kamiokande, (3,4) upward-going
muons in super-Kamiokande and Kamiokande, and (5) stop/through upward
muons in super-Kamiokande~\cite{combined}.} 
\end{figure}

\noindent
but is this compatible with the large mixing 
indicated (at least) for atmospheric
neutrinos? Indeed it is~\cite{ELLN}, and theoretically it is difficult to
see why any pair of
neutrinos should be almost degenerate. 
On the other hand, there are perfectly natural
$2\times 2$ light-neutrino mass matrices that are compatible with large
$\sin^22\theta_{\mu\tau}$ and the first mass 
hierarchy in (\ref{fourfifty}) if $m_{\nu_2}
\sim\sqrt{\Delta m^2_S} \sim 10^{-2^{1/2}}$ eV, 
particularly when it is observed~\cite{Tanimoto} that
renormalization-group effects below $M_{GUT}$ 
may enhance $\sin^2 2\theta_{\mu\tau}$, as
seen in Fig. 31~\cite{ELLN}. However, it is very difficult to understand
the much larger hierarchy that
would be needed for the vacuum solution to the 
solar neutrino problem with $m_{\nu_2}
\sim\sqrt{\Delta m^2_S} \sim 10^{-5}$ eV. 
It is a more model-dependent question whether the
large- or small-angle MSW solution is favoured. 
In one particular GUT model~\cite{ELLN}, we found the
large-angle MSW solution more plausible, but the 
small-angle MSW solution could not be excluded. We
still need more experimental information on 
neutrino masses and mixing, and this will
surely be an active experimental field for years to come.

\begin{figure}
\centerline{\includegraphics[height=2in]{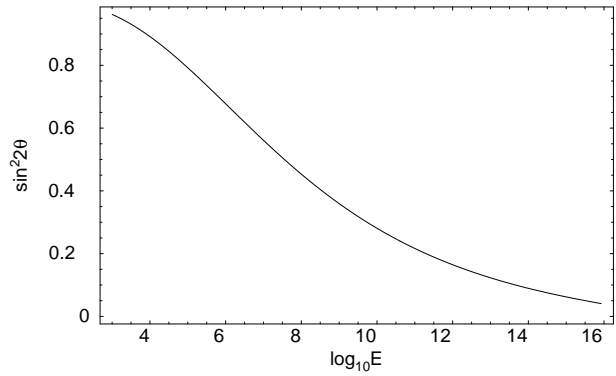}}
\caption[]{Possible renormalization-group effect on the mixing angle in a 
two-generation neutrino model with supersymmetry at large
$\tan\beta$~\cite{ELLN}.}
\end{figure}

\section{NONE OF THE ABOVE}

Now is a good moment to review the progress made 
in addressing the defects of the
Standard Model that were reviewed in Lecture 1, 
and to count how many parameters are
still free.  Grand unification reduces the three 
independent gauge couplings of the
Standard Model to just one, so that is certainly 
progress.  As far as the Standard
Model Higgs sector is concerned, supersymmetry 
fixes the quartic Higgs coupling, which
is progress, but the overall scale of electroweak 
gauge symmetry breaking is related to
the undetermined scale of supersymmetry 
breaking.  Indeed, in the absence of further
theoretical input, the soft supersymmetry-breaking 
mass introduces ${\cal O}(100)$ new
parameters unless (why?) one knows how to impose flavour universality.

Moreover, a complete treatment of neutrino 
masses involves additional parameters to
describe the GUT Higgs potential~\cite{BEGN,DG} and a see-saw mass
matrix~\cite{GRY}. What is more, we still need at
least one parameter to generate the cosmological 
baryon asymmetry and another to
generate cosmological inflation.  
Finally, we should not forget gravity, whose
parameters include the cosmological constant 
(if it vanishes, we need to understand
why) as well as Newton's constant.

Thus, although progress has been made, there remain big questions in the
supersymmetry-breaking sector and in quantum 
gravity, which are the main subjects
addressed in this Lecture.

\subsection{Supersymmetry Breaking}

It is clear that supersymmetry must be broken: $m_{\tilde{e}} \neq m_{e},
m_{\tilde{\gamma}} \neq m_{\gamma} = 0$, etc.  
Could it be explicit or must it be
spontaneous?  Explicit supersymmetry 
breaking is not only ugly and unlike what occurs
in gauge theories, it also induces inconsistencies 
at the quantum level when gravity is
introduced.  Therefore, attention has 
focused on spontaneous symmetry breaking. Since
the supersymmetry charge $Q$ is fermionic, 
this requires a non-zero matrix element for
$Q$ between the vacuum and some fermion 
$\chi$ called a Goldstone fermion or Goldstino:
\beq
< 0| Q | \chi > = f^2_{\chi} \not= 0
\label{fiveone}
\eeq
The non-zero matrix element (\ref{fiveone}) 
has the immediate consequence, within global
supersymmetry, that the vacuum energy is necessarily 
positive.  This follows from the
basic supersymmetry algebra:
\beq
\{ Q, Q \} \propto \gamma_{\mu} P^{\mu}
\label{fivetwo}
\eeq
Sandwiching (\ref{fivetwo}) between vacuum 
states and inserting the intermediate state 
$|\chi > < \chi |$, we find
\beq
| < 0 |Q| \chi >|^2 = f^4_{\chi} > 0
\label{fivethree}
\eeq
and hence the vacuum energy
\beq
< 0 | P_0 | 0 > \equiv E_0 > 0
\label{fivefour}
\eeq
How may this spontaneous symmetry breaking be achieved? 
We recall the general form of
the effective scalar potential in a globally supersymmetric theory:
\beq
V = \sum_{i} \left\vert \frac{\partial W}{\partial \phi^i} \right\vert^2 + 
\frac{1}{2} \sum_{\alpha} \; g^2_{\alpha} | \phi^* T^{\alpha} \phi |^2
\label{fivefive}
\eeq
where the former is called the $F$ term and 
the latter the $D$ term. In order to obtain
(\ref{fivefour}), we need either the first 
term to be non-zero ``$F$-breaking"~\cite{ORF} - or the
second term - ``$D$-breaking"~\cite{FI}.  The latter 
would require an extra U(1) gauge group factor
and many new matter fields, so we illustrate 
global supersymmetry breaking with the
simplest $F$-breaking model~\cite{ORF}.

Consider the superpotential
\beq
W = \alpha AB^2 + \beta C (B^2 - m^2)
\label{fivesix}
\eeq
where $A,B,C$ denote gauge-singlet matter 
supermultiplets.  It is easy to see that
(\ref{fivesix}) yields
\beq
F^{\dagger}_{A} = \alpha B^2, ~~~F^{\dagger}_{B} = 2B (\alpha A + \beta C),~~~
F^{\dagger}_C = \beta (B^2 - m^2)
\label{fiveseven}
\eeq
and hence an effective scalar potential
\beq
V = \sum_{i=A,B,C} \, |F_i|^2 = 
|2B (\alpha A + \beta C)|^2 + |\alpha B^2|^2 +
\beta (B^2 - m^2)|^2
\label{fiveeight}
\eeq
It is apparent that the last two terms cannot 
vanish simultaneously, so that $V > 0$
and supersymmetry is broken.

Recently there have been analyses of high-redshift supernovae~\cite{highz} 
and other astrophysical
and cosmological data~\cite{lambdacdm} that favour a non-zero 
cosmological constant, as would be
suggested by such positive vacuum energy.  
Unfortunately, the model (\ref{fivesix}) and
others like it  are too much of a good thing.  
The ``observed" cosmological constant,
if it is real at all, would correspond to
\beq
\Lambda \lappeq 10^{-123} m^4_P
\label{fivenine}
\eeq
whereas such a global model of supersymmetry breaking would correspond to 
\beq
\Lambda \sim (1 \; TeV)^4 \sim 10^{-64} \, m^4_P,
\label{fiveten}
\eeq
a discrepancy by some 60 orders of magnitude!
Even the QCD vacuum energy
\beq
E_{QCD} \sim (100 \, MeV)^4 \sim 10^{-80} \, m^4_P
\label{fiveeleven}
\eeq
is much larger than the "observed" value (\ref{fivenine}).
This discrepancy may be the biggest problem 
in theoretical physics, even bigger than
the hierarchy problem.  However, to address 
it requires a true quantum theory of
gravity, as is discussed later in this lecture.

However, before doing so, let us briefly 
review the latest incarnation of global
supersymmetry-breaking models, namely 
gauge-mediated or messenger models~\cite{GR}. The basic
idea is to hide the ugly origin of 
supersymmetry breaking in a hidden sector of the
theory that is coupled to observable particles 
via an intermediate set of ``messenger"
particles that share some of the gauge 
interactions of the Standard Model, as seen in Fig.~32.  Gauge
interactions then mediate the supersymmetry breaking needed in the observable
sector. These models were originally conceived 
in the early 1980's~\cite{early} because neither the
$F$-breaking scenario (\ref{fivesix}) 
nor $D$-breaking models fitted within the MSSM. They
have recently been reincarnated~\cite{new} with the idea that
supersymmetry breaking in the hidden
sector might originate from non-perturbative phenomena, which are much better
understood by now~\cite{GR}.

\begin{figure}
\centerline{\includegraphics[height=1.5in]{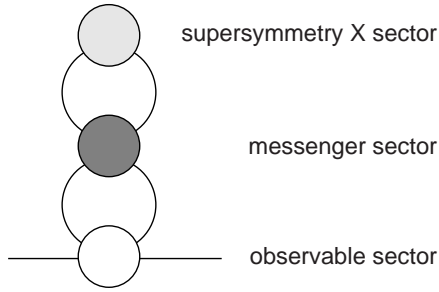}}
\caption[]{Sketch of the principle of gauge-mediated (messenger) models:
supersymmetry breaking in a hidden sector is communicated to the
observable sector via gauge interactions~\cite{GR}.} 
\end{figure}

There have been two principal motivations 
for this reincarnation.  One is that gauge
mediation naturally imposes flavour universality 
in the observable sector~\cite{new}.  All quarks
(or leptons) with the same charge acquire 
universal soft supersymmetry-breaking scalar
masses, avoiding any problems with 
flavour-changing neutral interactions~\cite{EN} and reducing
the effective number of parameters in the observable sector.  A feature of
gauge-mediated models is the appearance of a 
massless Goldstone fermion $\lambda$,
which would acquire a small mass when 
gravity is taken into account, as discussed in
the next section.  This implies that the 
lightest neutralino $\chi$ is unstable:
$\chi \to \lambda \gamma$ decay dominates.

This provides the second motivation for 
gauge-mediated models, which is the report by
the CDF collaboration~\cite{CDFevent} of an apparent $\bar{p}p \to e^+ e^-
\gamma \gamma +$
missing $p_T$ event.  It has been suggested~\cite{interpret} that this
might be due to 
$\tilde{e}^+ \tilde{e}^-$ pair production 
followed by $\tilde{e}^{\pm} \to e^{\pm} \chi
, \, \chi \to \lambda \gamma$ decays, or due 
to $\chi^+ \chi^-$ pair production
followed by $\chi^{\pm} \to e^{\pm} 
\nu \chi , \, \chi \to \lambda \gamma$ decays. However,
no other $\gamma \gamma +$ missing $p_T$ events have been observed either
at LEP~\cite{LEPC} or at
the FNAL Tevatron collider~\cite{noD0}, and most of the parameter spaces
for these interpretations have
now been excluded, as seen in Fig. 33~\cite{LEPC}.  Clearly, further
experimental input from future
collider runs is needed.

\begin{figure}
\centerline{\includegraphics[height=3in]{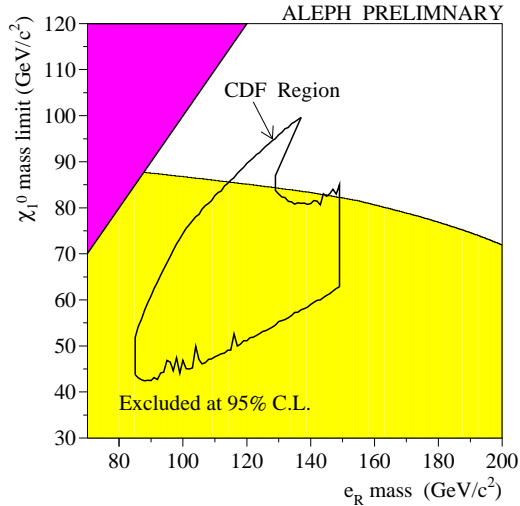}}
\caption[]{Region of the $(m_{\tilde e}, m_\chi)$ plane excluded 
by LEP~\cite{LEPC} in
models with a light gravitino, compared with the region favoured in the
$\bar pp\rightarrow \tilde e \tilde{\bar e} + X$
interpretation~\cite{interpret} of the
CDF $\bar pp\rightarrow e^+e^-\gamma\gamma p\llap{$/$}_T + X$
event~\cite{CDFevent}.}
\end{figure}

Some final comments on gauge-mediated models 
are in order.  The first is that it has proved
to be quite difficult to construct a model that 
is consistent with all the phenomenological
constraints, has a desirable stable vacuum, 
etc. . The other is that, to the extent that the
hidden scale $\Lambda_H \gg 1 \, \mbox{TeV}$, 
the apparent vacuum energy (cosmological
constant) $\sim \Lambda^4_H \gg 10^{-64} \, 
m^4_P$, worsening the discrepancy with the
astrophysical upper limit (or observation)  
(\ref{fivenine}).  However, in order to discuss
this issue seriously, one needs a supersymmetric 
quantum theory of gravity, to which we now
turn.

\subsection{Local Supersymmetry and Supergravity}

Why make a local theory of supersymmetry?  
One motivation is the analogy with gauge
theories, in which bosonic symmetries are 
made local.  Another is that local supersymmetry
necessarily involves the introduction of gravity.  
Since both gravity and (surely!)
supersymmetry exist, this seems an inevitable step.  
It also leads to the possibility of
unifying all the particle interactions including 
gravity, which was one of our original
motivations for supersymmetry.  Moreover, 
in the context of this Lecture, it is interesting
that local supersymmetry (supergravity) admits 
an elegant mechanism for supersymmetry
breaking~\cite{superHiggs}, analogous to the Higgs mechanism 
in gauge theories, which allows us to address
more seriously the possible existence of a cosmological constant.

The basic building block in a supergravity theory~\cite{FvF,DZ} is the
graviton supermultiplet of
(\ref{twoforteen}), which contains particles with 
helicities $(2, 3/2)$, the latter being the
gravitino of spin $3/2$.  Why is this 
required when one makes supersymmetry local?

We recall the basic global supersymmetry 
transformation laws (\ref{twoseventeen}) for bosons
and fermions.
  Consider
now the combination of two such global supersymmetry transformations:
\beq
[\delta_1 , \delta_2 ]~(\phi \,~ \mbox{or} ~\, \psi ) = 
- (\bar{\xi}_2 \gamma_{\mu} \xi_1 )~(i \, 
\partial_{\mu})~(\phi \, ~\mbox{or} ~\,
\psi ) + \ldots
\label{fivethirteen}
\eeq
The operator $(i \, \partial_{\mu})$ 
corresponds to the momentum $P_{\mu}$, and we see again
that the combination of two global supersymmetry 
transformations is a translation. Consider 
now what happens when we consider local 
supersymmetry transformations characterized
by a varying spinor $\xi(x)$.  It is evident 
that the infinitesimal translation 
$\bar{\xi}_2 \gamma^{\mu} \xi_1$ in 
(\ref{fivethirteen}) is now $x$-dependent,
and the previous global translation becomes 
a local coordinate transformation, as occurs in
General Relativity.

How do we make the theory invariant under 
such local supersymmetry transformations? 
Consider again the simplest globally 
supersymmetric model containing a free spin-1/2 fermion
and a free spin-0 boson (\ref{twosixteen}),
and make the local versions of the transformations 
(\ref{twoseventeen}).  Following the same
steps as in Lecture~2, we find that 
\beq
\delta {\cal L} = \partial_{\mu} (\cdots) + 2 \bar{\psi} \gamma_{\mu} \, 
\partial\llap{$/$} S(\partial^{\mu} \xi(x)) + \mbox{herm. conj.}
\label{fivefifteen}
\eeq
In contrast to the global case, the action 
$A = \int d^4 x {\cal L}$ is not invariant,
because of the second term in (\ref{fivefifteen}).  
To cancel it out and restore invariance,
we need more fields.

We proceed by analogy with gauge theories.  In order to make the kinetic term 
$(i \bar{\psi} \partial\llap{$/$} \psi )$ 
invariant under gauge transformations $\psi \to
e^{i \epsilon (x)} \psi$, we need to cancel a variation
\beq
- \bar{\psi} \partial_{\mu} \psi \partial^{\mu} \epsilon (x)
\label{fivesixteen}
\eeq
which is done by introducing a coupling to a gauge boson:
\beq
g \bar{\psi} \gamma_{\mu} \psi A^{\mu} (x)
\label{fiveseventeen}
\eeq
and the corresponding transformation:
\beq
\delta A_{\mu} (x) = \frac{1}{g} \partial_{\mu} \epsilon (x)
\label{fiveeighteen}
\eeq
In the supersymmetric case, we cancel the 
second term in (\ref{fivefifteen}) by a coupling:
\beq
\kappa \bar{\psi} \gamma_{\mu} \partial\llap{$/$} S \psi^{\mu} (x)
\label{fivenineteen}
\eeq
to a spin-3/2 spinor $\psi^{\mu} (x)$, 
representing a gauge fermion or gravitino, with the
corresponding transformation:
\beq
\delta \psi^{\mu} = - \frac{2}{\kappa} \, \partial^{\mu} \xi (x)
\label{fivetwenty}
\eeq
where $\kappa \equiv 8 \pi / m^2_P$.

For completeness, let us at least write 
down the Lagrangian for the graviton-gravitino
supermultiplet:
\beq
L = - \frac{1}{2\kappa^2} \, \sqrt{-g} R   - 
\frac{1}{2} \, \epsilon^{\mu \nu \rho \sigma}
\bar{\psi}_{\mu} \gamma_{5} \gamma_{\nu} {\cal D}_{\rho} \psi_{\sigma}
\label{fivetwentyone}
\eeq
where $g$ denotes the determinant of the metric tensor:
\beq
g_{\mu \nu} = \epsilon^m_{\mu} \eta_{mn} \epsilon^{\mu}_{\nu}
\label{fivetwentytwo}
\eeq
where $\epsilon^m_{\mu}$ is the vierbein and 
$\eta_{mn}$ the Minkowski metric tensor, and
${\cal D_{\rho}}$ is a covariant derivative
\beq
{\cal D_{\rho}} \equiv \partial_{\rho} + \frac{1}{4} \, \omega^{mn}_{\rho}
[\gamma_m , \gamma_n ]
\label{fivetwentythree}
\eeq
where $\omega^{mn}_{\rho}$ is the spin 
connection.  This is the simplest possible
generally-covariant model of a spin-3/2 field.  
It is remarkable that it is invariant under
the local supersymmetry transformations:
\bea
\delta \epsilon^m_{\mu} & = & \frac{x}{2} \, \bar{\xi} (x) \gamma^m
\psi_{\mu} (x), \nonumber \\
\delta \omega_{\mu}^{mn} & = & 0, 
\delta \psi_{\mu} = \frac{1}{x} \, {\cal D}_{\mu} \xi (x)
\label{fivetwentyfour}
\eea
just as the simplest possible $(1/2, 0)$ theory (\ref{twosixteen}) was globally
supersymmetric, and also the action of an 
adjoint spin-1/2 field in a gauge theory.

It is also remarkable that supergravity 
admits an elegant analogue of the Higgs mechanism of
spontaneous symmetry breaking~\cite{superHiggs}.  Just as one combines the
two polarization states of a
massless gauge field with the single state 
of a massless Goldstone boson to obtain the three
polarization states of a massive gauge boson, 
one may combine the two polarization states of
a massless gravitino $\psi_{\mu}$ with the 
two polarization states of a massless Goldstone
fermion $\lambda$ to obtain the four polarization 
states of a massive spin-3/2 particle
$\tilde{G}$.  This super-Higgs mechanism 
corresponds to a spontaneous breakdown of local
supersymmetry, since the massless graviton $G$ 
has a different mass from the gravitino
$\tilde{G}$:
\beq
m_G = 0 \not= m_{\tilde{G}}.
\label{fivetwentyfive}
\eeq
This is the only known consistent way of 
breaking local symmetry, just as the Higgs
mechanism is the only way to generate $m_W \not= 0$.

Moreover, this can be achieved while keeping zero 
vacuum energy (cosmological constant), at
least at the tree level.  The reason for 
this is the appearance in local supersymmetry
(supergravity) of a third term in the effective 
potential (\ref{fivefive}), which has a
{\it negative} sign~\cite{superHiggs}.  There is no time 
in these lectures to discuss this exciting feature
in detail:  the interested reader is referred 
to the original literature and the simplest
example~\cite{CFKN}.  In this latter case, $\Lambda = V = 0$ for {\it any}
value of the gravitino mass,
for which reason it was named no-scale supergravity~\cite{ELNT}.

Again, there is no time to discuss here details of 
the coupling of supergravity to matter~\cite{superHiggs}.
However, it is useful to have in mind the general 
features of the theory in the limit where
$\kappa \to 0$,  but the gravitino mass 
$m_{\tilde{G}} \equiv m_{3/2}$ remains fixed.  One
generally has non-zero gaugino masses $m_{1/2} 
\propto m_{3/2}$, and their universality is
quite generic.  One also has non-zero 
scalar masses $m_0 \propto m_{3/2}$, but their
universality is much more problematic, and 
even violated in generic string models.  It was
this failing that partly refuelled the renewed 
interest in the gauge-mediated models
mentioned in the previous section.  A 
generic supergravity theory also yields non-universal
trilinear soft supersymmetry-breaking couplings 
$A_{\lambda} \lambda \phi^3 : A_{\lambda}
\propto m_{3/2}$ and bilinear scalar 
couplings $B_{\mu} \mu \phi^2 : B_{\mu} \propto
m_{3/2}$.  Therefore, supergravity may generate the full menagerie of soft
supersymmetry-breaking terms:
\beq
- \frac{1}{2} \, \sum_a \, m_{1/2_a}\, \tilde{V}_a \tilde{V}_a - 
\sum_i \, m^2_{0_i} |\phi_i|^2 -
(\sum_{\lambda} A_{\lambda} \lambda \phi^3 + \mbox{h.c.})
- (\sum_{\mu} B_{\mu} \mu \phi^2 + \mbox{h.c.})
\label{fivetwentysix}
\eeq
 Since these are generated at the supergravity 
scale near $m_P \sim 10^{19}$ GeV, the soft
supersymmetry-breaking parameters are 
renormalized as discussed in Lecture 2.  The analogous
parameters in gauge-mediated models would 
also be renormalized, but to a different extent,
because the mediation scale $\ll m_P$. This 
difference may provide a signature of such
models, as discussed elsewhere~\cite{massdiff}.

Also renormalized is the vacuum energy 
(cosmological constant), which is a potential
embarassment.  Loop corrections in a 
non-supersymmetric theory are quartically divergent,
whereas those in a generic supergravity theory 
are only quadratically divergent, suggesting
a contribution to the cosmological constant 
of order $m^2_{3/2} m^2_P$, perhaps
$O(10^{-32})m^4_P$!  Particular models 
may have a one-loop quantum correction of order
$m^4_{3/2} = O(10^{-64})m^2_P$, but more 
magic (a new symmetry?) is needed to suppress the
cosmological constant to the required level 
(\ref{fivenine}).  This is one of the
motivations for tackling string theory, 
which is our only candidate for a fundamental
Theory of Everything including gravity.

\subsection{Problems of Gravity}

The greatest piece of unfinished business 
for twentieth-century physics is to reconcile
general relativity with quantum mechanics.  
There are aspects of this problem, one being
that of the cosmological constant, as 
discussed above.  Another is that of perturbative
quantum-gravity effects.  Tree-level graviton 
exchange in $2 \to 2$ scattering, such as 
$e^+e^- \to e^+e^-$ at LEP, has an 
amplitude $A_G \sim E^2/m^2_P$, and hence a cross section
\beq
\sigma_G \sim E^2/m^4_P
\label{fivetwentyseven}
\eeq
This is very small (negligible!) at LEP energies, 
reaching the unitarity limit only when $E
\sim m_P$.  However, when one calculates 
loop amplitudes involving gravitons, the rapid
growth with energy (\ref{fivetwentyseven}) 
leads to uncontrollable, non-renormalizable
divergences. These are of power type, and 
diverge faster and faster in higher orders of
perturbation theory.

There are also non-perturbative problems in 
the quantization of gravity, that first arose in
connection with black holes.  From the pioneering 
work of Bekenstein and Hawking~\cite{BekHawk} on
black-hole thermodynamics, we know that black 
holes have non-zero entropy $S$ and
temperature $T$, related to the Schwarzschild 
horizon radius.  This means that the quantum
description of a black hole should involve 
mixed states.  The intuition underlying this
feature is that information can be lost 
through the event horizon.  Consider, for example, a
pure quantum-mechanical pair state $|A,B> \equiv 
\sum_i c_i |A_i > |B_i >$ prepared near the
horizon, and what happens if one of the 
particles, say $A$, falls through the horizon while
$B$ escapes, as seen in Fig. 34.  In this case,
\beq
\sum_i \, c_i |A_i B_i > \to \sum_i |c_i|^2 |B_i > <B_i|
\label{fivetwentyeight}
\eeq
and $B$ emerges in a mixed state, as in Hawking's 
original treatment of the black-hole
radiation that bears his name~\cite{BekHawk}.

\begin{figure}
\centerline{\includegraphics[height=1.5in]{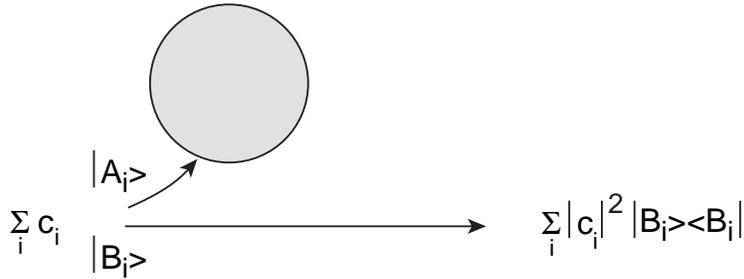}}
\caption[]{If a pair of particles $\vert A>~\vert B>$ is produced near
the horizon of a black hole, and one of them ($\vert A>$, say) falls in,
the remaining particle $\vert B>$ will appear to be in a mixed state.}
\end{figure}

The problem is that conventional quantum 
mechanics does not permit the evolution of a pure
initial state into a mixed final state.  This 
is an issue both for the quantum particles
discussed above and for the black hole 
itself.  We could imagine having prepared the black
hole by fusing massive or energetic particles 
in a pure initial state, e.g., by splitting a
laser beam and then firing the sub-beams at 
each other as in a laser device for inertial
nuclear fusion.

These problems point to a fundamental conflict 
between the proudest achievements of early
twentieth-century physics, quantum mechanics 
and general relativity.  One or the other
should be modified, and perhaps both.  
Since quantum mechanics is sacred to field theorists,
most particle physicists prefer to modify General 
Relativity by elevating it to string
theory~\cite{Green}.\footnote{It may be that this will eventually also
require a modification of the
{\it effective} quantum-mechanical space-time 
theory, even if the internal formulation of
string theory is fully quantum-mechanical, but that is another
story~\cite{EMN}.}

\subsection{Introduction to String Theory}

At the level of perturbation theory, the 
divergence difficulties of quantum gravity can be
related to the absence in a point-particle 
theory of a cutoff at short distances: for
example,
\beq
\int^{\Lambda \to \infty} \, d^4k \left (\frac{1}{k^2} \right )
\leftrightarrow \int_{1/ \Lambda \to 0} \, d^4x
\left ( \frac{1}{x^6} \right ) \sim \Lambda^2 \to \infty
\label{fivetwentynine}
\eeq
Such divergences can be alleviated or removed 
if one replaces point particles by extended
objects.  The simplest possibility is to 
extend in just one dimension, leading us to a
theory of strings.  In such a theory, instead 
of point particles describing one-dimensional
world lines, we have strings describing 
two-dimensional world sheets.  Most popular have
been closed loops of string, whose simplest 
world sheet is a tube.  The ``wiring diagrams"
generated by the Feynman rules of conventional 
point-like particle theories become the
``plumbing circuits" generated by the 
junctions and connections of these tubes of closed
string.  One could imagine generalizing this 
idea to higher-dimensional extended objects
such as membranes describing world volumes, 
etc., and we return later to this option.

On a historical note, string models first 
arose from old-fashioned strong-interaction
theory, before the advent of QCD.  The 
lowest-lying hadronic states were joined by a very
large number of excited states with increasing masses $m$ and spins $J$:
\beq
J = \alpha^\prime m^2
\nonumber
\eeq
where $\alpha^\prime$ was called the ``Regge slope".  
One interpretation of this spectrum was of
$\bar{q}q$ bound states in a linearly-rising 
potential, like an elastic string holding the
constituents together, with tension $\mu = 1/ 
\alpha^\prime$.  It was pointed out that such
an infinitely (?) large set of resonances in 
the direct $s$-channel of a scattering process
could be dual (equivalent) to the exchange of  
a similar infinite  set in the crossed
channel.  Mathematically, this idea was 
expressed by the Veneziano~\cite{Ven} amplitude for $2 \to 2$
scattering, and its generalizations to 
$2 \to n$ particle production processes.  Then it
was pointed out that these amplitudes 
could be derived formally from an underlying quantum
theory of string~\cite{Goddard}.  However, this first incarnation of
string theory was not able to
accommodate the point-like partons seen 
inside hadrons at this time - the converse of the
quantum-gravity motivation for string theory 
mentioned at the beginning of this section. 
Then along came QCD, which incorporated 
these point-like scaling properties and provided a
qualitative understanding of confinement, 
which has now become quantitative with the advent
of modern lattice calculations.  Thus string 
theory languished as a candidate model of the
strong interactions.

It was realized early on that unitarity 
required the existence of closed strings, even in an
{\it a priori} open-string theory.  Moreover, 
it was observed that the spectrum of a closed
string included a massless spin-2 particle, 
which was an embarrassment for a theory of the
strong interactions.  However, this led to the idea~\cite{TOE} of
reinterpreting string theory as a
Theory of Everything, with this massless 
spin-2 state interpreted as the graviton and the
string tension elevated to $\mu = O(m^2_P)$.

As already mentioned, one of the primary reasons 
for studying extended objects in connection
with quantum gravity is the softening of 
divergences associated with short-distance
behaviour.  Since the string propagates 
on a world sheet, the basic formalism is
two-dimensional.  Accordingly, string 
vibrations may be described in terms of left- and
right-moving waves:
\beq
\phi (r,t) \to \phi_L (r-t), \, \phi_R (r+t)
\label{fivethirty}
\eeq
If the string has no boundary, as for a 
closed string, the left- and right-movers are
independent.  When quantized, they may be 
described by a two-dimensional field theory.
Compared to a four-dimensional theory, it 
is relatively easy to make a two-dimensional field
theroy finite.  In this case, it has conformal 
symmetry, which has an infinite-dimensional
symmetry group in two dimensions.  However, 
as you already know from gauge theories, one must be careful to ensure
that this classical symmetry is not broken at 
the quantum level by anomalies.  If the quantum
string theory is to be consistent in a flat 
background space-time, the conformal anomaly
fixes the number of left- and right-movers 
each to be equivalent to 26 free bosons if the
theory has no supersymmetry, or 10 boson/fermion 
supermultiplets if the theory has $N = 1$
supersymmetry on the world sheet.  There 
are other important quantum consistency conditions,
and it was the demonstration by Green and Schwarz~\cite{GrSch} that
certain string theories are
completely anomaly-free that opened the 
floodgates of theoretical interest in string theory
as a Theory of Everything~\cite{Green}.

Among consistent string theories, one 
may enumerate the following.  The {\it Bosonic String}
exists in 26 dimensions, but this is not even its worst 
problem! It contains no fermionic matter
degrees of freedom, and the flat-space 
vacuum is intrinsically unstable. {\it Superstrings}
exist in 10 dimensions, have fermionic matter 
and also a stable flat-space vacuum.  On the
other hand, the ten-dimensional theory is 
left-right symmetric, and the incorporation of
parity violation in four dimensions is not 
trivial. The {\it Heterotic String}~\cite{heterotic} was originally
formulated in 10 dimensions, with parity 
violation already incorporated, since the left- and
right movers were treated differently.  
This theory also has a stable vacuum, but suffers
from the disadvantage of having too many 
dimensions. {\it Four-Dimensional Heterotic
Strings} may be obtained either by compactifying the six
surplus dimensions: $10 = 4 + 6$
compact dimensions with size $R \sim 1/m_P$~\cite{CY}, or by direct
construction in four dimensions,
replacing the missing dimensions by other 
internal degrees of freedom such as fermions~\cite{fourd} or
group manifolds or ...?  In this way it was 
possible to incorporate a GUT-like gauge group~\cite{AEHN}
or even something resembling the Standard Model~\cite{Faraggi}.

What are the general features of such string models? 
First, they predict there are no more
than 10 dimensions, which agrees with 
the observed number of 4! Secondly, they suggest that
the rank of the four-dimensional gauge group 
should not be very large, in agreement with the
rank 4 of the Standard Model!  Thirdly, the 
simplest four-dimensional string models do not
accommodate large matter representations~\cite{nobig}, 
such as an \underline{8} of SU(3) or a
\underline{3} of SU(2), again in agreement 
with the known representation structure of the
Standard Model! Fourthly, simple string models 
predict fairly successfully the mass of the
top quark.  This is because the maximum generic 
value of a Yukawa coupling $\lambda_t$ is of
the same order as the gauge coupling $g$.  
Applied to the top quark, this suggests that
\beq
m_t = \lambda_t <0 |H|0> = O(g) \times 250 \mbox{GeV}
\label{fivethirtyone}
\eeq
Moreover, the renormalization-group equation 
for $\lambda_t$ exhibits an approximate
infra-red fixed point, as seen in  Fig. 35.  
This means that a large range of Yukawa coupling
inputs at the Planck scale yield very similar 
physical values of $m_t \lappeq$ 190
GeV.  Fifthly, string theory makes a 
fairly successful prediction for the gauge
unification scale in terms of $m_P$.  If 
the intrinsic string coupling $g_s$ is weak, one
predicts~\cite{stringscale}

\begin{figure}
\centerline{\includegraphics[height=3in]{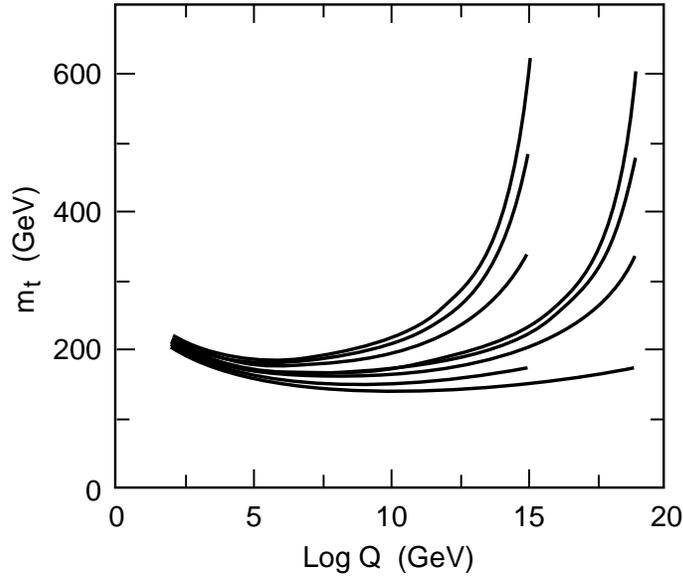}}
\caption[]{The approximate infra-red fixed point of the
renormalization-group equation for $m_t$ means that a wide range of input 
Yukawa couplings at the GUT or string scale lead to similar physical
values of $m_t \lappeq$ 190 GeV.} 
\end{figure}

\beq
M_{GUT} = O(g) \times \frac{m_P}{\sqrt{8 \pi}} \simeq
\mbox{few} \times 10^{17} \mbox{GeV}
\label{fivethirtytwo}
\eeq
where $g$ is the gauge coupling, which is 
${\cal O}(20)$ higher than the value calculated
from the bottom up in Lecture 4 on the basis 
of LEP measurement of the gauge couplings.  On
the one hand, it is impressive that the bottom-up 
extrapolation over 14 decades agrees to within 10
\% (on a logarithmic scale) with the top-down 
calculation (\ref{fivethirtyone}). 
Nevertheless, it would be nice to obtain 
closer agreement, and this provides the major
motivation for considering strongly-coupled 
string theory, which corresponds to a large
internal dimension $l > m^{-1}_{GUT}$, as we discuss next.

\subsection{Beyond String}

Current developments involve going beyond 
string to consider higher-dimensional extended
objects, such as generalized membranes with 
various numbers of internal dimensions.  These
may be obtained as solitons (non-perturbative 
classical solutions) of string theory, with
masses
\beq
m \propto {1\over g_s}
\label{fivethirtythree}
\eeq
analogously to monopoles in gauge theory.  
It is evident from (\ref{fivethirtythree}) that
such membrane-solitons become light in the 
limit of strong string coupling $g_s \to \infty$.

It was observed some time ago that there should 
be a strong-coupling/weak-coupling duality~\cite{MO} between
elementary excitations and monopoles in supersymmetric 
gauge theories.  These ideas have
recently been confirmed in a spectacular 
solution of $N = 2$ supersymmetric gauge theory in
four dimensions~\cite{SW}. Similarly, it has recently been shown that
there are analogous dualities in
string theory, whereby solitons in some 
strongly-coupled string theory are equivalent to
light string states in some other weakly-coupled 
string theory. Indeed, it appears that all
string theories are related by such dualities.  
A particularity of this discovery is that
the string coupling strength $g_s$ is 
related to an extra dimension, in such a way that its
size $R \to \infty$ as $g_s \to \infty$.  
This then leads to the idea of an underlying
11-dimensional framework called $M$ theory~\cite{Mtheory} 
that reduces to the different string theories in
different strong/weak-coupling linits, and reduces to eleven-dimensional
supergravity in the low-energy limit.

A particular class of string solitons called $D$-branes~\cite{Dbranes} 
offers a promising approach to the
black hole information paradox mentioned previously.  
According to this picture, black holes
are viewed as solitonic balls of string, 
and their entropy simply counts the number of
internal string states~\cite{EMN,Dcount}. These are in principle 
countable, so string theory may provide an
accounting system for the information contained 
in black holes.  Within this framework, the
previously paradoxical process (\ref{fivetwentyeight}) becomes
\beq
|A,B > + |BH > \to |B^\prime > + |BH^\prime >
\label{fivethirtyfour}
\eeq
and the final state is pure if the initial 
state was.  The apparent entropy of the final
state in (\ref{fivetwentyeight}) is now 
interpreted as entanglement.  The ``lost"
information is in principle encoded in 
the black-hole state, and this information could be
extracted if we measured all properties of this ball of string.

In practice, we do not know how to recover this 
information from macroscopic black holes, so
they appear to us as mixed states. What 
about microscopic black holes, namely fluctuations
in the space-time background with $\Delta 
E = O(m_P)$, that last for a period $\Delta  t =
O(1/m_P)$ and have a size $\Delta x = O(1/m_P)$?  
Do these steal information from us~\cite{Hawking}, or do
they give it back to us when they decay?  
Most people think there is no microscopic leakage
of information in this way, but not all of 
us~\cite{EMN} are convinced.  The neutral kaon system is
among the most sensitive experimental areas~\cite{EHNS,ELMN,CPLEAR} for
testing this speculative possibility.

A final experimental comment concerns the magnitude 
of the extra dimension in $M$ theory: LEP
data suggest that it may be relatively large, 
with size $L_{11} \gg 1/m_{GUT} \simeq 1/10^{16}~
\mbox{GeV} \gg 1/m_P$~\cite{Horava}.  Remember that the na\"\i ve string
unification scale
(\ref{fivethirtytwo}) is about 20 times 
larger than $m_{GUT}$.  This may be traced to the
fact that the gravitational interaction strength, 
although growing rapidly as a power of
energy (\ref{fivetwentyseven}), is 
still much smaller than the gauge coupling strength at
$E = m_{GUT}$.  However, if an extra space-time 
dimension appears at an energy $E <
m_{GUT}$, the gravitational interaction 
strength grows fast, as indicated in Fig. 36. 
Unification with gravity around $10^{16} \mbox{GeV}$ then becomes
possible, {\it if} the gauge couplings do not also acquire a
similar higher-dimensional kick.  Thus we are led to the 
startling capacitor-plate framework for fundamental physics shown in
Fig.~37.

\begin{figure}
\centerline{\includegraphics[height=3in]{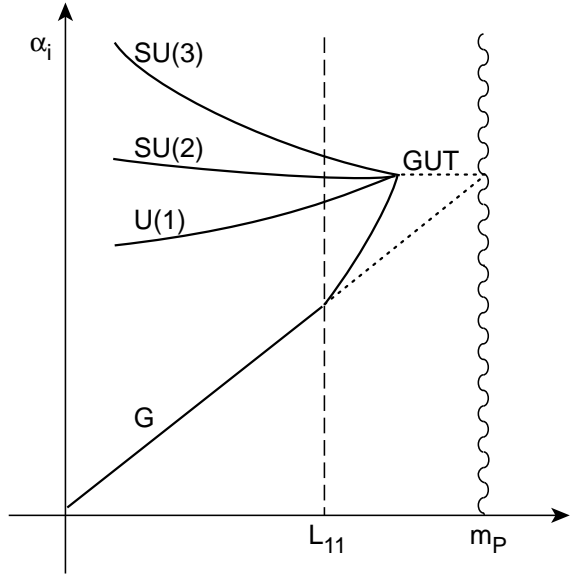}}
\caption[]{Sketch of the possible evolution of the gauge couplings and
the gravitational coupling $G$: if there is a large fifth dimension with
size $\gg m^{-1}_{GUT}$, $G$ may be unified with the gauge couplings at
the GUT scale.}
\end{figure}

\begin{figure}
\centerline{\includegraphics[height=2in]{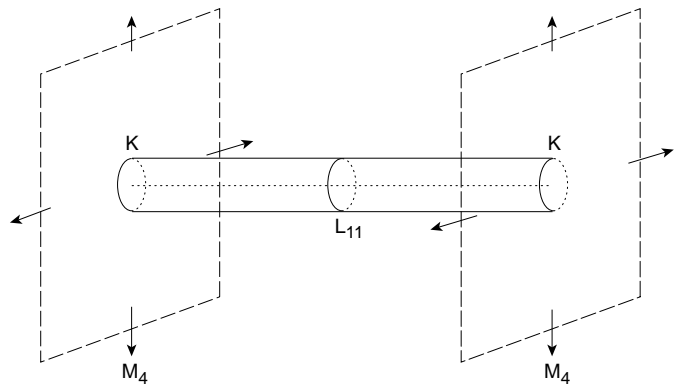}}
\caption[]{The capacitor-plate scenario favoured in
eleven-dimansional $M$ theory. The eleventh dimension has a size
$L_{11} \gg M_{GUT}^{-1}$, whereas dimensions $5, ... , 10$ are
compactified on a small manifold $K$ with characteristic size
$\sim M_{GUT}^{-1}$. the remaining four dimensions form
(approximately) a flat Minkowski space $M_4$.}
\end{figure}

Each plate is {\it a priori} ten-dimensional, 
and the bulk space between then is {\it a
priori} eleven-dimensional.  Six dimensions 
are compactified on a scale $L_6 \sim
1/m_{GUT}$, leaving a theory which is 
effectively five-dimensional in the bulk and
four-dimensional on the walls.  Conventional 
gauge interactions and observable matter
particles are hypothesized to live on one 
capacitor plate, and there are other hidden gauge
interactions and matter particles living on 
the other plate.  The fifth dimension has a
characteristic size which is estimated to 
be $O(10^{12}~ \mbox{to}~ 10^{13}~ \mbox{GeV}
)^{-1}$, and physics at large distances 
(smaller energies) looks effectively
four-dimensional.  Supersymmetry breaking 
is expected to originate on the hidden capacitor
plate in this scenario, and to be transmitted 
to the observable wall by gravitational
strength interactions in the bulk~\cite{ELPP}.

The phenomenological richness of this speculative $M$-theory 
approach is only beginning to be explored,
and it remains to be seen whether it offers a 
realistic phenomenological description. 
However, it does embody all the available 
theoretical wisdom as well as offering the
prospect of unifying all the observable gauge 
interactions with gravity at a single effective scale
$\sim m_{GUT}$.  As such, it constitutes our 
best contemporary guess about the Theory of
Everything within and Beyond the Standard Model.

\end{document}